\documentclass[twocolumn,aps,superscriptaddress,floatfix,prc]{revtex4-1}
\usepackage{mathrsfs}
\usepackage{amssymb}
\usepackage{amsmath}
\usepackage{graphicx}
\usepackage[normalem]{ulem}
\usepackage[dvips]{color}
\usepackage{bm}
\usepackage{longtable}
\usepackage{slashed}

\usepackage{times}

\renewcommand\sout{\bgroup \color{red} \ULdepth=-.5ex \ULset}

\renewcommand{\v}[1]{\textbf{#1}}
\renewcommand{\rm}[1]{\textrm{#1}}
\renewcommand{\d}{\mathrm{d}}

\begin{document}

\title{Relativistic self-energy decomposition of nuclear symmetry energy\\ and
equation of state of neutron matter within QCD sum rules}

\author{Bao-Jun Cai}\email{bjcai87@gmail.com}
\affiliation{School of Physics and Astronomy and
Shanghai Key Laboratory for Particle Physics and Cosmology, Shanghai
Jiao Tong University, Shanghai 200240, China}
\affiliation{Quantum Machine Learning Laboratory, Shadow Creator Inc., Shanghai 201206, China}

\author{Lie-Wen Chen}\email{Corresponding author: lwchen$@$sjtu.edu.cn} \affiliation{School of
Physics and Astronomy and Shanghai Key Laboratory for Particle
Physics and Cosmology, Shanghai Jiao Tong University, Shanghai
200240, China}

\date{\today}

\begin{abstract}
Properties of the relativistic nucleon self-energy
decomposition of the symmetry energy as well as the equation of state (EOS) of pure neutron matter (PNM)
are explored systematically within the QCD sum rules (QCDSR).
Our main conclusions are: 1). The five self-energy decomposition terms of the symmetry energy according to the nucleon Lorentz structure
are carefully studied, leading to the conclusion that the symmetry energy increases as the nucleon sigma term $\sigma_{\rm{N}}$
increases and the contributions to the symmetry energy due to the momentum dependence of the
self-energies in symmetric nuclear matter (SNM) are very small compared with those from other decomposition terms.
2). A smaller strange quark mass is found to generate a larger
symmetry energy, and this correlation is useful for understanding the origins of the uncertainties on the (nucleonic matter) symmetry energy from quark level.
3). The EOS of PNM at low densities can be effectively approximated by $E_{\rm{n}}(\rho)
\approx E_{\rm{n}}^{\rm{FFG}}(\rho)+({M\rho}/2{\langle\overline{q}q\rangle_{\rm{vac}}})[(1-\xi)({\sigma_{\rm{N}}}/{2m_{\rm{q}}})-5]$ which depends only on several physical quantities such as $m_{\rm{q}},\sigma_{\rm{N}}$ and $\langle\overline{q}q\rangle_{\rm{vac}}$,
and this formula already has predictive power and the results are found to be consistent with those from other celebrated microscopic many-body theories at low densities.
4). The higher order density terms in quark condensates are
shown to be important to describe the empirical EOS of PNM in the density region around and
above nuclear saturation density, and these higher order density terms are also found to hinder the appearance of chiral symmetry restoration in PNM
at high densities.
5). The symmetry energy is shown to depend strongly on the five-dimensional
condensate $\langle g_{\rm{s}}q^{\dag}\sigma \mathcal{G}q\rangle_{\rho,\delta}$, providing a useful approach to explore
the symmetry energy through knowledge on the condensates which can be extracted from hadronic physics.
6). The twist-four four-quark condensates are shown to have significant effects on the EOS of both SNM and PNM but have minor effects on the symmetry energy, and combined with the analyses on the effects of the higher order density terms in the chiral condensates,
three parameter sets of QCDSR are constructed and they are shown to be able to describe the EOS of PNM and the symmetry energy within a wide range of densities.
Our results in the present work demonstrate that the QCDSR approach can provide a useful way to understand the properties of dense nucleonic matter from non-perturbative QCD vacuum.
\end{abstract}

\maketitle


\setcounter{equation}{0}
\section{Introduction}

The investigation of the equation of state (EOS) of isospin asymmetric
nucleonic matter (ANM) from quantum
chromodynamics (QCD) or QCD-based effective theories is one of  longstanding issues in nuclear physics\,\cite{Bra14,ACCNP}.
The exact knowledge on the EOS of ANM provides important information on the
in-medium nucleon-nucleon effective interactions, which play a central role, for instance,
in understanding the structure and decay properties of finite nuclei\,\cite{Rin80,Bla80,Ben03}, the dynamical processes in nuclear
reactions\,\cite{LiBA98,Dan02,Bar05,Ste05,Che07a,
LCK08,Tsa12}, the structure and
evolution of neutron stars as well as the mechanism of core-collapse supernova
explosion\,\cite{Gle00,Lat04,Lat12,SNRev,Hem12,Oze16,Wat16,Oer17,Bal16,Cow19},
and the gravitational waves from binary neutron star merger or black-hole neutron star merger
in the multimessenger era\,\cite{Abb16, Abb17, Bau12,Tak14,Ber15,Bos18,Fat18,Abb18,Mos18,
De18,Abb19,Bru18,Bai17,Due19,NZhang18,NBZhang19,NBZhang19a,ZhouY19}.
Conventionally, the EOS of
ANM defined as the binding energy per nucleon, is expanded around the symmetric
nucleonic matter (SNM) as $E(\rho,\delta)\approx
E_0(\rho)+E_{\rm{sym}}(\rho)\delta^2+\mathcal{O}(\delta^4)$ where
$\rho=\rho_{\rm{n}}+\rho_{\rm{p}}$ and
$\delta=(\rho_{\rm{n}}-\rho_{\rm{p}})/\rho$ are respectively the nucleon density
and isospin asymmetry of the system in terms of the neutron and
proton densities $\rho_{\rm{n}}$ and $\rho_{\rm{p}}$.
In the above expansion, $E_0(\rho)\equiv E(\rho,0)$ is the EOS of
SNM, while $E_{\rm{sym}}(\rho)$ is the nuclear symmetry energy. The symmetry energy $E_{\rm{sym}}(\rho)$ can be generally decomposed into kinetic and potential parts, i.e.,
$E_{\rm{sym}}(\rho)=E_{\rm{sym}}^{\rm{kin}}(\rho)+E_{\rm{sym}}^{\rm{pot}}(\rho)$. Moreover, the potential part $E_{\rm{sym}}^{\rm{pot}}(\rho)$ can be further decomposed generally into several
terms originated from the Lorentz structure of the nucleon
self-energies\,\cite{Cai12}. A thorough understanding on the
origin and properties of each part of the symmetry energy is useful
in both nuclear physics and astrophysics\,\cite{EPJA}. For instance, in simulating
heavy-ion reactions using transport models one needs as an input the
potential symmetry energy of
quasi-nucleons\,\cite{LCK08,LiBA97,ChenLW05,Che14,Li15,Cai16,RWang19}, while the kinetic
symmetry energy is found to strongly affect the critical formation
density of the resonance state $\Delta(1232)$ in neutron
stars\,\cite{Cai15a,Dra14}.
Furthermore, it has been widely discussed recently that the kinetic symmetry energy is closely related to the nucleon-nucleon short range correlations (SRC)\,\cite{Hen17,Due18,Sch19,Tor18,Wei18,Wei19,Fom17,CXu11x,CXu13x,
Hen15b,CaiLi15a,CaiLi16a,CaiLi16b,CaiLi17,Yong17,Dut18}, indicating the very fundamental nature of the symmetry energy.
After hard efforts made in the last few decades, the magnitude of the total symmetry energy at the nuclear saturation density
$\rho_0\approx0.16\,\rm{fm}^{-3}$ is now best known to be around
$32\pm3\,\rm{MeV}$\,\cite{Che12a,LiBA13,Oer17,BALi17xx,Che17xx}.

The determination on the EOS of ANM based on the phenomenological approaches such as
the non-relativistic Skyrme--Hartree--Fock (SHF)
models\,\cite{Sto06,Cha09,Zha16xx,RWang18} and the relativistic mean field (RMF)
models\,\cite{Ser86,Ser97,Rei89,Rin96,Men06,ChenLW07} has made great success in recent years. However, the origin
of each individual term of the EOS in
these phenomenological approaches is usually blurry. For example, the contribution from the $\rho$ meson to the symmetry
energy largely depends on whether or not the $\delta$ meson and the cross interactions between the $\rho$ meson and isoscalar mesons are
included in the model for the nonlinear RMF model\,\cite{Cai12}, indicating the contribution of the
$\rho$ meson to the symmetry energy is only effective. On the other hand, from the viewpoint
of many body theories\,\cite{Ser86}, the decomposition and analyses
of a quantity via the Green's functions or more precisely, according to
the Lorentz structure of the quantity itself, is physical.
In this sense, any
effective method, especially that based on the QCD, encapsulating the proper Lorentz structure of the nucleon self-energies to investigate the EOS of ANM will be appealing.

The QCD sum rules (QCDSR) method\,\cite{Shi79a,Shi79b,Shi79c} provides an important
non-perturbative QCD approach to explore the properties of nucleonic matter (see, e.g., refs.\,\cite{Dru90}).
Intuitively, when the QCD coupling constant is small at high
energies/small distances, the theory becomes asymptotically free,
guaranteeing the applicability of perturbative calculations. As the
energy scale decreases, the coupling constant of the theory becomes
large, perturbative methods break down eventually and
non-perturbative effects emerge. Among these effects, the most
important is the appearance of the quark/gluon condensates. The QCDSR is actually based on some duality relations. More specifically,
the basic idea of QCDSR for nucleonic
matter calculations\,\cite{Dru90,Coh91,Coh92,Fur92,Jin93,Jin94,Dru04,
Dru12,Dru17a,Dru17b,Dru19,Jeo13,Jeo16,Mar18,Son18,
Coh95,Rei85,Nar89,Shi92,Dos94,Lei97,Nar04,Iof06,Iof11,Gub13} is to relate the condensates to the nucleon
self-energies using the operator product expansion (OPE) technique\,\cite{Wil69},
where information on the nucleon self-energies is introduced via nucleon-nucleon
correlation functions. On the other hand, the EOS of ANM can be obtained through the
self-energy decomposition by analyzing the general Lorentz
structure of the nucleon self-energies\,\cite{Cai12}.
Within the QCDSR {method}, the exact information on the nucleon self-energies and
nucleonic matter EOS can thus provide useful constraints on the in-medium quark/gluon
condensates and vice verse. It should be noted that the in-medium quark
condensates provide an order parameter of spontaneous
chiral symmetry breaking in QCD.
Practically, the QCDSR method is expected to work well at lower
densities/momenta where effects of the complicated and largely unknown high mass-dimensional
condensates as well as continuum effects are not important.

In this work,
we use the QCDSR method to investigate the EOS of isospin asymmetric nucleonic matter through the nucleon self-energy decomposition formulae\,\cite{Cai12} based on the Hugenholtz--Van Hove (HVH) theorem\,\cite{Hug58}, and especially focus on the nuclear symmetry energy and the EOS of pure neutron matter (PNM).
It is necessary to point out that the symmetry energy was also studied recently by the QCDSR through a specific approach mapping the nucleon self-energies to the symmetry energy\,\cite{Jeo13}. Moreover, some main results on the EOS of PNM via the QCDSR were already reported in ref.\,\cite{Cai17QCDSR}, and more details of the calculations in ref.\,\cite{Cai17QCDSR} will be given in the present work.

Before going to present the details on the QCDSR method and its application in EOS of ANM, we would like to give a brief overview on several celebrated or potentially useful QCD-based approaches to the study on the EOS of ANM.
These approaches include:
1). Chiral perturbation theory (ChPT)\,\cite{Wei90,Wei91,Wei92,Gas84,Gas85,Gas88,Mei93,
Ber95,Eck95,Pic95,Bed02,Epe06,Ber07,Bij07,
Bur07,Ber08,Epe09,Bog10,Mac12,Sch12,Hol13,Ham13,Heb15,Hol16}.
The basic idea of the ChPT is based on the general effective field
theory\,\cite{Geo93,Wei79,Pol92} and the ChPT has become a well developed tool for
systematically dealing with the EOS, e.g., of PNM\,\cite{Lut02,Kai02,Heb10,Lac11,Tew13,
Kru13,Kru13a,Dri14,Dri16,Hag14,Hol17,Eks18,Dri19}.
However, due to the large uncertainties, e.g., on the many-nucleon
forces\,\cite{Ham13,Heb15} and the nature of the method itself, the applicability of this approach is essentially
limited to the relatively low density region.
2). Purturbative
QCD (pQCD)\,\cite{Col11,Dok91,Ste95}. The pQCD approach is often used to explore the
high-density behavior of the dense nucleonic/quark
matter\,\cite{Gro81,Shu84,Sch98,Alf98,Rap98,Sho03,Hua03,Alf08,Fuk11,Fre77a,
Fre77b,Fre77c,Bal78,Kap79,Kur10,Jeo16b},
where the QCD coupling constant
$\alpha_{\rm{s}}=g_{\rm{s}}^2/4\pi$ is generally small, indicating the possibility of the
perturbative schemes applied in dense matter at extremely high densities\,\cite{Kap06,Bel96}.
3). The large $N_{\rm{c}}$
method\,\cite{Hoo74,Hoo74a,Wit79,Wit79a,Col85,Man98}. The QCD
becomes solvable when the color number $N_{\rm{c}}$ is large. This is an effective field theory based on the expansion
of $N_{\rm{c}}^{-1}$, and by using the
techniques developed for large $N_{\rm{c}}$ theories,
information on the in-medium nucleon self-energies can
be obtained\,\cite{Coh11}, which is potentially useful for the further study on
nuclear matter EOS.
4). Method of Dyson-Schwinger
equation (DSE)\,\cite{Dys49,Sch51,Sch51a,Rob94,Rob00,Liu01,LiuY02,Yua06,Bas12,Zon05,Zon08}.
The DSE method is one of non-perturbative approaches, in which the resummation
techniques\,\cite{Bra90,Bra90a,Fen90} are adopted to make effective
approximations to the QCD Lagrangian.
5). Functional renormalization group (FRG) approach\,\cite{Wet93,Pol84}.
The FRG method has been applied to investigate the EOS of ANM\,\cite{Dre13,Dre14,Dre15,Dre17}, where
the quantum fluctuations are included
non-perturbatively\,\cite{Bag01,Ber02,Pol03,Ros12} via the
renormalization group techniques\,\cite{Pol84,Dre13,Dre14,Dre15,Dre17}.
6). The Skyrme model\,\cite{Sky61}.
In the Skyrme model, the nucleons are described as a topological solitons (i.e., skyrmions) in a meson field theory and thus the nucleon properties in nuclear matter can be obtained\,\cite{Kal97,Rak98}, which can be further used to obtain nuclear matter EOS\,\cite{Lee11,Ma17}.
7). Last but not least, numerical approaches such as lattice
QCD\,\cite{EPJA1,Lee08,Lee17}, quantum Monte Carlo (QMC)
simulations\,\cite{Car15,Gan15} and several different types of
ab-initia methods\,\cite{Dru10,Her16,Naz16} are important for the
understanding of the root properties of finite nuclei and the dense
nucleonic matter.

Early successes of QCDSR in nucleonic matter calculations can be traced back to the
prediction on the large nucleon Lorentz covariant self-energies\,\cite{Coh91}. The present work is a natural generalization to the
investigation on the EOS of ANM.
Besides the prediction on the EOS of PNM by the QCDSR method reported earlier\,\cite{Cai17QCDSR}, as we shall see, the present
results on the nuclear symmetry energy obtained via the QCDSR method are also found to be consistent
with predictions by other state-of-the-art microscopic many body theories,
indicating that QCDSR can be applied to explore the EOS of ANM
in a quantitative manner.
The QCDSR method thus establishes a connection between the EOS of ANM and the non-perturbative QCD vacuum.

The paper is organized as follows.
Section~\ref{SEC_EOS}
gives a general description of the EOS of ANM and the Lorentz structure of nucleon self-energies in ANM.
In Section~\ref{SEC_PhysFou},
the physical foundation of the QCDSR is introduced. In particular, the techniques on the calculation of the Wilson's coefficients and the
properties of the condensates used in this work are given.
Section~\ref{SEC_MassVacuum}
is devoted to the nucleon mass in vacuum within the QCDSR method, and the major task of this section is to determine the Ioffe parameter $t$ through the physical nucleon mass in vacuum, which is a starting point of the following investigation on the EOS of ANM.
In Section~\ref{SEC_EsymStru},
the self-energy structure of the symmetry energy is analyzed via the simplified QCDSR, and the high order symmetry energy effects on the EOS of ANM are briefly discussed.
Then Section~\ref{SEC_HOEsti}
studies these high order symmetry energy effects in some detail.
Section~\ref{SEC_MassVacuum}, Section \ref{SEC_EsymStru}, and Section \ref{SEC_HOEsti} are mainly qualitative,
and are included to reveal some important features on the symmetry energy and nuclear matter EOS from QCDSR. Several important analytical
expressions are given in these three sections.
In Section~\ref{SEC_FullQSR},
a full calculation on the EOS of SNM, the symmetry energy and the EOS of PNM is given, where the first parameter set QCDSR-1 (naive QCDSR) is constructed.
In Section~\ref{SEC_HighCond},
the effects of the higher order density terms in the chiral condensates on the symmetry energy and the PNM EOS are explored, and as a result, the second parameter set, i.e., QCDSR-2, is given.
In Section~\ref{SEC_TWIST4},
the contribution from the twist-four four-quark condensates to nucleonic matter EOS is studied with the third parameter set QCDSR-3 constructed.
Section~\ref{SEC_Summary}
gives the summaries and outlook of the work.

\setcounter{equation}{0}
\section{EOS of ANM and Nucleon self-energies}\label{SEC_EOS}

In this section, we give a general description of the EOS of ANM
and its relation to the in-medium nucleon self-energy structure in
its relativistic form.

\subsection{Definition of EOS of ANM}

The EOS of ANM can be obtained through the total energy density
$\varepsilon(\rho,\delta)$ by $ E(\rho ,\delta )={\varepsilon(\rho
,\delta )}/{\rho }-M$ where $M=0.939\,$GeV is the nucleon rest mass. Moreover, the $E(\rho, \delta)$ can be
expanded as a power series of even-order terms in $\delta$ as
\begin{equation}
E(\rho ,\delta )\approx E_{0}(\rho )+E_{\text{sym}}(\rho )\delta ^{2}+%
\mathcal{O}(\delta ^{4}), \label{EOSpert}
\end{equation}%
where $E_{0}(\rho )$ is the EOS of SNM,
and the symmetry energy $E_{\mathrm{sym}}(\rho )$
is expressed as
\begin{eqnarray}
E_{\mathrm{sym}}(\rho ) &=&\left. \frac{1}{2!}\frac{\partial ^{2}E(\rho
,\delta )}{\partial \delta ^{2}}\right\vert _{\delta =0}.
\label{Esym}
\end{eqnarray}
Around the saturation density $\rho _{0}$,
the $E_{0}(\rho )$ can be expanded, e.g., up to 2nd-order in
density, as,
\begin{equation}
E_{0}(\rho )\approx E_{0}(\rho _{0})+\frac{1}{2}K_0\chi
^{2}+\mathcal{O}(\chi ^{3}), \label{DenExp0}
\end{equation}%
where $\chi =(\rho -\rho _{0})/3\rho _{0} $ is a dimensionless
variable characterizing the deviations of the density from the
saturation point $\rho _{0}$. The first term $E_{0}(\rho _{0})$ on
the right-hand-side of Eq.\,(\ref{DenExp0}) is the binding energy per
nucleon in SNM at $\rho_{0}$ and the coefficient $K_{0}$ of the second term
is
\begin{equation}
K_{0} =\left. 9\rho _{0}^{2}\frac{\text{d} ^{2}E_{0}(\rho
)}{\text{d} \rho ^{2}}\right\vert _{\rho =\rho _{0}},
\end{equation}%
which is the so-called incompressibility coefficient of SNM.

Similarly, one can expand the $E_{\mathrm{sym}}(\rho )$ around an
arbitrary reference density $\rho _{\text{r}}$ as (see, e.g., ref.\,\cite{Zha13})
\begin{equation}
E_{\text{sym}}(\rho )\approx E_{\text{sym}}({\rho
_{\text{r}}})+L(\rho
_{\text{r}})\chi_{\text{r}}+\mathcal{O}(\chi_{\text{r}} ^{2}),
\label{EsymLKr}
\end{equation}
with $\chi_{\text{r}}=(\rho -\rho _{\text{{r}}})/3\rho _{\text{r}}$,
and the slope parameter of the nuclear symmetry energy at $\rho
_{\text{r}}$ is expressed as
\begin{equation}\label{def_L}
L(\rho_{\text{r}}) =\left.3\rho_{\text{r}}
\frac{\text{d}E_{\mathrm{sym}}(\rho )}{\text{d}\rho } \right|_{\rho
=\rho_{\text{r}} }.
\end{equation}
For $\rho_{\text{r}} = \rho_0$, the $L(\rho_{\text{r}})$ is reduced
to the conventional slope parameter $L\equiv 3\rho_0
{\text{d}E_{\mathrm{sym}}(\rho )}/{\text{d}\rho }|_{\rho =\rho_0}$.

\subsection{Lorentz structure of nucleon self-energies}

For the translational and rotational invariance, parity conservation,
time-reversal invariance, and hermiticity in the rest frame of
infinite nucleonic matter, the nucleon self-energy may be
written generally in the relativistic case
as\,\cite{Wal74,Mil74,Jam81,Hor83,Hor87,Ser86},
\begin{align}
\Sigma(|\v{k}|,k^0)=&\Sigma_{\rm{S}}(|\v{k}|,k^0)
-\gamma_{\mu}\Sigma^{\mu}(|\v{k}|,k^0)\notag\\
=&\Sigma_{\rm{S}}(|\v{k}|,k^0)+\gamma^0\Sigma_{\rm{V}}(|\v{k}|,k^0)\notag\\
&+\vec{\gamma}\cdot\v{k}^0\Sigma_{\rm{K}}(|\v{k}|,k^0),\label{DefSE2}
\end{align}
where the isospin and density dependence of the nucleon self-energy are suppressed. The
quantities $\Sigma_{\rm{K}}(|\v{k}|,k^0)$,
$\Sigma_{\rm{S}}(|\v{k}|,k^0)$ and
$\Sigma_{\rm{V}}(|\v{k}|,k^0)\equiv-\Sigma^0(|\v{k}|,k^0)$ are
Lorentz (rotational) scalar functions of $|\v{k}|$ and $k^0$ (the
Minkovski metric is selected as $g_{\mu\nu}=(+,-,-,-)$ in the
present work), $\v{k}^0=\v{k}/|\v{k}|$ is the unit vector along the
direction of the momentum $\v{k}$. In the rest frame of infinite
nucleonic matter, these invariants can be expressed in terms of $k^0$,
$|\v{k}|$ (and $\rho$ as well as isospin $\delta$). The general proof of the decomposition
Eq.~(\ref{DefSE2}) can be found in ref.\,\cite{Ser86}.

The effects of interactions between nucleons on the propagation of a
nucleon in the medium can be included to all orders via Dyson's
equation\,\cite{Fet71}, i.e.,
\begin{equation}\label{DysonEq}
G(k)=G^0(k)+G^0(k)\Sigma(k)G(k)
\end{equation} where $G^0(k)$ is the noninteracting nucleon Green's function (propagator) and $\Sigma$ is the proper self-energy.
Eq.\,(\ref{DefSE2}) and Eq.\,(\ref{DysonEq}) are completely general
in the rest frame of infinite matter (in this case, the
nucleon current-density four-vector has only a time-like
non-vanishing component) and in principle could be used to determine $G$
exactly. Dyson's equation Eq.\,(\ref{DysonEq}) can be solved
formally, yielding
\begin{equation}\label{SolDysonEq}
G^{-1}(k)=\gamma_{\mu}[k^{\mu}+\Sigma^{\mu}(k)]-[M+\Sigma_{\rm{S}}(k)].
\end{equation}
The location of the poles in $G(k)$ may be specified using the
modified Feynman diagrams approach.

By defining the (Dirac) effective mass as well as the effective
four-momentum of a nucleon, i.e.,
\begin{align}
M^{\ast}=&M+\Sigma_{\rm{S}}(k),\label{DefEffMass}\\
\v{k}^{\ast}=&\v{k}\left[1+{\Sigma_{\rm{K}}(k)}/{|\v{k}|}\right],\label{DefEffMom}\\
e^{\ast}=&[{\v{k}^{\ast,2}+M^{\ast,2}}]^{1/2},\label{DefEffSiEn}\\
k^{\ast\mu}=&k^{\mu}+\Sigma^{\mu}(k)\equiv\left[k^0+\Sigma^0(k),\v{k}^{\ast}\right],
\label{DefEffDis}
\end{align}
one can rewrite the solution of Eq.\,(\ref{DysonEq}) in a compact
form, i.e., $ G(k)=G_{\rm{F}}(k)+G_{\rm{D}}(k) $, where
\begin{equation}
G_{\rm{F}}(k)=\frac{\gamma^{\mu}k_{\mu}^{\ast}+M^{\ast}}{k^{\ast,2}-M^{\ast,2}+i0^+},\label{SolDysonEqF}
\end{equation}and
\begin{equation}
G_{\rm{D}}(k)=\frac{i\pi}{e^{\ast}}(\gamma^{\mu}k_{\mu}^{\ast}+M^{\ast})
\delta(k^0-e)\Theta(k_{\rm{F}}-|\v{k}|).\label{SolDysonEqD}
\end{equation}
$G_{\rm{F}}$ and $G_{\rm{D}}$ are two parts originated from the
Pauli exclusion principle and the propagation of real nucleons in
the Fermi sea in the interacting nucleonic matter,
respectively\,\cite{Hor83, Hor87}.

The total single particle energy $e$ can be obtained from the dispersion relation,
i.e., the solution of the following transcendental equation
\begin{align}\label{DisRela}
e=&[e^{\ast}+\Sigma_{\rm{V}}(k)]_{k^0=e}\notag\\
=&\left[\left[\v{k}+\v{k}^0\Sigma_{\rm{K}}(|\v{k}|,e)\right]^2
+\left[M+\Sigma_{\rm{S}}(|\v{k}|,e)\right]^2\right]^{1/2}\notag\\
&+\Sigma_{\rm{V}}(|\v{k}|,e),
\end{align}
which evidently depends on $|\v{k}|$, the density $\rho$ and the
energy $e$ itself. The above results are valid for any
approximations to the self-energy in infinite matter. In order to
arrive the Hartree--Fock approximation, for example, we include in $\Sigma$ only
the contributions from tadpole and exchange diagrams\,\cite{Fet71,Hor83, Hor87,Ser86}.
Moreover, if the self-energy has no explicit energy dependence, then one obtains
\begin{align}
e(|\v{k}|)=&\left[\left[\v{k}+\v{k}^0\Sigma_{\rm{K}}(|\v{k}|)\right]^2
+\left[M+\Sigma_{\rm{S}}(|\v{k}|)\right]^2\right]^{1/2}\notag\\
&+\Sigma_{\rm{V}}(|\v{k}|).
\end{align} When the above
expression is generalized to ANM with any isospin asymmetry $\delta$, we then have
\begin{align}\label{DisRelaANM1}
e_J(\rho,\delta,|\v{k}|)=&\Big[\left[\v{k}+\v{k}^0\Sigma_{\rm{K}}^J(\rho,\delta,|\v{k}|)\right]^2\notag\\
&+\left[M+\Sigma_{\rm{S}}^J(\rho,\delta,|\v{k}|)\right]^2\Big]^{1/2}
+\Sigma_{\rm{V}}^J(\rho,\delta,|\v{k}|),
\end{align}
where the isospin and density dependence of the quantity is recovered. Due to
the general smallness of
$\Sigma_{\rm{K}}^J(\rho,\delta,|\v{k}|)$\,\cite{Plo06}, we will
neglect this term in the following study. Consequently, the single nucleon energy is given by
\begin{align}\label{DisRelaANM}
e_J(\rho,\delta,|\v{k}|)=&\sqrt{|\v{k}|^2
+\left[M+\Sigma_{\rm{S}}^J(\rho,\delta,|\v{k}|)\right]^2}\notag\\
&+\Sigma_{\rm{V}}^J(\rho,\delta,|\v{k}|).
\end{align}

In the present work, the EOS of ANM is obtained by the formulae
based on the Hugenholtz--Van Hove (HVH) theorem\,\cite{Hug58}. More specifically\,\cite{Cai12},
\begin{align}
E_0(\rho)=&\frac{1}{\rho}\int_0^{\rho}\d\rho\left(e_{\rm{F}}^{\ast}+\Sigma_{\rm{V}}^0\right)-M\label{sec1_SNM},\\
E_{\rm{n}}(\rho)=&\frac{1}{\rho}\int_0^{\rho}\d\rho\left(e_{\rm{F,n}}^{\ast}+\Sigma_{\rm{V}}^{\rm{n}}\right)-M\label{sec1_PNM},\\
E_{\rm{sym}}(\rho)=&\frac{k_{\rm{F}}^2}{6e_{\rm{F}}^{\ast}}
+\frac{k_{\rm{F}}}{6}\left(\frac{M_0^{\ast}}{e_{\rm{F}}^{\ast}}\frac{\d\Sigma_{\rm{S}}^0}{\d|\v{k}|}
+\frac{\d\Sigma_{\rm{V}}^0}{\d|\v{k}|}\right)_{|\v{k}|=k_{\rm{F}}}\notag\\
&+\frac{1}{2}\left(\frac{M_0^{\ast}}{e_{\rm{F}}^{\ast}}\Sigma^{\rm{S}}_{\rm{sym}}+
\Sigma^{\rm{V}}_{\rm{sym}}\right),\label{sec1_Esym}
\end{align}
where ``0'' denotes SNM, i.e.,
$\Sigma_{\rm{S(V)}}^{0/\rm{n}}$ is the scalar (vector) self-energy in SNM/PNM,
$e_{\rm{F}}^{\ast}=({M_0^{\ast,2}+k_{\rm{F}}^2})^{1/2}=[(M+\Sigma_{\rm{S}}^0)^2+k_{\rm{F}}^2]^{1/2}$, $E_{\rm{n}}$ is the EOS of PNM with
$e_{\rm{F,n}}^{\ast}=(M_{\rm{n}}^{\ast,2}+k_{\rm{F,n}}^2)^{1/2}$, here $M_{\rm{n}}^{\ast}$ is the neutron effective mass in PNM.
Moreover, $k_{\rm{F}}=\left({3\pi^2\rho}/{2}\right)^{1/3}$ ($k_{\rm{F,n}}=2^{1/3}k_{\rm{F}}$) is the
Fermi momentum in SNM (PNM),
$\Sigma_{\rm{sym}}^{\rm{S/V}}\equiv \Sigma_{\rm{sym},1}^{\rm{S/V}}$
is the first-order symmetry self-energy\,\cite{Cai12}.
The main task of the present work is to explore the density/momentum dependence of the $\Sigma_{\rm{S/V}}^J(\rho,\delta,|\v{k}|)
$ via the QCDSR method, and obtain the EOS of SNM, the EOS of PNM and the symmetry energy through Eqs.\,(\ref{sec1_SNM}), (\ref{sec1_PNM}), and (\ref{sec1_Esym}).
Furthermore, the slope parameter of symmetry energy could be obtained by Eq.\,(\ref{def_L}).

\setcounter{equation}{0}
\section{Foundation of QCDSR}\label{SEC_PhysFou}

In this section, we briefly describe the physical foundation of QCDSR\,\cite{Coh91,Coh92,Fur92,Jin93,Jin94,Dru17a,Dru17b,Dru19,Dru04,Dru12,Jeo13,Coh95}.
We first discuss the QCDSR in vacuum, which is relatively
simple but contains all the important ingredients of the method. The generalization
of the QCDSR in vacuum to finite densities is then followed. The quark/gluon condensates used
in this work and the fitting scheme are finally given.

\subsection{QCDSR in Vacuum}

We start our discussions first by introducing the QCDSR in vacuum. In this work,
$A_{\lambda}^{A}$ denotes the gluon field where $A=1\sim8$ is the
color index and $\lambda=0\sim3$ is the space-time
index\,\cite{Coh95,Pes95,Wei95,Zee11,Sre07}. The matrix form of the
gluon field has the following structure,
\begin{equation}
\mathcal{A}_{ab}^{\mu}=A^{A\mu}t_{ab}^A,\end{equation} where
$t^{A}=\lambda^A/2$ and $\lambda^A$'s are the Gell-Mann matrices
which have the following basic properties
\begin{equation}
[t^A,t^B]=if^{ABC}t^C,~~\rm{tr}(t^A)=0,~~\rm{tr}(t^At^B)=\frac{1}{2}\delta^{AB}
,\end{equation}
with $f^{ABC}$ being the structure constant of the
group SU(3). The strength tensor for gluon field is given by
\begin{equation}
\mathcal{G}_{\mu\nu}=G_{\mu\nu}^At^A\equiv
D_{\mu}\mathcal{A}_{\mu}-D_{\nu}\mathcal{A}_{\mu} ,\end{equation}
where $D_{\mu}=\partial_{\mu}-ig_{\rm{s}}A_{\mu}$ is covariant
derivative and $g_{\rm{s}}$ is the coupling constant. Another form
of the above equation is,
\begin{equation}
\mathcal{G}_{\mu\nu}=\frac{i}{g_{\rm{s}}}[D_{\mu},D_{\nu}],\end{equation}
and \begin{equation}G_{\mu\nu}^A=\partial_{\mu}A_{\mu}^A
-\partial_{\nu}A_{\nu}^A+g_{\rm{s}}f^{ABC}A_{\mu}^BA_{\nu}^C.
\end{equation}

In order to discuss QCDSR in vacuum, one should introduce
appropriate correlation functions of nucleons, here we
adopt\,\cite{Coh95}
\begin{equation} \Pi_{\alpha\beta}(q)\equiv
i\int\d^4xe^{iqx}\langle0|\rm{T}\eta_{\alpha}(x)\overline{\eta}_{\beta}(0)|0\rangle,
\end{equation}
where $q$ is the momentum transfer between nucleons, $|0\rangle$ is
the non-perturbative physical vacuum, $\eta_{\alpha}$ is the
interpolation field of nucleons, and $\alpha,\beta$ are Dirac spinor
index. The interpolation field for proton is given by\,\cite{Iof81},
\begin{equation}\label{ProtonInField}
\eta_{\rm{p}}=\varepsilon_{abc}(\rm{u}_a^{\rm{T}}\rm{C}\gamma_{\mu}\rm{u}_b)\gamma_5\gamma^{\mu}\rm{d}_c,\end{equation}
where u and d are quark fields and $a,b$ and $c$ are the color
index. T represents transpose of the quark field in Dirac space, and
C is the charge conjugation operator. The central quantity in QCDSR are the spectral functions (densities),
\begin{align}
\rho_{\alpha\beta}(q)=&\frac{1}{2\pi}\int\d^4xe^{iqx}\langle0|\eta_{\alpha}(x)\overline{\eta}_{\beta}(0)|0\rangle,\\
\widetilde{\rho}_{\alpha\beta}(q)=&\frac{1}{2\pi}\int\d^4xe^{iqx}\langle0|\overline{\eta}_{\beta}(0)\eta_{\alpha}(x)|0\rangle.
\end{align}
Using the spectral functions $\rho_{\alpha\beta}(q)$ and
$\widetilde{\rho}_{\alpha\beta}(q)$, we can rewrite the correlation
function as
\begin{equation}
\Pi_{\alpha\beta}(q)=-\int_{-\infty}^{\infty}\d
q_0'\left[\frac{\rho_{\alpha\beta}(q')}{q_0-q_0'+i0^+}+\frac{\widetilde{\rho}_{\alpha\beta}(q')}{q_0-q_0'-i0^+}\right]
\end{equation}
with $q'_{\mu}=({q}_0',\v{q})$. In fact, the spectral functions can
always be written in the following form after inserting a set of
intermediate states,
\begin{equation}
\rho_{\alpha\beta}=(2\pi)^3\sum_n\delta^4(q-P_n)\langle0|\eta_{\alpha}(0)|n\rangle\langle
n|\overline{\eta}_{\beta}(0)|0\rangle.\end{equation} Using
$\delta^4(q+P_n)$ instead of $\delta^4(q-P_n)$ gives a similar
formula for $\widetilde{\rho}_{\alpha\beta}$ where $P_{n}^{\mu}$ is
four momentum of state $n$.

Lorentz symmetry and parity invariance together mean that the general
structure of $\rho_{\alpha\beta}$ is
\begin{equation}\label{sec3_spectral_vacuum_decomp}
\rho_{\alpha\beta}(q)=\rho_{\rm{s}}(q^2)\delta_{\alpha\beta}+\rho_{\rm{q}}(q^2)\slashed{q}_{\alpha\beta}
,\end{equation} where $\rho_{\rm{s}}$ and $\rho_{\rm{q}}$ are scalar
functions of $q$. Correspondingly, we have
\begin{equation}
\Pi_{\alpha\beta}(q)=\Pi_{\rm{s}}(q^2)\delta_{\alpha\beta}+\Pi_{\rm{q}}(q^2)\slashed{q}_{\alpha\beta}
.\end{equation} In the vacuum, we only need to contain integral for
positive energy (which shall be modified in the finite density
case), where the coefficients are\,\cite{Coh95}
\begin{equation}\label{Poly_for}
\Pi_i(q^2)=\int_0^{\infty}\d
s\frac{\rho_i(s)}{s-q^2}+\rm{polynomials},~~i=\rm{s,~q},
\end{equation}
with $s$ the threshold parameter ($\sim M^2$ for a
nucleon). For example, the simplest phenomenological nucleon spectral densities take the
form $\rho_{\rm{s}}^{\rm{phen}}(s)=M\delta(s-M^2)$ and $\rho_{\rm{q}}^{\rm{phen}}(s)
=\delta(s-M^2)$, corresponding to $\Pi(q)=-(\slashed{q}+M)/(q^2-M^2+i0^+)$, which is the standard nucleon
propagator in vacuum, i.e., the two-point nucleon-nucleon correlation function.

The other important aspect of QCDSR is the OPE. For two local operators $A$ and $B$, we
have\,\cite{Wil69,Pes95,Wei95}
\begin{equation}
\rm{T}A(x)B(0)=\sum_nC_n^{AB}(x,\mu)\mathcal{O}_n(0,\mu),~~x\to0,\end{equation}
where $C_n^{AB}$'s are the Wilson's coefficients which can be
calculated by standard perturbative methods, and $\mu$ is the
renormalization energy scale. In the momentum space, we then have the
correlation function from the OPE side as,
\begin{equation}
\Pi(Q^2)=\sum_nC_n^i(Q^2)\langle\mathcal{O}_n\rangle,
\end{equation}
where $Q^2=-q^2$ and $\langle\mathcal{O}_n\rangle$'s are the
different types of quark/gluon condensates. Notice that
OPE is only applicable in the large $Q^2$ region, i.e., in the deep
space-like region.

Physically, it is no prior that the correlation functions from
OPE side should be same as these from the phenomenological side, and
they could also be very different from each other. The basic assumption
of QCDSR is that in some range of $q^2$ these different
correlation functions are physically equivalent. This range of $q^2$, or
equivalently, the range of applicability, is called QCDSR
window. At this point, it should be pointed out that the QCDSR approach is usually expected work well
at lower densities/momenta. The nucleon spectral
functions in nuclear medium are very complicated, and at low
densities/momenta there exists a very narrow resonance state
corresponding to the nucleon which can be described as a delta
function. As density/momentum
increases, continuum excitations emerge and these high
density/momentum states will have increasing importance at high
densities/momenta.
However, in QCDSR,
the contributions from high order states (e.g., high density/momentum momentum states) are significantly suppressed by the Borel
transformation of correlation functions, leading to that QCDSR shall be mainly applicable in the low
density/momentum region.

According to the above general analysis of spectral functions, one can write out
the general structure of the nucleon spectral densities as
\begin{align}
\rho_{\rm{s}}^{\rm{phen}}(s)=&\lambda_{}^2M_{}\delta(s-M_{}^2)+\cdots,\label{sd-1}\\
\rho_{\rm{q}}^{\rm{phen}}(s)=&\lambda_{}^2\delta(s-M_{}^2)+\cdots.\label{sd-2}\end{align}
Delta function indicates that it is a
resonance (the nucleon) and the ellipsis denotes high order states.
In the above expressions,
$\lambda_{}$ is the constant related to two physical states,
$|0\rangle$ and $|q\rangle$ connected through $\langle0|\eta(0)|q\rangle=\lambda_{}u(q) $ with $q^2=M^2$
and $u(q)$ the Dirac spinor, $s$ is a threshold parameter.
Correspondingly, the correlation functions are given by
\begin{align}
\Pi_{\rm{s}}^{\rm{phen}}(q^2)=&-\lambda_{}^2\frac{M_{}}{q^2-M_{}^2+i0^+}+\cdots,\\
~~\Pi_{\rm{q}}^{\rm{phen}}(q^2)=&-\lambda_{}^2\frac{1}{q^2-M_{}^2+i0^+}+\cdots,
\end{align}
which shall be rewritten in a unified form,
\begin{equation}
\Pi^{\rm{phen}}(q)=-\lambda_{}^2\frac{\slashed{q}+M_{}}{q^2-M_{}^2+i0^+}+\cdots.
\end{equation}
When the Borel transformation is made on the correlation functions
from both the phenomenological side and the OPE side, one obtains the QCDSR equations, which connect the nucleon self-energies
appearing on the phenomenological side and the quark/gluon condensates on the OPE side. Before giving the Borel
transformation of the correlation functions, we discuss the
essential procedures on QCDSR calculations:

1. Firstly, we determine the interpolation field to be studied, for
example, the interpolation field for proton, Eq.\,(\ref{ProtonInField}),
or the more general expression,
\begin{equation}\label{for_genIoffe}
\eta_{\rm{p}}(x)=2[t\eta_1^{\rm{p}}(x)+\eta_2^{\rm{p}}(x)]\end{equation}
with two independent terms,
\begin{align}
\eta_1^{\rm{p}}(x)=&\varepsilon_{abc}[\rm{u}_a^{\rm{T}}\rm{C}\gamma_5\rm{d}_b(x)]\rm{u}_c(x),\\
\eta_2^{\rm{p}}(x)=&\varepsilon_{abc}[\rm{u}_a^{\rm{T}}\rm{C}\rm{d}_b(x)]\gamma_5\rm{u}_c(x).
\end{align}
In Eq.\,(\ref{for_genIoffe}), $t$ is a parameter whose natural value
is around $-1$\,\cite{Iof81}. The interpolation field for proton
with $t=-1$ is called the Ioffe interpolation field. In Section~\ref{SEC_MassVacuum}, we determine the $t$ in a
self-consistent manner instead of using $t=-1$. In order to obtain the
interpolation field for the neutron, one just needs to exchange u and d  in Eq.\,(\ref{for_genIoffe}).

2. The second step is to determine the tensor structure of the
spectral functions. For instance, there are only ``s'' and ``q'' parts
for the nucleon correlation functions in vacuum while a new term will
emerge at finite densities.

3. Then one writes down the dispersion relations for the
nucleon correlation functions on the phenomenological side, e.g., Eq.\,(\ref{Poly_for}), which is a fundamental step in QCDSR, and in the
next subsection we give the general correlation functions for nucleon at finite densities.

4. At the same time, one writes down the OPE for the
interpolation fields in terms of  quark/gluon condensates,
where the central quantities in this step are the Wilson's
coefficients. Using perturbative method from standard quantum field theories
will furnish this calculation.

5. Finally, one makes the Borel transformation on two types of
the correlation functions, i.e., one from the phenomenological side and the other from the OPE side,
and then obtains the QCDSR equations. Solving the QCDSR
equations and analyzing the results are the next
procedure in the whole program.

At last of this subsection, we discuss the Borel transformation. For any function of the momentum transfer, $f(Q^2)$, the Borel
transformation
\begin{equation}
\mathcal{B}[f(Q^2)]\equiv \widehat{f}(\mathscr{M}^2)\end{equation}
is defined through
\begin{equation}
\widehat{f}(\mathscr{M}^2)\equiv\lim_{\substack{Q^2,n\to\infty\\
Q^2/n=\mathscr{M}^2}}\frac{(Q^2)^{n+1}}{n!}\left(-\frac{\d}{\d
Q^2}\right)^nf(Q^2),\end{equation} where $\mathscr{M}$ is the Borel
mass. For instance, Borel transformation of some typical functions
are given by
\begin{align}
\mathcal{B}\left[\frac{1}{(Q^2)^k}\right]=&\frac{1}{(k-1)!(\mathscr{M}^2)^{k-1}},\\
\mathcal{B}[(Q^2)^m]=&0,\\
\mathcal{B}\left[\frac{1}{(Q^2)^k}\ln
Q^2\right]=&\frac{1}{(k-1)!(\mathscr{M}^2)^{k-1}}\notag\\
&\times\left[\ln \mathscr{M}^2-\frac{1}{k}-\gamma_{\rm{E}}+\sum_{j=1}^k\frac{1}{j}\right],\\
\mathcal{B}[(Q^2)^m\ln
Q^2]=&(-1)^{m+1}m!(\mathscr{M}^2)^{m+1},\cdots,
\end{align}
where $m$ equals to $0,1,2,\cdots$, $k$ equals to $1,2,3,\cdots$,
and $\gamma_{\rm{E}}\approx0.577$ is the Euler constant.

Under Borel transformation, the correlation function~(\ref{Poly_for}) becomes
\begin{equation}
\widehat{\Pi}_i(\mathscr{M}^2)=\int_0^{\infty}\d
se^{-s/\mathscr{M}^2}\rho_i(s),~~i=\rm{s,~q},
\end{equation}
where the polynomials in Eq.~(\ref{Poly_for}) disappear. The disappearance of the polynomials is the most
important approximation made by the Borel transformation. As
discussed above, the high order states, and/or, the continuum excitations
including polynomials in Eq.~(\ref{Poly_for}), become more and more
important at high densities/momenta, the disappearance of the polynomials under Borel transformation would make the physical predictions
eventually incredible at high densities/momenta. This is the main reason why QCDSR should mainly be used
in low density/momentum region.
Moreover, the high-$s$
states become unimportant due to the suppression factor
$e^{-s/\mathscr{M}^2}$, and they can be even removed (as the polynomials in Eq.\,(\ref{Poly_for})).
As a rough example on the density region above which the QCDSR should be broken down,
we consider the formation of the $\Delta$ resonance as an excited state in dense nucleonic matter.
As shown in ref.\,\cite{Cai15a}, the formation density of the first charged state of $\Delta(1232)$
could be smaller than 2$\rho_0$, even to be around the saturation density. Thus it is
conservative to expect that the QCDSR should not be applied at densities around or larger than 2$\rho_0$.
However, a comprehensive analysis of the applicable region of the QCDSR deserve more further work.

\subsection{QCDSR at Finite Densities}

In this subsection, we generalize the QCDSR in vacuum to finite densities. The nucleon propagator in medium at finite densities is given by\,\cite{Ser86,Hor83,Hor87,Rin96,Ser97},
\begin{equation}
G(q)=-i\int\d^4xe^{iqx}\langle\Psi_0|\rm{T}\psi(x)\overline{\psi}(0)|\Psi_0\rangle,\end{equation}
where $\Phi_0$ is the physical ground state for the infinite nucleonic matter and
$\psi$ is the corresponding nucleon field. The nucleon self-energy
$\Sigma(q)$ is defined through Dyson's equation in the following
form (with isospin index suppressed),
\begin{equation}
[G(q)]^{-1}=\slashed{q}-M-\Sigma(q),
\end{equation}
which can be decomposed into
\begin{equation}
G(q)=G_{\rm{s}}(q^2,qu)+G_{\rm{q}}(q^2,q
u)\slashed{q}+G_{\rm{u}}(q^2,qu)\slashed{u}\end{equation} by
symmetry principles, where $u_{\mu}$ is the nucleon four-velocity and $qu=q_{\mu}u^{\mu}$. The ``u'' term, i.e.,
$G_{\rm{u}}(q^2,qu)\slashed{u}$, is new at finite densities.
Similarly, we decompose the self-energy into the corresponding
terms,
\begin{equation}
\Sigma(q)=\widetilde{\Sigma}_{\rm{s}}(q^2,q
u)+\widetilde{\Sigma}_{\rm{v}}^{\mu}(q)\gamma_{\mu},\end{equation}
with
\begin{equation}
\widetilde{\Sigma}_{\rm{v}}^{\mu}(q)=\Sigma_{\rm{u}}(q^2,qu)u^{\mu}+\Sigma_{\rm{q}}(q^2,qu)q^{\mu}.\end{equation}

Defining the scalar self-energy in nuclear medium as $
\Sigma_{\rm{S}}=M^{\ast}-M$ with
\begin{equation}M^{\ast}=\frac{M+\widetilde{\Sigma}_{\rm{s}}}{1-\Sigma_{\rm{q}}},
\end{equation}
and the vector self-energy as
\begin{equation}
\Sigma_{\rm{V}}=\frac{\Sigma_{\rm{u}}}{1-\Sigma_{\rm{q}}},\end{equation}
we can then rewrite the propagator of a nucleon as
\begin{equation}
G(q)=\frac{1}{\slashed{q}-M-\Sigma(q)}\longrightarrow\lambda^{\ast,2}\frac{\slashed{q}+M^{\ast}
-\slashed{u}\Sigma_{\rm{V}}}{(q_0-{e})(q_0-\overline{{e}})}
,\end{equation}
where $\lambda^{\ast,2}$ is the residual factor\,\cite{Coh95}, and ${e}$ and $\overline{{e}}$ are
the poles of positive energy branch and negative energy branch,
i.e.,
\begin{align}
{e}={e}(\rho,\v{q})=&
\Sigma_{\rm{V}}(\rho,\v{q})+{e}^{\ast}(\rho,\v{q}),\\
\overline{{e}}=\overline{{e}}(\rho,\v{q}) =&
\Sigma_{\rm{V}}(\rho,\v{q})-{e}^{\ast}(\rho,\v{q}),
\end{align}
with
\begin{equation}
{e}^{\ast}(\rho,\v{q})=\left[{\v{q}^2+M^{\ast,2}(\rho,\v{q})}\right]^{1/2}.
\end{equation}
The discontinuity passing through the real axis of $q$ represents
the spectral function of the correlation function, i.e.,
\begin{align}
\Delta G_{\rm{s}}(q_0)=&-2\pi
i\frac{\lambda^{\ast,2}M^{\ast}}{2{e}^{\ast}}[\delta(q_0-{e})-\delta(q_0-\overline{{e}})],\\
\Delta G_{\rm{q}}(q_0)=&-2\pi
i\frac{\lambda^{\ast,2}}{2{e}^{\ast}}[\delta(q_0-{e})-\delta(q_0-\overline{{e}})],\\
\Delta G_{\rm{u}}(q_0)=&-2\pi
i\frac{\lambda^{\ast,2}\Sigma_{\rm{V}}}{2{e}^{\ast}}[\delta(q_0-{e})-\delta(q_0-\overline{{e}})].
\end{align}

Based on the nucleon propagator given above, the nucleon correlation functions can be obtained correspondingly,
\begin{equation}
\Pi(q)\equiv
i\int\d^4xe^{iqx}\langle\Psi_0|\rm{T}\eta(x)\overline{\eta}(0)|\Psi_0\rangle,\end{equation}
where $\eta$ is proton's interpolation field. Very similarly, we
decompose $\Pi(q)$ into three parts,
\begin{equation}
\Pi(q)=\Pi_{\rm{s}}(q^2,qu)+\Pi_{\rm{q}}(q^2,qu)\slashed{q}+\Pi_{\rm{u}}(q^2,qu)\slashed{u},\end{equation}
with
\begin{align}
\Pi_{\rm{s}}(q^2,qu)=&\frac{1}{4}\rm{tr}(\Pi),\\
\Pi_{\rm{q}}(q^2,qu)=&\frac{1}{q^2-(qu)^2}\left[\frac{1}{4}\rm{tr}(\slashed{q}\Pi)-\frac{1}{4}qu\rm{tr}(\slashed{u}\Pi)\right],\\
\Pi_{\rm{u}}(q^2,qu)=&\frac{1}{q^2-(qu)^2}\left[\frac{1}{4}q^2\rm{tr}(\slashed{u}\Pi)-\frac{1}{4}q
u\rm{tr}(\slashed{q}\Pi)\right].
\end{align}
Furthermore, from the discussions on the propagator above, we have
\begin{align}
\Pi_{\rm{s}}(q_0,\v{q})=&-\lambda^{\ast,2}\frac{M^{\ast}}{(q_0-{e})(q_0-\overline{{e}})},\\
\Pi_{\rm{q}}(q_0,\v{q})=&-\lambda^{\ast,2}\frac{1}{(q_0-{e})(q_0-\overline{{e}})},\\
\Pi_{\rm{u}}(q_0,\v{q})=&-\lambda^{\ast,2}\frac{\Sigma_{\rm{V}}}{(q_0-{e})(q_0-\overline{{e}})},
\end{align}
and their Borel transformations are $
\lambda^{\ast,2}M^{\ast}e^{-({e}^2-\v{q}^2)/\mathscr{M}^2}$,
$\lambda^{\ast,2}e^{-({e}^2-\v{q}^2)/\mathscr{M}^2}$, and $
\lambda^{\ast,2}\Sigma_{\rm{V}}e^{-({e}^2-\v{q}^2)/\mathscr{M}^2}
$, respectively.

On the other hand, the correlation functions constructed from quark/gluon
condensates are
\begin{equation}
\Pi_i(q^2,qu)=\sum_nC_n^i(q^2,qu)\langle\mathcal{O}_n\rangle_{\rho}
,~~i=\rm{s,~q,~u},\end{equation} where $
\langle\mathcal{O}_n\rangle_{\rho}=\langle\Psi_0|\mathcal{O}_n|\Psi_0\rangle$
are quark/gluon condensates at finite densities.
In this work, the quark/gluon condensates at finite densities up to mass dimension-6
are included in the QCDSR equations, i.e., $\langle\overline{q}q\rangle$,
$\left\langle({\alpha_{\rm{s}}}/{\pi})G^2\right\rangle$,
$\langle
g_{\rm{s}}\overline{q}\sigma\mathcal{G}q\rangle$, $\langle
g_{\rm{s}}{q}^{\dag}\sigma\mathcal{G}q\rangle$, $\langle\overline{q}\Gamma_1q\overline{q}\Gamma_2q\rangle
$ and $
\langle\overline{q}\Gamma_1\lambda^Aq\overline{q}\Gamma_2\lambda^Aq\rangle$.
Properties of them will be given in the following subsections.

\subsection{OPE Coefficients}

Wilson's coefficients (OPE coefficients) $C_n^i(q^2,qu)$ could be calculated
by standard perturbative
method\,\cite{Coh91,Coh92,Fur92,Jin93,Jin94,Jeo13,Coh95}. To make
the discussions simpler, we calculate the OPE coefficients in
SNM\,\cite{Coh95} based on Fock--Schwinger gauge in background field
method\,\cite{Nov84}. The generalizations to the ANM  are
straightforward without difficulty, which will be given in the last
of this subsection. The Fock--Schwinger gauge is
\begin{equation}\label{FSG}
x_{\mu}\mathcal{A}^{\mu}(x)=0,
\end{equation}
with
$\mathcal{A}^{\mu}\equiv A^{A\mu}t^A$.
Eq.\,(\ref{FSG}) indicates
\begin{equation}
0=\partial_{\mu}(y^{\nu}A_{\nu}^A(y))=A_{\mu}^A(y)+y^{\nu}\partial_{\mu}A_{\nu}^A(y).\end{equation}
Moreover, with
\begin{equation}
y^{\nu}\partial_{\mu}A_{\nu}^A=y^{\nu}G_{\mu\nu}^A+y^{\nu}\partial_{\nu}\partial_{\mu}^A,\end{equation}
one then has
\begin{equation}
A_{\mu}^A(y)+y^{\nu}\partial_{\nu}A_{\mu}^A=y^{\nu}G_{\nu\mu}^A.\end{equation}
Using $y^{\nu}=\alpha x^{\nu}$, we then have
\begin{equation}
\frac{\d}{\d\alpha}[\alpha A_{\mu}(\alpha x)]=\alpha
x^{\nu}G_{\nu\mu}^A(\alpha x) ,\end{equation} so
\begin{equation}
A_{\mu}^A(x)=\int_0^1\d\alpha\alpha x^{\nu}G_{\nu\mu}^A(\alpha x)
,\end{equation} then
\begin{align}
\mathcal{A}_{\nu}(x)=&\int_0^1\d\alpha\alpha
x^{\mu}\mathcal{G}_{\mu\nu}(\alpha
x)\notag\\
=&\frac{1}{2}x^{\mu}\mathcal{G}_{\mu\nu}(0)+\frac{1}{2}x^{\lambda}x^{\mu}(D_{\lambda}\mathcal{G}_{\mu\nu})_{\lambda=0}
+\cdots.
\end{align}

\begin{figure}[h!]
\centering
  \includegraphics[width=8.5cm]{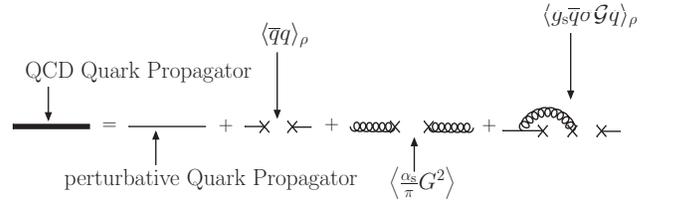}
  \caption{Quark propagator in nuclear medium. Straight line denotes quark condensates
  and the wave line represents gluon condensates.}
  \label{fig_QuarkPro}
\end{figure}

In the background field method, the non-perturbative effects of
quarks are represented by Grassmann background fields,
$\chi_{a\alpha}^q,\overline{\chi}_{a\alpha}^q $, while the effects
of gluons are represented by their classical fields, $F_{\mu\nu}^A$.
The propagator of two quarks in coordinate space reads\,\cite{Nov84}
\begin{align}
S_{ab,\alpha\beta}^q(x,0)\equiv
&\langle\rm{T}q_{a\alpha}(x)\overline{q}_{b\beta}(0)\rangle_{\rho}\notag\\
=&\frac{i}{2\pi^2}\delta_{ab}\frac{1}{(x^2)^2}[\slashed{x}]_{\alpha\beta}
-\frac{im_{\rm{q}}}{4\pi^2}\delta_{ab}\frac{\delta_{\alpha\beta}}{x^2}\notag\\
&+\chi_{a\alpha}^q(x)\overline{\chi}_{b\beta}^q(0)\notag\\
&-\frac{ig_{\rm{s}}}{32\pi^2}F_{\mu\nu}^A(0)t_{ab}^A\frac{1}{x^2}
[\slashed{x}\sigma^{\mu\nu}+\sigma^{\mu\nu}\slashed{x}]_{\alpha\beta}\notag\\
&+\cdots.
\end{align}
The products of Grassmann fields and classical fields can be written
as products of matrix elements of the ground states of quarks and
gluons, i.e.,\begin{align}
\chi_{a\alpha}^q(x)\overline{\chi}_{b\beta}^q(0)=&\langle
q_{a\alpha}(x)\overline{q}_{b\beta}(0)\rangle_{\rho},\\
F_{\kappa\lambda}^AF_{\mu\nu}^B=&\langle
G_{\kappa\lambda}^AG_{\mu\nu}^B\rangle_{\rho},\\
\chi_{a\alpha}^q\overline{\chi}_{b\beta}^qF_{\mu\nu}^A=&\langle
q_{a\alpha}\overline{q}_{b\beta}G_{\mu\nu}^A\rangle_{\rho},\\
\chi_{a\alpha}^q\overline{\chi}_{b\beta}^q\chi_{c\gamma}^q\overline{\chi}_{d\delta}^q
=&\langle
q_{a\alpha}\overline{q}_{b\beta}q_{c\gamma}\overline{q}_{d\delta}\rangle_{\rho},
\end{align}
etc., where all the values of fields are calculated at $x=0$. Then
we can write the propagator of quarks as
\begin{align}
S_{ab,\alpha\beta}^q(x,0)=&\frac{i}{2\pi^2}\delta_{ab}\frac{1}{(x^2)^2}[\slashed{x}]_{\alpha\beta}
-\frac{im_{\rm{q}}}{4\pi^2}\delta_{ab}\frac{\delta_{\alpha\beta}}{x^2}\notag\\
&+\langle
q_{a\alpha}(x)\overline{q}_{b\beta}(0)\rangle_{\rho}\notag\\
&-\frac{ig_{\rm{s}}}{32\pi^2}G_{\mu\nu}^A(0)t_{ab}^A\frac{1}{x^2}
[\slashed{x}\sigma^{\mu\nu}+\sigma^{\mu\nu}\slashed{x}]_{\alpha\beta}\notag\\
&-\frac{1}{2^23}\delta_{\alpha\beta}\delta^{ab}\langle\overline{q}q\rangle_{\rho}
+\frac{im_{\rm{q}}}{2^43}[\slashed{x}]_{\alpha\beta}\delta^{ab}\langle\overline{q}q\rangle_{\rho}\notag\\
&-\frac{x^2}{2^63}\delta_{\alpha\beta}\delta^{ab}\langle
g_{\rm{s}}\overline{q}\sigma\mathcal{G}q\rangle_{\rho}\notag\\
&+\frac{im_{\rm{q}}x^2}{2^73^2}[\slashed{x}]_{\alpha\beta}\delta^{ab}
\langle
g_{\rm{s}}\overline{q}\sigma\mathcal{G}q\rangle_{\rho}\notag\\&
-\frac{\pi^2x^4}{2^83^3}\delta_{\alpha\beta}\delta^{ab}\langle\overline{q}q\rangle_{\rho}\left\langle
\frac{\alpha_{\rm{s}}}{\pi}G^2\right\rangle_{\rho}+\cdots.
\end{align}
In Fig.\,\ref{fig_QuarkPro}, we show the quark propagator in
nuclear medium graphically, where the last three terms represent
non-perturbative effects.

The matrix element $\langle
q_{a\alpha}(x)\overline{q}_{b\beta}(0)\rangle_{\rho}$ can be
projected as,
\begin{align}
\langle q_{a\alpha}(x)\overline{q}_{b\beta}(0)\rangle_{\rho}=
-\frac{\delta_{ab}}{12}\big[&\langle\overline{q}(0)q(x)\rangle_{\rho}\delta_{\alpha\beta}\notag\\
&
+\langle\overline{q}(0)\gamma_{\lambda}q(x)\rangle_{\rho}\gamma_{\alpha\beta}^{\lambda}\big]
,\end{align} where parity symmetry and color neutrality of the
ground state of nucleonic matter are taken into account when writing
down the above expression. At short distance, we expand the
quark field as
\begin{align}
q(x)=&q(0)+x^{\mu}(\partial_{\mu}q)_{x=0}\notag\\
&+\frac{1}{2}x^{\mu}x^{\nu}(\partial_{\mu}\partial_{\nu}q)_{x=0}+\cdots
.\end{align}

\noindent
Then using the Fock--Schwinger gauge, one obtains an identity
for gluons,
\begin{align}
x^{\nu}\mathcal{A}_{\nu}(0)+&x^{\mu}x^{\nu}(\partial_{\mu}\mathcal{A}_{\nu})_{x=0}\notag\\
&+\frac{1}{2}x^{\lambda}x^{\mu}x^{\nu}(\partial_{\lambda}\partial_{\mu}\mathcal{A}_{\nu})_{x=0}+\cdots
=0.\end{align}
 Each term above is zero, and we thus have
\begin{equation}
x^{\mu}(D_{\mu}q)_{x=0}=x^{\mu}(\partial_{\mu}q)_{x=0},
\end{equation} and
similarly, we have
\begin{equation}x^{\mu}x^{\nu}(D_{\mu}D_{\nu}q)_{x=0}
=x^{\mu}x^{\nu}(\partial_{\mu}\partial_{\nu}q)_{x=0},
\end{equation} etc. After
these simplifications on gluon fields, the quark fields then can be
re-expanded as
\begin{align}
q(x)=&q(0)+x^{\mu}(D_{\mu}q)_{x=0}\notag\\
&+\frac{1}{2}x^{\mu}x^{\nu}(D_{\mu}D_{\nu}q)_{x=0}+\cdots.
\end{align}  Finally,
we obtain
\begin{widetext}
\begin{align}\label{dfge}
\langle q_{a\alpha}(x)\overline{q}_{b\beta}(0)\rangle_{\rho}
=-\frac{\delta_{ab}}{12}\Bigg[&\left(\langle\overline{q}q\rangle_{\rho}+x^{\mu}
\langle\overline{q}D_{\mu}q\rangle_{\rho}+\frac{1}{2}x^{\mu}x^{\nu}
\langle\overline{q}D_{\mu}D_{\nu}q\rangle_{\rho}+\cdots\right)\delta_{\alpha\beta}\notag\\
&+\left(\langle\overline{q}\gamma_{\lambda}q\rangle_{\rho}+x^{\mu}
\langle\overline{q}\gamma_{\lambda}D_{\mu}q\rangle_{\rho}+\frac{1}{2}x^{\mu}x^{\nu}
\langle\overline{q}\gamma_{\lambda}D_{\mu}D_{\nu}q\rangle_{\rho}+\cdots\right)\gamma_{\alpha\beta}^{\lambda}\Bigg]
,\end{align} where all the derivatives are calculated at $x=0$.

The calculations of Wilson's coefficients of condensates
$\langle\overline{q}D_{\mu_1}D_{\mu_2}\cdots
D_{\mu_n}q\rangle_{\rho}[\langle\overline{q}\gamma_{\mu}D_{\mu_1}D_{\mu_2}\cdots
D_{\mu_n}q\rangle_{\rho}]$ are very similar to that of
$\langle\overline{q}q\rangle_{\rho}[\langle\overline{q}\gamma_{\mu}
q\rangle_{\rho}]$. For example, in the coordinate space, we have
\begin{equation}
C_{\overline{q}D_{\mu_1}D_{\mu_2}\cdots
D_{\mu_n}q}(x)=\frac{1}{n!}x^{\mu_1}\cdots
x^{\mu_n}C_{\overline{q}q}(x),~~
C_{\overline{q}\gamma_{\mu}D_{\mu_1}D_{\mu_2}\cdots
D_{\mu_n}q}(x)=\frac{1}{n!}x^{\mu_1}\cdots
x^{\mu_n}C_{\overline{q}\gamma_{\mu}q}(x).
\end{equation}
In the momentum space, we have
\begin{align}
C_{\overline{q}D_{\mu_1}D_{\mu_2}\cdots
D_{\mu_n}q}(q)=&\frac{(-i)^n}{n!}\left(\frac{\partial}{\partial
q_{\mu_1}}\cdots\frac{\partial}{\partial
q_{\mu_n}}\right)C_{\overline{q}q}(q),\\
C_{\overline{q}\gamma_{\mu}D_{\mu_1}D_{\mu_2}\cdots
D_{\mu_n}q}(q)=&\frac{(-i)^n}{n!}\left(\frac{\partial}{\partial
q_{\mu_1}}\cdots\frac{\partial}{\partial
q_{\mu_n}}\right)C_{\overline{q}\gamma_{\mu}q}(q).
\end{align}
The condensate terms in Eq.~(\ref{dfge}) can be furnished by writing
out all the possible terms required by Lorentz symmetry, while the
coefficients of these terms can be obtained through their traces of
the decomposition, i.e.,
\begin{align}
\langle\overline{q}\gamma_{\mu}q\rangle_{\rho}
=&\langle\overline{q}\slashed{u}q\rangle_{\rho}u_{\mu},\\
\langle\overline{q}D_{\mu}q\rangle_{\rho}=&
\langle\overline{q}u\cdot
Dq\rangle_{\rho}u_{\mu}=-im_{\rm{q}}\langle\overline{q}\slashed{u}q\rangle_{\rho}u_{\mu},\\
\langle\overline{q}\gamma_{\mu}D_{\nu}q\rangle_{\rho}
=&\frac{4}{3}\langle\overline{q}\slashed{u}u\cdot
Dq\rangle_{\rho}\left(u_{\mu}u_{\nu}-\frac{1}{4}g_{\mu\nu}\right)-\frac{1}{3}\langle\overline{q}\slashed{D}q
\rangle_{\rho}(u_{\mu}u_{\nu}-g_{\mu\nu})\notag\\
=&\frac{4}{3}\langle\overline{q}\slashed{u}u\cdot
Dq\rangle_{\rho}\left(u_{\mu}u_{\nu}-\frac{1}{4}g_{\mu\nu}\right)
+\frac{1}{3}im_{\rm{q}}\langle\overline{q}q\rangle_{\rho}(u_{\mu}u_{\nu}-g_{\mu\nu}),\\
\langle\overline{q}D_{\mu}D_{\nu}q\rangle_{\rho}=&
\frac{4}{3}\langle\overline{q}u\cdot Du\cdot
Dq\rangle_{\rho}\left(u_{\mu}u_{\nu}-\frac{1}{4}g_{\mu\nu}\right)
-\frac{1}{3}\langle\overline{q}D^2q
\rangle_{\rho}(u_{\mu}u_{\nu}-g_{\mu\nu})\notag\\
=&\frac{4}{3}\langle\overline{q}u\cdot Du\cdot
Dq\rangle_{\rho}\left(u_{\mu}u_{\nu}-\frac{1}{4}g_{\mu\nu}\right)
-\frac{1}{6}\langle
g_{\rm{s}}\overline{q}\sigma\cdot\mathcal{G}q\rangle_{\rho}
(u_{\mu}u_{\nu}-g_{\mu\nu}),
\end{align}where $g_{\mu\nu}=(+,-,-,-)$ and $u\cdot
D=uD=u_{\mu}D^{\mu}$, $\sigma\cdot\mathcal{G}=\sigma\mathcal{G}$.

Equations of motion are useful when deriving
these expressions, e.g., for the second identity in the
second line, we have used the relation $
D_{\mu}\equiv2^{-1}(\gamma_{\mu}\slashed{D}+\slashed{D}\gamma_{\mu})$,
combined with the translation invariance gives $
\langle\overline{q}i\slashed{D}\slashed{u}q\rangle_{\rho}=-
\langle\overline{q}i\overleftarrow{\slashed{D}}\slashed{u}q\rangle_{\rho}$.
Very similarly, we have
\begin{align}
\langle\overline{q}\gamma_{\lambda}D_{\mu}D_{\nu}q\rangle_{\rho}=&
2\langle\overline{q}\slashed{u}u\cdot Du\cdot Dq\rangle_{\rho}
\left[u_{\lambda}u_{\mu}u_{\nu}-\frac{1}{6}(u_{\lambda}
g_{\mu\nu}+u_{\mu}g_{\lambda\nu}+u_{\nu}g_{\lambda\mu})\right]
-\frac{1}{3}\langle\overline{q}\slashed{u}D^2q\rangle_{\rho}(u_{\lambda}u_{\mu}u_{\nu}-u_{\lambda}g_{\mu\nu})\notag\\
&-\frac{1}{3}\langle\overline{q}\slashed{u}\cdot
D\slashed{D}q\rangle_{\rho}(u_{\lambda}u_{\mu}u_{\nu}-u_{\mu}g_{\lambda\nu})
-\frac{1}{3}\langle\overline{q}\slashed{D}u\cdot
Dq\rangle_{\rho}(u_{\lambda}u_{\mu}u_{\nu}-u_{\nu}g_{\lambda\mu}),\\
=&2\langle\overline{q}\slashed{u}u\cdot Du\cdot Dq\rangle_{\rho}
\left[u_{\lambda}u_{\mu}u_{\nu}-\frac{1}{6}(u_{\lambda}
g_{\mu\nu}+u_{\mu}g_{\lambda\nu}+u_{\nu}g_{\lambda\mu})\right]
-\frac{1}{6}\langle
g_{\rm{s}}\overline{q}\slashed{u}\sigma\mathcal{G}q\rangle_{\rho}(u_{\lambda}u_{\mu}u_{\nu}
-u_{\lambda}g_{\mu\nu}),
\end{align}
where the relation $
\langle\overline{q}\slashed{u}D^2q\rangle_{\rho} =2^{-1}\langle
g_{\rm{s}}\overline{q}\slashed{u}\sigma\mathcal{G}q\rangle_{\rho}$
is used. Expressions for other condensates with different dimensions
are similarly obtained, i.e.,
\begin{align}
\langle
g_{\rm{s}}q_{a\alpha}\overline{q}_{b\beta}G_{\mu\nu}^A\rangle_{\rho}
=&-\frac{t_{ab}^A}{96}\Big\{ \langle
g_{\rm{s}}\overline{q}\sigma\mathcal{G}q\rangle_{\rho}\left[\sigma_{\mu\nu}+i(u_{\mu}\gamma_{\nu}
-u_{\nu}\gamma_{\mu})\slashed{u}\right]_{\alpha\beta} +\langle
g_{\rm{s}}\overline{q}\slashed{u}\sigma\mathcal{G}q\rangle_{\rho}\left[\sigma_{\mu\nu}\slashed{u}
+i(u_{\mu}\gamma_{\nu}
-u_{\nu}\gamma_{\mu})\right]_{\alpha\beta}\notag\\
&-4\left(\langle\overline{q}u\cdot Du\cdot
Dq\rangle_{\rho}+im_{\rm{q}}\langle\overline{q}\slashed{u}u\cdot
Dq\rangle_{\rho}\right)\left[\sigma_{\mu\nu}+2i(u_{\mu}\gamma_{\nu}
-u_{\nu}\gamma_{\mu})\slashed{u}\right] \Big\},\\
\left\langle\frac{\alpha_{\rm{s}}}{\pi}
G_{\kappa\lambda}^AG_{\mu\nu}^B\right\rangle_{\rho}=&
\frac{\delta^{AB}}{96}\Bigg\{
\left\langle\frac{\alpha_{\rm{s}}}{\pi}
G^2\right\rangle_{\rho}(g_{\kappa\mu}g_{\lambda\nu}-g_{\kappa\nu}g_{\lambda\mu})
-2\left\langle\frac{\alpha_{\rm{s}}}{\pi}\left[(u\cdot
G)^2+\left(u\cdot\widetilde{G}\right)^2\right]\right\rangle_{\rho}\notag\\
&\hspace*{2cm}\times\left[g_{\kappa\mu}g_{\lambda\nu}-g_{\kappa\nu}g_{\lambda\mu}-2(
g_{\kappa\mu}u_{\lambda}u_{\nu}-g_{\kappa\nu}u_{\lambda}u_{\mu}
-g_{\lambda\mu}u_{\kappa}u_{\nu}+g_{\lambda\nu}u_{\kappa}u_{\mu})\right]
\Bigg\},
\end{align}
where we have $ (u\cdot G)^2=u^{\lambda}G_{\lambda\nu}^Au_{\mu}G^{A\mu\nu},
\widetilde{G}^{A\mu\nu}=2^{-1}\varepsilon^{\mu\nu\kappa\lambda}G_{\kappa\lambda}^A$.
Furthermore, the decomposition of the four-quark condensates has the following
form,
\begin{align}
\langle\overline{\rm{u}}_{a\alpha}\rm{u}_{b\beta}\overline{\rm{u}}_{c\gamma}\rm{u}_{d\delta}\rangle_{\rho}
&\approx\langle\overline{\rm{u}}_{a\alpha}\rm{u}_{b\beta}\rangle_{\rho}\langle\overline{\rm{u}}_{c\gamma}\rm{u}_{d\delta}\rangle_{\rho}
-\langle\overline{\rm{u}}_{a\alpha}\rm{u}_{d\delta}\rangle_{\rho}\langle\overline{\rm{u}}_{c\gamma}\rm{u}_{b\beta}\rangle_{\rho},\\
\langle\overline{\rm{u}}_{a\alpha}\rm{u}_{b\beta}\overline{\rm{d}}_{c\gamma}\rm{d}_{d\delta}\rangle_{\rho}
&\approx\langle\overline{\rm{u}}_{a\alpha}\rm{u}_{b\beta}\rangle_{\rho}\langle\overline{\rm{d}}_{c\gamma}\rm{d}_{d\delta}\rangle_{\rho}.
\end{align}

In order to translate the above expressions from coordinate space
into momentum space, we use the following formulae,
\begin{equation}
\int\frac{\d^4x}{x^2}e^{iqx}=-\frac{4\pi^2i}{q^2} ,~~
\int\frac{\d^4x}{(x^2)^n}e^{iqx}=\frac{i(-1)^n2^{4-2n}\pi^2}{\Gamma(n-1)\Gamma(n)}
(q^2)^{n-2}\ln(-q^2)+P_{n-2}(q^2),~~n\geq2,
\end{equation}
where $P_m(q^2)$ is the polynomial of $q^2$ of order $m$. We
decompose the correlation functions into their odd and even parts,
i.e.,
\begin{equation}
\Pi_i(q_0,\v{q})=\Pi_i^{\rm{E}}(q_0,\v{q})+q_0\Pi_i^{\rm{O}}(q_0,\v{q}),~~i=\rm{s,~q,~u}
,\end{equation}
with
\begin{align}
\Pi_{\rm{s}}^{\rm{E}}(q^2)=&\frac{c_1}{16\pi^2}q^2\ln(-q^2)\langle\overline{q}q\rangle_{\rho}
+\frac{3c_2}{16\pi^2}\ln(-q^2)\langle
g_{\rm{s}}\overline{q}\sigma\mathcal{G}q\rangle_{\rho}\notag\\
&+\frac{2c_3}{3\pi^2}\frac{q_0^2}{q^2}\left(\langle\overline{q}iD_0iD_0q\rangle_{\rho}
+\frac{1}{8}\langle
g_{\rm{s}}\overline{q}\sigma\mathcal{G}q\rangle_{\rho}\right),\\
\Pi_{\rm{s}}^{\rm{O}}(q^2)=&-\frac{c_1}{8\pi^2}\ln(-q^2)\langle\overline{q}iD_0q\rangle_{\rho}
-\frac{c_1}{3q^2}\langle\overline{q}q\rangle_{\rho}\langle
q^{\dag}q\rangle_{\rho},\\
\Pi_{\rm{q}}^{\rm{E}}(q^2)=&-\frac{c_4}{512\pi^4}(q^2)^2\ln(-q^2)\notag\\
&+\frac{c_4}{72\pi^2}\left[
5\ln(-q^2)-\frac{8q_0^2}{q^2}\right]\langle
q^{\dag}iD_0q\rangle_{\rho}
-\frac{c_4}{256\pi^2}\ln(-q^2)\left\langle\frac{\alpha_{\rm{s}}}{\pi}G^2\right\rangle_{\rho}\notag\\
&-\frac{c_4}{1152}\left[\ln(-q^2)-\frac{4q_0^2}{q^2}\right]\left\langle
\frac{\alpha_{\rm{s}}}{\pi}\left[(u\cdot
G)^2+\left(u\cdot\widetilde{G}\right)^2\right]\right\rangle_{\rho}\notag\\
&-\frac{c_1}{6q^2}\langle\overline{q}q\rangle_{\rho}^2-\frac{c_4}{6q^2}\langle
q^{\dag}q\rangle_{\rho}^2,\\
\Pi_{\rm{q}}^{\rm{O}}(q^2)=&\frac{c_4}{24\pi^2}\ln(-q^2)\langle
q^{\dag}q\rangle_{\rho}+\frac{c_5}{72\pi^2q^2}\langle
g_{\rm{s}}q^{\dag}\sigma\mathcal{G}q\rangle_{\rho}\notag\\
&-\frac{c_4}{12\pi^2q^2}\left(1+\frac{2q_0^2}{q^2}\right)\left(\langle{q}^{\dag}iD_0iD_0q\rangle_{\rho}
+\frac{1}{12}\langle
g_{\rm{s}}{q}^{\dag}\sigma\mathcal{G}q\rangle_{\rho}\right),\\
\Pi_{\rm{u}}^{\rm{E}}(q^2)=&\frac{c_4}{24\pi^2}\ln(-q^2)\langle
q^{\dag}q\rangle_{\rho}-\frac{c_5}{48\pi^2}\ln(-q^2)\langle
g_{\rm{s}}q^{\dag}\sigma\mathcal{G}q\rangle_{\rho}\notag\\
&+\frac{c_4}{2\pi^2}\frac{q_0^2}{q^2}\left(\langle{q}^{\dag}iD_0iD_0q\rangle_{\rho}
+\frac{1}{12}\langle
g_{\rm{s}}{q}^{\dag}\sigma\mathcal{G}q\rangle_{\rho}\right),\\
\Pi_{\rm{u}}^{\rm{O}}(q^2)=&-\frac{5c_4}{18\pi^2}\ln(-q^2)\langle
q^{\dag}iD_0q\rangle_{\rho}-\frac{c_4}{3q^2}\langle
q^{\dag}q\rangle_{\rho}^2,
\end{align}
where we have $q_0^2=(q\cdot u)^2,u=(1,\v{0})$, and $
c_1=7t^2-2t-5,
c_2=1-t^2,
c_3=2t^2-2t-1,
c_4=5t^2+2t+5$ and $
c_5=7t^2+10t+7$\,\cite{Jin94}. In this work, the following
condensates are included for the QCDSR equations,
\begin{align}
d=3:&~~\langle\overline{q}q\rangle_{\rho},\label{xx1}\\
d=4:&~~\left\langle({\alpha_{\rm{s}}}/{\pi})G^2\right\rangle_{\rho},\label{xx2}\\
d=5:&~~\langle
g_{\rm{s}}\overline{q}\sigma\mathcal{G}q\rangle_{\rho},~\langle
g_{\rm{s}}{q}^{\dag}\sigma\mathcal{G}q\rangle_{\rho},\label{xx3}\\
d=6:&~~\langle\overline{q}\Gamma_1q\overline{q}\Gamma_2q\rangle_{\rho},~
\langle\overline{q}\Gamma_1\lambda^Aq\overline{q}\Gamma_2\lambda^Aq\rangle_{\rho}\label{xx4}.
\end{align}

\subsection{QCDSR Equations in ANM}

Now we generalize the above results of SNM to the case of ANM, and the QCDSR equations for the proton are given by\,\cite{Jeo13},
\begin{align}
\lambda_{\rm{p}}^{\ast,2}M_{\rm{p}}^{\ast}e^{-({e}_{\rm{p}}^{2}(\rho,\v{q})-\v{q}^2)/\mathscr{M}^2}=&
-\frac{c_1}{16\pi^2}\mathscr{M}^4E_1\langle\overline{\rm{d}}\rm{d}\rangle_{\rho,\delta}\notag\\
&-\mathscr{C}_{5}\cdot\Bigg\{\frac{3c_2}{16\pi^2}\mathscr{M}^2E_0\langle
g_{\rm{s}}\overline{\rm{d}}\sigma\mathcal{G}\rm{d}\rangle_{\rho,\delta}L^{-4/9}\notag\\
&+\frac{2c_3}{3\pi^2}\v{q}^2\left[
\langle\overline{\rm{d}}iD_0iD_0\rm{d}\rangle_{\rho,\delta}
+\frac{1}{8}\langle
g_{\rm{s}}\overline{\rm{d}}\sigma\mathcal{G}\rm{d}\rangle_{\rho,\delta}\right]
L^{-4/9}\Bigg\}\notag\\
&-\mathscr{C}_{\rm{H}}\cdot\left[\frac{c_1}{8\pi^2}\overline{{e}_{\rm{p}}}\mathscr{M}^2E_0\langle\overline{\rm{d}}iD_0\rm{d}\rangle_{\rho,\delta}L^{-4/9}
{+\frac{c_1}{3}\overline{{e}_{\rm{p}}}\langle\overline{\rm{d}}\rm{d}\rangle_{\rho,\delta}
\langle\rm{d}^{\dag}\rm{d}\rangle_{\rho,\delta}}\right],\label{QSR_EQ_1a}\\
\lambda_{\rm{p}}^{\ast,2}e^{-({e}_{\rm{p}}^{2}(\rho,\v{q})-\v{q}^2)/\mathscr{M}^2}=&
\frac{c_4}{256\pi^4}\mathscr{M}^6E_2L^{-4/9}\notag\\
&-\mathscr{C}_{5}\cdot\Bigg[\frac{c_4}{72\pi^2}\mathscr{M}^2\left(E_0-\frac{4\v{q}^2}{\mathscr{M}^2}\right)\langle\rm{d}^{\dag}iD_0\rm{d}
\rangle_{\rho,\delta}L^{-4/9}\notag\\
&+\frac{c_4}{72\pi^2}\mathscr{M}^2\left(4E_0-\frac{4\v{q}^2}{\mathscr{M}^2}\right)
\langle\rm{u}^{\dag}iD_0\rm{u}\rangle_{\rho,\delta}L^{-4/9}\Bigg]\notag\\
&+\mathscr{C}_{4}\cdot\Bigg[\frac{c_4}{256\pi^2}\mathscr{M}^2E_0\left\langle\frac{\alpha_{\rm{s}}}{\pi}
G^2\right\rangle_{\rho,\delta}L^{-4/9}\notag\\
&-\frac{c_4}{1152\pi^2}\mathscr{M}^2\left(E_0-\frac{4\v{q}^2}{\mathscr{M}^2}\right)\left\langle\frac{\alpha_{\rm{s}}}{\pi}(\v{E}^2+\v{B}^2)\right\rangle_{\rho,\delta}
L^{-4/9}\Bigg]\notag\\
&+\mathscr{C}_{\rm{H}}\cdot\overline{{e}_{\rm{p}}}L^{-4/9}\Bigg\{
\frac{c_4}{48\pi^2}\mathscr{M}^2E_0\left[\langle\rm{u}^{\dag}\rm{u}\rangle_{\rho,\delta}+\langle\rm{d}^{\dag}\rm{d}\rangle_{\rho,\delta}\right]\notag\\
&\hspace*{0.5cm}-\frac{c_4}{12\pi^2}\left(2-\frac{\v{q}^2}{\mathscr{M}^2}\right)
\left[\langle\rm{u}^{\dag}
iD_0iD_0\rm{u}\rangle_{\rho,\delta}+\frac{1}{12}\langle
g_{\rm{s}}\rm{u}^{\dag}\sigma\mathcal{G}\rm{u}\rangle_{\rho,\delta}\right]\notag\\
&\hspace*{0.5cm}-\frac{c_4}{12\pi^2}\left(1-\frac{\v{q}^2}{\mathscr{M}^2}\right)\left[\langle\rm{d}^{\dag}
iD_0iD_0\rm{d}\rangle_{\rho,\delta}+\frac{1}{12}\langle
g_{\rm{s}}\rm{d}^{\dag}\sigma\mathcal{G}\rm{d}\rangle_{\rho,\delta}\right]
+\frac{c_5}{72\pi^2}\langle
g_{\rm{s}}\rm{d}^{\dag}\sigma\mathcal{G}\rm{d}\rangle_{\rho,\delta}
\Bigg\}\notag\\
&+\mathscr{C}_{\rm{B}}\cdot\frac{c_1}{6}
\widetilde{\langle\overline{\rm{d}}\rm{d}\rangle_{\rho,\delta}^2}L^{4/9}
+\mathscr{C}_{\rm{D}}\cdot\frac{c_4}{6}
\langle\rm{d}^{\dag}\rm{d}\rangle_{\rho,\delta}^2L^{-4/9}+B_{\rm{tw4}}^{\rm{II}},\label{QSR_EQ_2a}\\
\lambda_{\rm{p}}^{\ast,2}\Sigma_{\rm{V}}^{\rm{p}}e^{-({e}_{\rm{p}}^{2}(\rho,\v{q})-\v{q}^2)/\mathscr{M}^2}=&
\frac{c_4}{96\pi^2}\mathscr{M}^4E_1\left[7\langle\rm{u}^{\dag}\rm{u}\rangle_{\rho,\delta}+\langle\rm{d}^{\dag}\rm{d}\rangle_{\rho,\delta}\right]\notag\\
&+\mathscr{C}_{5}\cdot\Bigg\{\frac{3c_4}{8\pi^2}\v{q}^2\left[\langle\rm{u}^{\dag}
iD_0iD_0\rm{u}\rangle_{\rho,\delta}+\frac{1}{12}\langle
g_{\rm{s}}\rm{u}^{\dag}\sigma\mathcal{G}\rm{u}\rangle_{\rho,\delta}\right]L^{-4/9}\notag\\
&+\frac{c_4}{8\pi^2}\v{q}^2\left[\langle\rm{d}^{\dag}
iD_0iD_0\rm{d}\rangle_{\rho,\delta}+\frac{1}{12}\langle
g_{\rm{s}}\rm{d}^{\dag}\sigma\mathcal{G}\rm{d}\rangle_{\rho,\delta}\right]L^{-4/9}\notag\\
&-\frac{c_5}{24\pi^2}\mathscr{M}^2E_0\langle
g_{\rm{s}}\rm{u}^{\dag}\sigma\mathcal{G}\rm{u}\rangle_{\rho,\delta}L^{-4/9}+\frac{c_5}{48\pi^2}\mathscr{M}^2E_0\langle
g_{\rm{s}}\rm{d}^{\dag}\sigma\mathcal{G}\rm{d}\rangle_{\rho,\delta}L^{-4/9}\Bigg\}\notag\\
&+\mathscr{C}_{\rm{H}}\cdot\overline{{e}_{\rm{p}}}\mathscr{M}^2E_0L^{-4/9}\Bigg\{
\frac{c_4}{18\pi^2}\langle\rm{d}^{\dag}iD_0iD_0\rm{d}\rangle_{\rho,\delta}
+\frac{2c_4}{9\pi^2}\langle\rm{u}^{\dag}iD_0iD_0\rm{u}\rangle_{\rho,\delta}\notag\\
&\hspace*{0.5cm}+\frac{c_4}{288\pi^2}\left\langle\frac{\alpha_{\rm{s}}}{\pi}(\v{E}^2+\v{B}^2)\right\rangle_{\rho,\delta}
+\frac{c_4}{3}\langle\rm{d}^{\dag}\rm{d}\rangle_{\rho,\delta}^2\Bigg\}+B_{\rm{tw4}}^{\rm{III}}
,\label{QSR_EQ_3a}
\end{align}
\end{widetext}
where $\mathscr{M}$ is the Borel Mass and $\lambda_{\rm{p}}^{\ast}$
is the residual for quasi-proton, and
\begin{align}
{e}_{\rm{p}}(\rho,\v{q})=\left[{\v{q}^2+M_{\rm{p}}^{\ast,2}(\rho,\v{q})}\right]^{1/2}+\Sigma_{\rm{V}}^{\rm{p}}(\rho,\v{q})
\end{align}
together with
\begin{align}
\overline{{e}_{\rm{p}}}(\rho,\v{q})=-\left[{\v{q}^2+M_{\rm{p}}^{\ast,2}(\rho,\v{q})}\right]^{1/2}+\Sigma_{\rm{V}}^{\rm{p}}(\rho,\v{q}),
\end{align}
are the total effective single particle energy for proton (proton-hole).
In order to write down the QCDSR equations for the neutron, one
can exchange the quark fields of d and u. In the above
equations, $\langle\cdots\rangle_{\rho,\delta}$ represents the
condensates at finite density $\rho$ and isospin asymmetry
$\delta$. $B_{\rm{tw4}}^{\rm{II}}$ and $B_{\rm{tw4}}^{\rm{III}}$ are
contributions from twist-four four-quark condensates\,\cite{Jeo13}, which will be omitted
until Section~\ref{SEC_TWIST4} where we discuss their physical
effects in some detail. The wave line in the four-quark condensates means\,\cite{Coh95}
\begin{equation}\label{f4}
\widetilde{\langle\overline{q}q\rangle_{\rho,\delta}^2}=(1-f)\langle\overline{q}q\rangle_{\rm{vac}}^2
+f\langle\overline{q}q\rangle_{\rho,\delta}^2,\end{equation} with
$f$ an effective parameter introduced\,\cite{Cai17QCDSR}, and
\begin{equation}\label{def_L_a}
L^{-2\Gamma_{\eta}+\Gamma_{\mathcal{O}_n}}=\left[\frac{\ln(\mathscr{M}/\Lambda_{\rm{QCD}})}{\ln(\mu/\Lambda_{\rm{QCD}})}\right]^{-2\Gamma_{\eta}+\Gamma_{\mathcal{O}_n}}
\end{equation}
characterizes the anomalous dimension of the interpolation fields,
$\Lambda_{\rm{QCD}}\approx0.17\,\rm{GeV}$ is the QCD energy scale,
and $\mu\approx0.5\,\rm{GeV}$ is the corresponding renormalization
scale. The energy dependence of the QCD coupling constant, i.e.,
$\alpha_{\rm{s}}\equiv g_{\rm{s}}^2/4\pi$ is given by $
\alpha_{\rm{s}}(\v{q}^2)={4\pi}/{9\ln(\v{q}^2/\mu^2)}$. The factor
$L$ contains radiation effects effectively\,\cite{Jeo13}. Furthermore, the contributions from
continuum excitations whose physical origin will be discussed in
Section~\ref{SEC_MassVacuum}, are included by the following
functions,
\begin{align}
E_0=&1-e^{-s_0^{\ast}/\mathscr{M}^2},\label{def_E0}\\
E_1=&1-e^{-s_0^{\ast}/\mathscr{M}^2}\left(\frac{s_0^{\ast}}{\mathscr{M}^2}+1\right),\label{def_E1}\\
E_2=&1-e^{-s_0^{\ast}/\mathscr{M}^2}\left(\frac{s_0^{\ast,2}}{2\mathscr{M}^4}+\frac{s_0^{\ast}
}{\mathscr{M}^2}+1\right),\label{def_E2}
\end{align}
where we have $s_0^{\ast}=\omega_0^2-\v{q}^2$ with $\omega_0$ representing
the effect of continuum excitations.

In Eq.\,(\ref{QSR_EQ_1a}), Eq.\,(\ref{QSR_EQ_2a}) and
Eq.\,(\ref{QSR_EQ_3a}), the five parameters, i.e.,
$\mathscr{C}_{4},\mathscr{C}_5,\mathscr{C}_{\rm{H}},\mathscr{C}_{\rm{B}},\mathscr{C}_{\rm{D}}
$ are introduced. If they take the value ``+1'' then the
corresponding contributions are included, otherwise if they take the
value ``0'' then the corresponding contributions are absent. More specifically, $\mathscr{C}_{4}$
characterizes the four-dimensional condensates, $\mathscr{C}_5$
for the five-dimensional condensates, $\mathscr{C}_{\rm{H}}$ for
the contributions from quasi-hole effects, while
$\mathscr{C}_{\rm{B}}$ and $\mathscr{C}_{\rm{D}}$ characterize the
four-quark condensates of the types of
$\langle\overline{q}q\rangle_{\rho,\delta}^2$ and $\langle
q^{\dag}q\rangle_{\rho,\delta}^2$, respectively. In the following
analysis, effects from the anomalous dimension and the continuum
excitations denoted by the two parameters $\mathscr{C}_{\rm{A}}$ and
$\mathscr{C}_{\rm{C}}$, will be also studied. If
$\mathscr{C}_{\rm{A}}=0$, then $L=1$ and if
$\mathscr{C}_{\rm{C}}=0$, the functions $E_0,E_1$ and $E_2$ take the
value 1.

\subsection{Quark and Gluon Condensates Used in This Work}\label{sb_QG}

In this subsection, we discuss the properties of quark/gluon condensates with their density dependence, which are used
as input for QCDSR equations. For the
three-dimensional chiral condensate
$\langle\overline{q}q\rangle_{\rho,\delta}$, one could introduce\,\cite{Jeo13},
\begin{equation}
\langle\overline{q}q\rangle_{\rm{asym}}^{\rm{p}}=\frac{1}{2}\left[\langle\rm{p}|\overline{\rm{u}}\rm{u}|\rm{p}\rangle
-\langle\rm{p}|\overline{\rm{d}}\rm{d}|\rm{p}\rangle\right]
,\end{equation} where $|\rm{p}\rangle$ is the proton state. The mass
of a nucleon in vacuum can be represented through the trace of
energy-momentum tensor, i.e., $
M\langle\rm{N}|\overline{\psi}\psi|\rm{N}\rangle=\langle\rm{N}|\mathcal{T}|\rm{N}\rangle$,
where
\begin{align}
\mathcal{T}=&g_{\mu\nu}\mathcal{T}^{\mu\nu}=m_{\rm{u}}\overline{\rm{u}}\rm{u}+
m_{\rm{d}}\overline{\rm{d}}\rm{d}+m_{\rm{s}}\overline{\rm{s}}\rm{s}+\sum_{h=\rm{c,b,t}}m_h\overline{h}h\notag\\
\approx&\frac{\overline{\beta}}{4\alpha_{\rm{s}}}G^2+m_{\rm{u}}\overline{\rm{u}}\rm{u}+
m_{\rm{d}}\overline{\rm{d}}\rm{d}+m_{\rm{s}}\overline{\rm{s}}\rm{s}\end{align}
with $\mathcal{T}^{\mu\nu}$ the energy momentum tensor, $h$ denotes
the heavy quark field, $\overline{\beta}=-9\alpha_{\rm{s}}^2/2\pi$
is the reduced Gell-Mann--Low functions. For the baryon octet, we
have
\begin{align}
M_{\rm{p}}=&A+m_{\rm{u}}B_{\rm{u}}+m_{\rm{d}}B_{\rm{d}}+m_{\rm{s}}B_{\rm{s}},\\
M_{\rm{n}}=&A+m_{\rm{u}}B_{\rm{d}}+m_{\rm{d}}B_{\rm{u}}+m_{\rm{s}}B_{\rm{s}},\\
M_{\Sigma^+}=&A+m_{\rm{u}}B_{\rm{u}}+m_{\rm{d}}B_{\rm{s}}+m_{\rm{s}}B_{\rm{d}},\\
M_{\Sigma^-}=&A+m_{\rm{u}}B_{\rm{s}}+m_{\rm{d}}B_{\rm{u}}+m_{\rm{s}}B_{\rm{d}},\\
M_{\Xi^0}=&A+m_{\rm{u}}B_{\rm{d}}+m_{\rm{d}}B_{\rm{s}}+m_{\rm{s}}B_{\rm{u}},\\
M_{\Xi^-}=&A+m_{\rm{u}}B_{\rm{s}}+m_{\rm{d}}B_{\rm{d}}+m_{\rm{s}}B_{\rm{u}},\end{align}
with
$A=\langle(\overline{\beta}/4\alpha_{\rm{s}})G^2\rangle_{\rm{p}},B_{\rm{u}}=\langle\overline{\rm{u}}\rm{u}\rangle_{\rm{p}}
$, etc.  After straightforward calculations, we obtain
\begin{equation} \langle\rm{p}|\overline{\rm{u}}\rm{u}|\rm{p}\rangle
-\langle\rm{p}|\overline{\rm{d}}\rm{d}|\rm{p}\rangle
=\frac{(M_{\Xi^0}+M_{\Xi^-})-(M_{\Sigma^+}+M_{\Sigma^-})}{2m_{\rm{s}}-2m_{\rm{q}}}
\end{equation}
with $m_{\rm{q}}=2^{-1}({m_{\rm{u}}+m_{\rm{d}}})$.  On the other
hand, we have
\begin{equation}
\langle\overline{q}q\rangle_{\rm{sym}}^{\rm{p}}=\frac{1}{2}\left[\langle\rm{p}|\overline{\rm{u}}\rm{u}|\rm{p}\rangle
+\langle\rm{p}|\overline{\rm{d}}\rm{d}|\rm{p}\rangle\right]=\frac{\sigma_{\rm{N}}}{2m_{\rm{q}}},\end{equation}
in which the nucleon sigma term is $ \sigma_{\rm{N}}\equiv m_{\rm{q}}{\d M}/{\d
m_{\rm{q}}}\approx45\,\rm{MeV}$\,\cite{Gas91} (see also ref.~\cite{Ala12}). Introducing
$\langle\rm{p}|\overline{\rm{u}}\rm{u}|\rm{p}\rangle
\pm\langle\rm{p}|\overline{\rm{d}}\rm{d}|\rm{p}\rangle=a_{\pm}\langle\rm{p}|\overline{\rm{u}}\rm{u}|\rm{p}\rangle$,
then we can rewrite the asymmetric part in terms of the symmetric
part as
\begin{equation}
\langle\overline{q}q\rangle_{\rm{asym}}^{\rm{p}}=\frac{a_-}{a_+}\langle\overline{q}q\rangle_{\rm{sym}}^{\rm{p}}
,\end{equation} where\,\cite{Jeo13}
\begin{equation}\label{zz_apm}
a_{\pm}=1\pm\frac{\displaystyle\frac{\sigma_{\rm{N}}}{m_{\rm{q}}}-\frac{(m_{\Xi^0}+m_{\Xi^-})-(m_{\Sigma^+}+m_{\Sigma^-})}{2m_{\rm{s}}-2m_{\rm{q}}}
}{\displaystyle\frac{\sigma_{\rm{N}}}{m_{\rm{q}}}+\frac{(m_{\Xi^0}+m_{\Xi^-})-(m_{\Sigma^+}+m_{\Sigma^-})}{2m_{\rm{s}}-2m_{\rm{q}}}}
.\end{equation}  The masses
of the quarks and the baryons are given by\,\cite{PDG},
\begin{align}
m_{\rm{q}}\approx&3.5^{+0.7}_{-0.2}\,\rm{MeV},~~m_{\rm{s}}\approx95\pm5\,\rm{MeV},\\
M_{\Xi^0}\approx&1315\,\rm{MeV},~~M_{\Xi^-}\approx1321\,\rm{MeV},\\
M_{\Sigma^+}\approx&1190\,\rm{MeV},~~M_{\Sigma^-}\approx1197\,\rm{MeV},\end{align}
according to the particle data group (PDG), we then have
\begin{equation}\label{def_alphabeta}
a_{\pm}(\sigma_{\rm{N}})=1\pm\frac{{\sigma_{\rm{N}}}/{3.5}-{249}/{183}}{{\sigma_{\rm{N}}}/{3.5}+{249}/{183}}
.\end{equation}

In the following studies we denote $\alpha\equiv a_-$ and
$\beta\equiv a_+$ and $\xi=\alpha/\beta\approx0.1$. Collecting all the
elements discussed above we finally obtain the chiral condensate
at finite densities,
\begin{equation}\label{chiral_cond-1}
\langle\overline{q}q\rangle_{\rho,\delta}^{\rm{u,d}} \approx \langle\overline{q}q\rangle_{\rm{vac}}
+\left(1\mp\frac{\alpha}{\beta}\delta\right)\langle\overline{q}q\rangle_{\rm{sym}}^{\rm{p}}\rho
,\end{equation} where ``$-$'' is for the u quark and ``+'' for the d
quark. The corresponding condensate in the vacuum takes the
following value\,\cite{Jeo13,Coh95}
\begin{equation}
\langle\overline{q}q
\rangle_{\rm{vac}}\approx-(252\,\rm{MeV})^3.\end{equation} In Eq.\,(\ref{chiral_cond-1}), only the linear term in density
$\rho$ is considered. In Section~\ref{SEC_HighCond}, we will
consider possible higher order terms in density and study how
these higher order terms affect the nucleonic matter EOS.

Finally, we list other condensates of quarks and gluons. Another
three-dimensional condensate is $\langle
q^{\dag}q\rangle_{\rho,\delta}^{\rm{u,d}}$\,\cite{Coh95},
\begin{equation}\label{zz_rhodagrho}
\langle
q^{\dag}q\rangle_{\rho,\delta}^{\rm{u,d}}=\left(\frac{3}{2}\mp\frac{1}{2}\delta\right)\rho
,\end{equation}which is the actually the quark density\,\cite{Coh92,Fur92,Jin93,Jin94,Coh95,Jeo13}.
Other quark/gluon condensates include\,\cite{Coh92,Fur92,Jin93,Jin94,Coh95,Jeo13}:

1. Four-dimensional condensates,
\begin{align}
\langle\overline{q}iD_0q\rangle_{\rho,\delta}\approx&0,\\
\langle q^{\dag}iD_0q\rangle_{\rho,\delta}=&\langle
q^{\dag}iD_0q\rangle_{\rm{vac}}+\langle
q^{\dag}iD_0q\rangle_{\rm{fin}}\rho\notag\\
\approx&\left(1\mp\vartheta_1\right)\frac{1}{2}M\varphi_1\rho,
\end{align}
where $\vartheta_1\approx0.35,\varphi_1\approx0.55$\,\cite{Jeo13}. The lowest
order of gluon condensates have mass dimension four, i.e.,
\begin{align}
\left\langle\frac{\alpha_{\rm{s}}}{\pi}(\v{E}^2-\v{B}^2)\right\rangle_{\rho,\delta}
=&-\frac{1}{2}\left\langle\frac{\alpha_{\rm{s}}}{\pi}G^2\right\rangle_{\rm{vac}}\notag\\
&
+\left\langle\frac{\alpha_{\rm{s}}}{\pi}(\v{E}^2-\v{B}^2)\right\rangle_{\rm{fin}}\rho,\notag\\
\approx&-\frac{1}{2}(330\pm30\,\rm{MeV})^4\notag\\
&+(325\pm75\,\rm{MeV})\rho,\\
\left\langle\frac{\alpha_{\rm{s}}}{\pi}(\v{E}^2+\v{B}^2)\right\rangle_{\rho,\delta}
=&\left\langle\frac{\alpha_{\rm{s}}}{\pi}(\v{E}^2+\v{B}^2)\right\rangle_{\rm{fin}}\rho\notag\\
\approx&(100\pm10\,\rm{MeV})\rho.
\end{align}
where $G^2=2(\v{B}^2-\v{E}^2)$ with $\v{B}$ and $\v{E}$ the magnetic
field and electrical fields of QCD, respectively.

2. Five-dimensional condensates. In dimension five, we have
several condensates constructed from quarks and gluons,
\begin{align}
\langle\overline{q}iD_0iD_0q\rangle_{\rho,\delta}\approx&
\langle\overline{q}iD_0iD_0q\rangle_{\rm{fin}}\rho\notag\\
\approx&\left(1\mp\frac{\alpha}{\beta}\delta\right)M^2\varphi_2\rho
\end{align}
where $\varphi_2\approx0.34$, and,
\begin{align}
\langle
g_{\rm{s}}\overline{q}\sigma\mathcal{G}q\rangle_{\rho,\delta}\approx&\left(1\mp\frac{\alpha}{\beta}\delta\right)\langle
g_{\rm{s}}\overline{q}\sigma\mathcal{G}q\rangle_{\rm{sym}}^{\rm{p}}\rho,
\end{align}
with $(620\,\rm{MeV})^2\leq\langle
g_{\rm{s}}\overline{q}\sigma\mathcal{G}q\rangle_{\rm{sym}}^{\rm{p}}\leq(3\,\rm{GeV})^2$,
and
\begin{align}
\langle{q}^{\dag}iD_0iD_0q\rangle_{\rho,\delta}=&\langle{q}^{\dag}iD_0iD_0q\rangle_{\rm{fin}}\rho\notag\\
\approx& (1\mp\vartheta_3\delta)\frac{1}{2}M^2\varphi_3\rho
\end{align}
with $\vartheta_3\approx0.51,\varphi_3\approx0.145$. Similarly, we have
\begin{equation}\label{yy1} \langle
g_{\rm{s}}{q}^{\dag}\sigma\mathcal{G}q\rangle_{\rho,\delta}\approx\left(1\mp\vartheta_3\delta\right)\langle
g_{\rm{s}}{q}^{\dag}\sigma\mathcal{G}q\rangle_{\rm{sym}}^{\rm{p}}\rho
\end{equation}
where $-(330\,\rm{MeV})^2\leq\langle
g_{\rm{s}}{q}^{\dag}\sigma\mathcal{G}q\rangle_{\rm{sym}}^{\rm{p}}\leq(660\,\rm{MeV})^2$.

3. For the six-dimensional condensates, we consider the
effective four-quark condensates defined in Eq.\,(\ref{f4}).
And in Section~\ref{SEC_TWIST4}, we study the twist-four four-quark condensates effects on the quantities we are interested in.

For more detailed physical discussions on the condensates, see, e.g.,
refs.\,\cite{Coh92,Fur92,Jin93,Jin94} and ref.\,\cite{Coh95}.

\subsection{Fitting Scheme}\label{sb_FS}

In this work, the quark/gluon condensates at finite densities up to mass dimension-six
are included in the QCDSR equations, see Eqs.\,(\ref{xx1}), (\ref{xx2}), (\ref{xx3}), and (\ref{xx4}),
and the default central values are listed in the last subsection, see Eqs.\,(\ref{chiral_cond-1}) to (\ref{yy1}).
Moreover, the quark masses are taken to be $m_{\rm{q}}=3.5\,\rm{MeV}$, $m_{\rm{s}}=95\,\rm{MeV}$, the $\sigma$-N term $\sigma_{\rm{N}}=45\,\rm{MeV}$,
the Borel mass $\mathscr{M}^2=1.05\,\rm{GeV}^2$\,\cite{Iof84}, and the threshold parameter defined in Eqs.\,(\ref{def_E0}), (\ref{def_E1}), and (\ref{def_E2}), as $\omega_0=1.5\,\rm{GeV}$\,\cite{Coh95} except a slight different $\omega_0=1.4\,\rm{GeV}$ in Section~\ref{SEC_EsymStru} and  Section~\ref{SEC_HOEsti}.

Starting from Section~\ref{SEC_HighCond}, we consider the following quark chiral condensates,
\begin{equation}\label{chiral_cond}
\langle\overline{q}q\rangle_{\rho,\delta}\approx\langle\overline{q}q\rangle_{\rm{vac}}
+\frac{\sigma_{\rm{N}}}{2m_{\rm{q}}}\left(1\mp\xi\delta\right)\rho
+\Phi(1\mp g\delta)\rho^2 ,
\end{equation}
The motivation for including the last term ``$\Phi(1\mp
g\delta)\rho^2$'' in Eq.\,(\ref{chiral_cond}) is as follows\,\cite{Cai17QCDSR}: As the
density increases, the linear approximation for the chiral
condensates becomes worse eventually, {and}
higher order terms in density should be included in the $\langle \overline{q}q\rangle_{\rho,\delta}$. However, the density dependence of the chiral condensates is extremely complicated, and there {is} no general power counting scheme to incorporate these higher order density terms. Besides the $\rho^2$ term we adopted here, for instance,
based on the chiral effective theories\,\cite{Kai09,Kru13a}, a term proportional to $\rho^{5/3}$ was found
in the perturbative expansion of $\langle\overline{q}q\rangle_{\rho,\delta}$ in
$\rho$. On the other hand, using the chiral Ward identity\,\cite{God13}, a $\rho^{4/3}$ term
was found in the density expansion in the chiral condensates.
In our work, including the higher-order $\rho^2$ term is mainly for the improvement
of describing the empirical EOS of PNM {around and above saturation density,
for which we use the celebrated Akmal--Pandharipande--Ravenhall (APR) EOS\,\cite{APR}.
In this sense, the $\Phi$-term adopted here is an effective
correction to the chiral condensates beyond the linear leading-order.
Two aspects related to the $\Phi$-term should be pointed out: 1).
Without the higher-order $\rho^2$ term, the EOS of PNM {around and above saturation density} can not
be adjusted to be consistent with that APR EOS, i.e., there exists
systematic discrepancy between the QCDSR EOS and the APR EOS;
2). Using an effective correction with a different power in density, e.g., a $\rho^{5/3}$ term,
the conclusion does not change, i.e., the EOS
of PNM around and above saturation density can still be adjusted to fit the APR EOS,
and the sign of the coefficients $\Phi$ and $g$ will not change although their absolute values change, see the results in Fig.\,\ref{fig_R2PNM}.
In the following, we abbreviate the
QCDSR using the chiral condensate without the last term in
Eq.\,(\ref{chiral_cond}) as the ``naive QCDSR''\,\cite{Cai17QCDSR}, which will be explored in detail in Subsection~\ref{ss_4}.

In carrying out the QCDSR calculations, we fix the central
value of the $E_{\rm{n}}(\rho)$ at a very low density
$\rho_{\rm{vl}}=0.02\,\rm{fm}^{-3}$ to be consistent with the
prediction by the ChPT\,\cite{Tew13,Kru13},
i.e., $E_{\rm{n}}(\rho_{\rm{vl}})\approx4.2\,\rm{MeV}$,
the central value of the symmetry energy $E_{\rm{sym}}(\rho)$ at a critical density
$\rho_{\rm{c}}=0.11\,\rm{fm}^{-3}$ to be $E_{\rm{sym}}(\rho_{\rm{c}}) \approx26.65$ MeV\,\cite{Zha13},
and fit the EOS of PNM be close to the APR EOS as much as possible,
via varying $\Phi$, $g$ and $f$.
{We note that} the parameter $f$ defined in Eq.\,(\ref{f4}) is {essentially} determined by
$E_{\rm{n}}(\rho_{\rm{vl}})$, and the overall fitting of the EOS of PNM to the APR EOS and the symmetry energy at $\rho_{\rm{c}}$
determines the other two parameters $\Phi$ and $g$\,\cite{Cai17QCDSR}.

\setcounter{equation}{0}
\section{Nucleon Mass in Vacuum}\label{SEC_MassVacuum}

In this section, we use the QCDSR to study the nucleon mass in vacuum.
The motivation is twofold: firstly the connection between the nucleon mass and the quark/gluon condensates
in QCDSR is explored, and most importantly the scheme for determining the Ioffe parameter $t$\,\cite{Iof81}
is given here, i.e., via the relation $M_{\rm{QCDSR}}^{\rm{vac,static}}(t)=939\,\rm{MeV}$.

Conventionally, the high energy states or the continuum states are not given by the QCDSR method itself,
and in order to model these states in the nucleon spectral densities (\ref{sd-1}) and (\ref{sd-2}),
one usually adopts the corresponding results from the OPE calculations\,\cite{Coh95}.
More specifically, the nucleon correlation functions from the OPE side including only the lowest order terms
are given by
\begin{align}
\Pi_{\rm{s}}^{\rm{OPE}}(q^2)=&\frac{q^2}{4\pi}\ln(-q^2)\langle\overline{q}q\rangle_{\rm{vac}},\\
\Pi_{\rm{q}}^{\rm{OPE}}(q^2)=&-\frac{(q^2)^2}{64\pi^4}\ln(-q^2)
-\frac{1}{32\pi^2}\ln(-q^2)\left\langle\frac{\alpha_{\rm{s}}}{\pi}G^2\right\rangle_{\rm{vac}}\notag\\
&-\frac{2}{3q^2}\langle\overline{q}q\rangle_{\rm{vac}}^2,\end{align}
where the last term in $\Pi_{\rm{q}}^{\rm{OPE}}(q^2)$ is the four-quark condensates expressed in terms of
the square of the chiral condensate. Moreover, $t=-1$ is adopted here for simplicity. The Borel transformations of them are given by
\begin{align}
\widehat{\Pi}_{\rm{s}}^{\rm{OPE}}(\mathscr{M}^2)=&-\frac{1}{4\pi^2}\mathscr{M}^4\langle\overline{q}q\rangle_{\rm{vac}},\label{OPE-ss}\\
\widehat{\Pi}_{\rm{q}}^{\rm{OPE}}(\mathscr{M}^2)=&\frac{1}{32\pi^4}\mathscr{M}^6+\frac{1}{32\pi^2}\mathscr{M}^4\left\langle\frac{\alpha_{\rm{s}}}{\pi}G^2\right\rangle_{\rm{vac}}\notag\\
&+\frac{2}{3}\langle\overline{q}q\rangle_{\rm{vac}}^2.\label{OPE-qq}
\end{align}
Consequently, the high energy/continuum states are approximated by the equivalent OPE terms, starting at
the sharp threshold $s_0=\omega_0^2$, i.e.\,\cite{Coh95},
\begin{align}
\rho_{\rm{s}}^{\rm{phen}}(s)=&\lambda_{}^2M_{}\delta(s-M_{}^2)
-\frac{1}{4\pi^2}s\langle\overline{q}q\rangle_{\rm{vac}}\Theta(s-s_0),\label{z_1}\\
\rho_{\rm{q}}^{\rm{phen}}(s)=&\lambda_{}^2\delta(s-M_{}^2)\notag\\
&+\left[
\frac{1}{64\pi^4}s^2+\frac{1}{32\pi^2}\left\langle\frac{\alpha_{\rm{s}}}{\pi}G^2\right\rangle_{\rm{vac}}
\right]\Theta(s-s_0),\label{z_2}\end{align} see Fig.\,\ref{fig_HighSpec_12}
for the sketch of the effects of $s_0$.
\begin{figure}[h!]
\centering
\includegraphics[width=7.5cm]{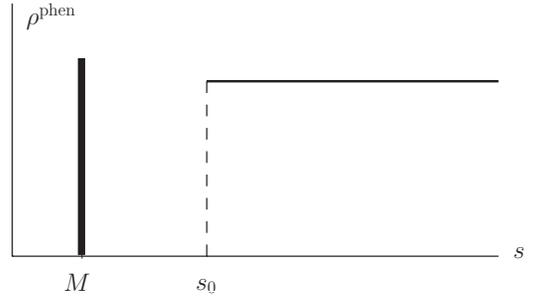}
  \caption{Sketch of the spectral densities (\ref{z_1}) and (\ref{z_2}).}\label{fig_HighSpec_12}
\end{figure}

After putting these spectral functions into Eq.\,(\ref{Poly_for}) and
finishing the Borel transformation, we obtain the correlation
functions at the phenomenological side in QCDSR,
\begin{align}
\widehat{\Pi}_{\rm{s}}^{\rm{phen}}(\mathscr{M}^2)
=&\lambda_{}^2M_{}e^{-M_{}^2/\mathscr{M}^2}\notag\\
&-\frac{\mathscr{M}^4}{4\pi^2}\langle\overline{q}q\rangle_{\rm{vac}}\left(1+\frac{s_0}{\mathscr{M}^2}
\right)e^{-s_0/\mathscr{M}^2},\\
\widehat{\Pi}_{\rm{q}}^{\rm{phen}}(\mathscr{M}^2)
=&\lambda_{}^2e^{-M_{}^2/\mathscr{M}^2}\notag\\
&+\frac{\mathscr{M}^6}{32\pi^4}\left(1+\frac{s_0}{\mathscr{M}^2}
+\frac{s_0^2}{2\mathscr{M}^4}\right)e^{-s_0/\mathscr{M}^2} \notag\\
&+\frac{\mathscr{M}^2}{32\pi^2}\left\langle\frac{\alpha_{\rm{s}}}{\pi}G^2\right\rangle_{\rm{vac}}
e^{-s_0/\mathscr{M}^2}.
\end{align}
Then according to the
QCDSR, i.e.,
\begin{align}
\widehat{\Pi}_{\rm{s}}^{\rm{phen}}(\mathscr{M}^2)=&\widehat{\Pi}_{\rm{s}}^{\rm{OPE}}(\mathscr{M}^2)
,\\
\widehat{\Pi}_{\rm{q}}^{\rm{phen}}(\mathscr{M}^2)=&\widehat{\Pi}_{\rm{q}}^{\rm{OPE}}(\mathscr{M}^2)
,\end{align}
where $\widehat{\Pi}_{\rm{s}}^{\rm{OPE}}(\mathscr{M}^2)$ and $\widehat{\Pi}_{\rm{q}}^{\rm{OPE}}(\mathscr{M}^2)$
are given by Eq.\,(\ref{OPE-ss}) and Eq.\,(\ref{OPE-qq}), respectively, one obtains the expression for nucleon mass in vacuum as\,\cite{Footnotexx},
\begin{equation}
M_{}=\frac{2\phi\mathscr{M}^4E_1}{\displaystyle
\mathscr{M}^6E_2+b\mathscr{M}^2E_0+{4\phi^2}/{3}},\label{for_Ioffe_1}
\end{equation}
where $E_0,E_1$ and $E_2$ are defined in Eq.\,(\ref{def_E0}),
Eq.\,(\ref{def_E1}) and Eq.\,(\ref{def_E2}), and the following
abbreviations are introduced,
\begin{equation}
\phi=-(2\pi)^2\langle\overline{q}q\rangle_{\rm{vac}},~~b=\pi^2\left\langle
\frac{\alpha_{\rm{s}}}{\pi} G^2\right\rangle_{\rm{vac}}
.\end{equation} Furthermore, if we neglect all the high-dimensional
except the three-dimensional chiral
condensates, a very simple formula for the nucleon mass in
vacuum is obtained, i.e.,
\begin{equation}
M=-\frac{8\pi^2}{\mathscr{M}^2}\langle\overline{q}q\rangle_{\rm{vac}},
\end{equation}
demonstrating that the nucleon mass in vacuum is roughly determined by
the the chiral condensates in vacuum.

\begin{figure}[h!]
\centering
\includegraphics[width=8.5cm]{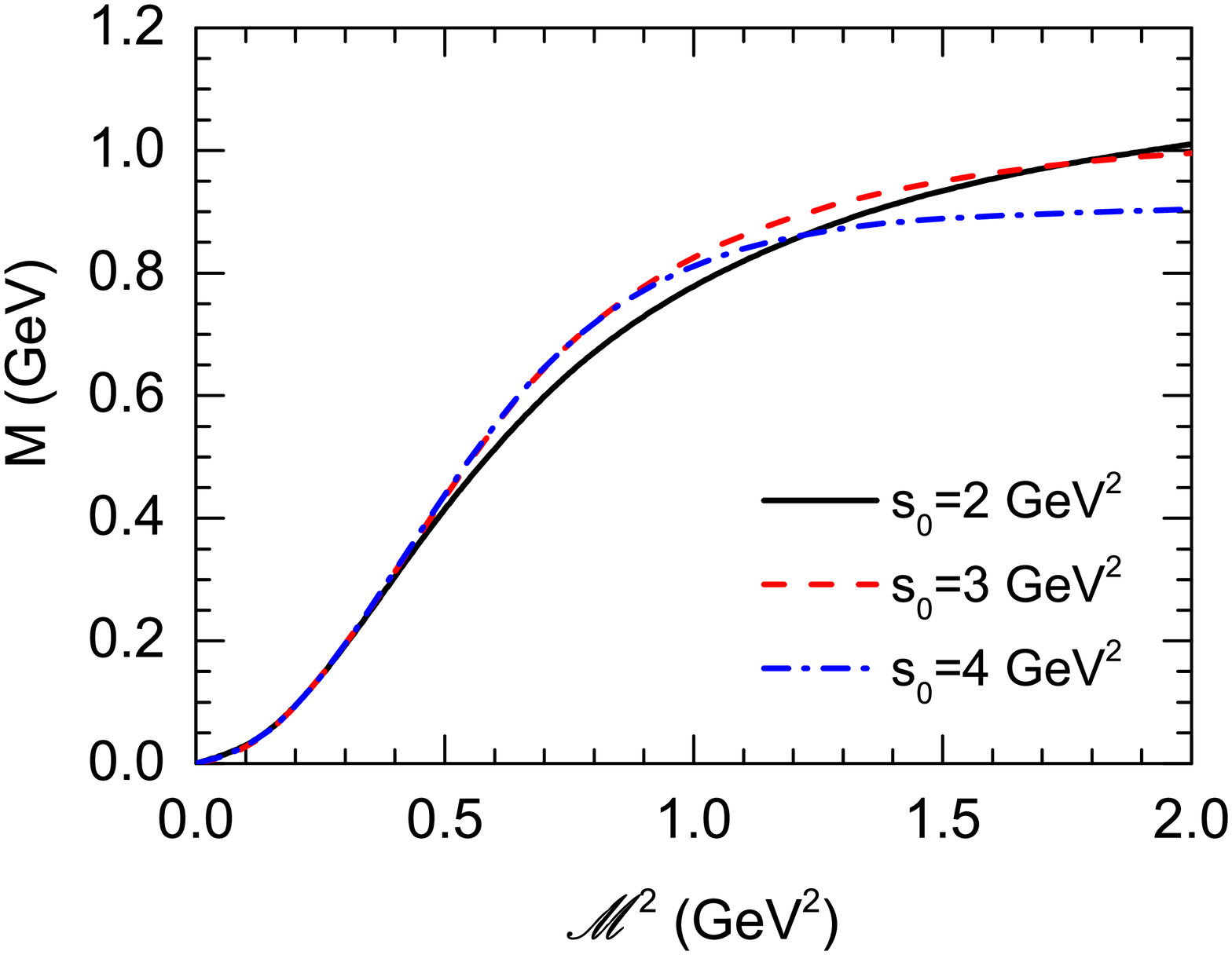}
  \caption{(Color Online) Nucleon mass in vacuum as a function of Borel mass squared by Eq.\,(\ref{for_Ioffe_1}).}\label{fig_MBM}
\end{figure}

Moreover, the above discussion clearly shows that the nucleon mass in
vacuum obtained in QCDSR is not necessarily about its realistic value 939\,MeV.
For instance, taking $\mathscr{M}^2=1.05\,\rm{GeV}^2$, $\langle\overline{q}q\rangle_{\rm{vac}}=-(0.252\,\rm{GeV})^3$, $\langle (\alpha_{\rm{s}}/\pi)G^2\rangle_{\rm{vac}}
=(0.33\,\rm{GeV})^4$, and $s_0=\omega_0^2=2.25\,\rm{GeV}^2$, one obtains the mass $M\approx0.847\,\rm{GeV}$, which has about a 92\,MeV deviation from 939\,MeV (roughly an effect
about 10\%),
and in Fig.\,\ref{fig_MBM}, the nucleon mass in vacuum as
a function of the Borel mass (squared) obtained by Eq.\,(\ref{for_Ioffe_1}) is shown.
In this work, we put an extra
constraint on the QCDSR method, i.e., the nucleon mass in vacuum
is set to be 939\,MeV. Consequently the Ioffe parameter $t$ should be determined
uniquely once the other parameters are fixed.
Specifically, when we rewrite the QCDSR equations in vacuum under the static condition, i.e., taking $\rho=0$ and $\v{q}=\v{0}$, one obtains,
\begin{align}
\lambda^{2}M_{}e^{-({e}^{2}(\rho,\v{0}))/\mathscr{M}^2}=&
-\frac{c_1}{16\pi^2}\mathscr{M}^4E_1\langle\overline{q}q\rangle_{\rm{vac}},\\
\lambda^{2}e^{-({e}^{2}(\rho,\v{0}))/\mathscr{M}^2}=&
\frac{c_4}{256\pi^4}\mathscr{M}^6E_2L^{-4/9}\notag\\
&+\frac{c_4}{256\pi^2}
\mathscr{M}^2E_0\left\langle\frac{\alpha_{\rm{s}}}{\pi}G^2
\right\rangle_{\rm{vac}}\notag\\
&\times L^{-4/9}
+\frac{c_1}{6}\langle\overline{q}q\rangle_{\rm{vac}}^2L^{4/9},
\end{align}
where $c_1=7t^2-2t-5$ and $c_4=5t^2+2t+5$. Then the Ioffe parameter $t$ could be expressed as
\begin{align}\label{def_Ioffe}
t=&\frac{\sqrt{12(F+H+1)(3F+3H-2)}}{7{F}+7{H}-5}\notag\\
&+\frac{{F}+{H}+1}{7{F}+7{H}-5}
\end{align}
with
\begin{align}
{F}=&-\frac{16}{{M}}\frac{\displaystyle\mathscr{M}^2E_1\langle\overline{q}q\rangle_{\rm{vac}}L^{4/9}}{\displaystyle
\frac{\mathscr{M}^4E_2}{\pi^2}+E_0\left\langle\frac{\alpha_{\rm{s}}}{\pi}G^2
\right\rangle_{\rm{vac}}},\\
{H}=&-\frac{128\pi^2}{3\mathscr{M}^2}\frac{\displaystyle
\langle\overline{q}q\rangle_{\rm{vac}}^2L^{8/9}}{\displaystyle
\frac{\mathscr{M}^4E_2}{\pi^2}+E_0\left\langle\frac{\alpha_{\rm{s}}}{\pi}G^2
\right\rangle_{\rm{vac}}}.
\end{align}

\begin{figure}[h!]
\centering
  \includegraphics[width=8.5cm]{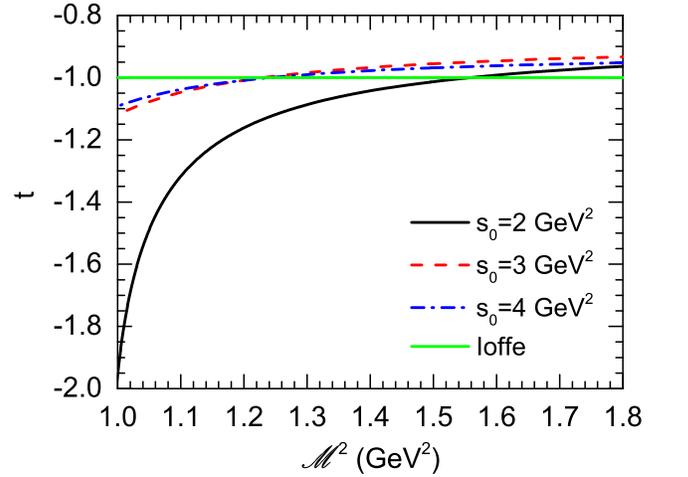}
  \caption{(Color Online) Ioffe Parameter $t$ as a function of the Borel mass squared with different $s_0$.}
  \label{fig_t_Borel}
\end{figure}

In Fig.\,\ref{fig_t_Borel}, the Ioffe parameter $t$ as a function
of the Borel mass squared $\mathscr{M}^2$ with different threshold parameter $s_0$ is shown.
The Borel mass
squared was extensively studied in the literature and the value of it was found to fall within the range
$0.8\,\rm{GeV}^2\lesssim\mathscr{M}^2\lesssim1.4\,\rm{GeV}^2$\,\cite{Iof84,Coh95}.
In this work, we choose $ \mathscr{M}^2\approx1.05\,\rm{GeV}^2 $ as
our default value for the Borel mass squared, and the Ioffe parameter obtained via Eq.\,(\ref{def_Ioffe}) is found to be around $-1.2$ considering $2\,\rm{GeV}^2\lesssim
s_0\lesssim3\,\rm{GeV}^2$, which is close to the Ioffe value ($t_{\rm{Ioffe}}=-1$).
For example, taking $s_0=3\,\rm{GeV}^2$, the Ioffe parameter at $\mathscr{M}^2=1.05\,\rm{GeV}^2$ is found to be about $t\approx-1.04$.
Moreover, the Ioffe parameter will be set to its default value ($t_{\rm{Ioffe}}=-1$) in Section~\ref{SEC_EsymStru} and Section~\ref{SEC_HOEsti}
mainly for qualitative discussions,
and starting from Section~\ref{SEC_FullQSR}, $t$ would be self-consistently determined by the scheme given here, i.e., Eq.\,(\ref{def_Ioffe}).
The treatment on the Ioffe parameter given here is natural, thus one can obtain the physical nucleon mass in vacuum when the QCDSR equations
are applied at zero density.

\setcounter{equation}{0}
\section{Main Structure of The Symmetry Energy}\label{SEC_EsymStru}

In this section, we discuss the structure of the symmetry energy through
the simplified QCDSR equations, which will be given shortly. The main purpose of this section is to qualitatively analyze the nucleon self-energy structure of the symmetry energy
through Eq.\,(\ref{sec1_Esym}), and leave the detailed/quantitative calculation on the $E_{\rm{sym}}(\rho)$ to the following sections.

In the simplified version, the QCDSR equations for proton have the following form,
\begin{align}
\lambda_{\rm{p}}^{\ast,2}M_{\rm{p}}^{\ast}e^{-({e}_{\rm{p}}^2(\rho,\v{q})-\v{q}^2)/\mathscr{M}^2}=&-\frac{\mathscr{M}^4}{4\pi^2}
\langle\overline{\rm{d}}\rm{d}\rangle_{\rho,\delta},\label{msQCDSRp1}\\
\lambda_{\rm{p}}^{\ast,2}e^{-({e}_{\rm{p}}^2(\rho,\v{q})-\v{q}^2)/\mathscr{M}^2}=&\frac{\mathscr{M}^6}{32\pi^4},\label{msQCDSRp2}\\
\lambda_{\rm{p}}^{\ast,2}\Sigma^{\rm{p}}_{\rm{V}}e^{-({e}_{\rm{p}}^2(\rho,\v{q})-\v{q}^2)/\mathscr{M}^2}=&\frac{1}{12\pi^2}\mathscr{M}^4\notag\\
&\times\left(7\langle\rm{u}^{\dag}\rm{u}\rangle_{\rho,\delta}+\langle\rm{d}^{\dag}\rm{d}\rangle_{\rho,\delta}\right),\label{msQCDSRp3}\end{align}
i.e., they only include the three-dimensional chiral
condensates, and all the high-dimensional condensates as well as
the contributions from continuum excitations, quasi-hole effects are
neglected. In the above expressions, $M_{\rm{p}}^{\ast}$ is the
proton's Dirac effective mass at finite densities introduced through the following dispersion relation,\begin{equation}\label{zz_1}
{e}_{\rm{p}}(\rho,\v{q})=[M_{\rm{p}}^{\ast,2}(\rho,\v{q})+\v{q}^2]^{1/2}+\Sigma_{\rm{V}}^{\rm{p}}(\rho,\v{q})
.\end{equation}
The discussion in this section on the symmetry energy from QCDSR is more qualitative and thus the Ioffe parameter $t$ is set at its natural value, i.e., $t=-1$, as mentioned in the last section.
Via exchanging the d quark and
the u quark, one can obtain the corresponding sum rule equations for
neutron,
i.e.,
\begin{align}
\lambda_{\rm{n}}^{\ast,2}M_{\rm{n}}^{\ast}e^{-({e}_{\rm{n}}^2(\rho,\v{q})-\v{q}^2)/\mathscr{M}^2}=&-\frac{\mathscr{M}^4}{4\pi^2}
\langle\overline{\rm{u}}\rm{u}\rangle_{\rho,\delta},\label{msQCDSRn1}\\
\lambda_{\rm{n}}^{\ast,2}e^{-({e}_{\rm{n}}^2(\rho,\v{q})-\v{q}^2)/\mathscr{M}^2}=&\frac{\mathscr{M}^6}{32\pi^4},\label{msQCDSRn2}\\
\lambda_{\rm{n}}^{\ast,2}\Sigma^{\rm{n}}_{\rm{V}}e^{-({e}_{\rm{n}}^2(\rho,\v{q})-\v{q}^2)/\mathscr{M}^2}=&\frac{1}{12\pi^2}\mathscr{M}^4\notag\\
&\times\left(7\langle\rm{d}^{\dag}\rm{d}\rangle_{\rho,\delta}+\langle\rm{u}^{\dag}\rm{u}\rangle_{\rho,\delta}\right),\label{msQCDSRn3}\end{align}
and the total single particle energy for neutron is defined similarly as Eq.\,(\ref{zz_1}).
The QCDSR with Eqs.\,(\ref{msQCDSRp1}), (\ref{msQCDSRp2}),  (\ref{msQCDSRp3}), (\ref{msQCDSRn1}), (\ref{msQCDSRn2}), and
(\ref{msQCDSRn3}) and the dispersion relation for proton/neutron, is called the mostly simplified QCDSR (``msQCDSR'').
It is necessary to point out that in ``msQCDSR'', the nucleon self-energies have no momentum dependence,
thus the corresponding momentum-dependent terms in Eq.\,(\ref{sec1_Esym}) are absent naturally.
The momentum-dependent terms and their consequences will be explored in some detail in Section~\ref{SEC_FullQSR},
and the corresponding contributions to the symmetry energy are found to be small (see the blue and green lines of Fig.\,\ref{fig_EsymDecomp}).

The nucleon Dirac effective mass $M_J^{\ast}$ and the vector self-energy $\Sigma_{\rm{V}}^J$ could be
obtained immediately in msQCDSR as,
\begin{align}
M_J^{\ast}(\rho)=&-\frac{8\pi^2}{\mathscr{M}^2}\langle\overline{q}q\rangle_{\rho,\delta}^{\rm{u,d}},\label{zz_MJ}\\
\Sigma_{\rm{V}}^{J}(\rho)=&\frac{8\pi^2}{3\mathscr{M}^2}\left(8\langle{q}^{\dag}q\rangle_{\rho,\delta}^{\rm{u,d}}\right)\label{zz_SVJ}.
\end{align}
The nucleon Dirac effective mass in SNM is then given by
\begin{equation}\label{zz_M0ast}
M_0^{\ast}(\rho)=-\frac{8\pi^2}{\mathscr{M}^2}\langle\overline{q}q\rangle_{\rho}
\approx-\frac{8\pi^2}{\mathscr{M}^2}\left(\langle\overline{q}q\rangle_{\rm{vac}}+\frac{\sigma_{\rm{N}}\rho}{2m_{\rm{q}}}\right),
\end{equation}
and the vector-self energy by \begin{equation}
\Sigma_{\rm{V}}^0(\rho)=\frac{64\pi^2}{3\mathscr{M}^2}\langle
q^{\dag}q\rangle_{\rho},\end{equation}
with
$\langle q^{\dag}q\rangle_{\rho}=3\rho/2$.

Similarly, the nucleon Dirac effective mass and the vector self-energy in ANM are given by
\begin{align}
M_{\rm{p}}^{\ast}(\rho)=&-\frac{8\pi^2}{\mathscr{M}^2}\left[\langle\overline{q}q\rangle_{\rm{vac}}+\frac{\sigma_{\rm{N}}\rho}{2m_{\rm{q}}}\left(1+\frac{\alpha}{\beta}\delta\right)\right],\label{xy-1}\\
M_{\rm{n}}^{\ast}(\rho)=&-\frac{8\pi^2}{\mathscr{M}^2}\left[\langle\overline{q}q\rangle_{\rm{vac}}+\frac{\sigma_{\rm{N}}\rho}{2m_{\rm{q}}}\left(1-\frac{\alpha}{\beta}\delta\right)\right],\label{xy-2}
\end{align}
and
\begin{align}
\Sigma_{\rm{V}}^{\rm{p}}(\rho)=&\frac{8\pi^2}{3\mathscr{M}^2}\left(7\langle\rm{u}^{\dag}\rm{u}\rangle_{\rho,\delta}+\langle\rm{d}^{\dag}\rm{d}\rangle_{\rho,\delta}\right),\label{xy-3}\\
\Sigma_{\rm{V}}^{\rm{n}}(\rho)=&\frac{8\pi^2}{3\mathscr{M}^2}\left(7\langle\rm{d}^{\dag}\rm{d}\rangle_{\rho,\delta}+\langle\rm{u}^{\dag}\rm{u}\rangle_{\rho,\delta}\right),\label{xy-4}
\end{align}
respectively. The definitions of $\alpha$ and $\beta$ are given in Eq.\,(\ref{def_alphabeta}).
Consequently, the first-order symmetry scalar/vector self-energy is obtained as
\begin{align}
\Sigma_{\rm{sym}}^{\rm{S}}(\rho)=&\frac{\Sigma_{\rm{S}}^{\rm{n}}-\Sigma_{\rm{S}}^{\rm{p}}}{2\delta}=\frac{4\pi^2\sigma_{\rm{N}}}{\mathscr{M}^2m_{\rm{q}}}\frac{\alpha}{\beta}\rho,
\label{zz_ESsym}\\
\Sigma_{\rm{sym}}^{\rm{V}}(\rho)=&\frac{\Sigma_{\rm{V}}^{\rm{n}}-\Sigma_{\rm{V}}^{\rm{p}}}{2\delta}=\frac{8\pi^2}{\mathscr{M}^2}\rho,\end{align}
where the symmetry self-energy is generally defined as\,\cite{Cai12}
\begin{align}
\Sigma^{\rm{S/V}}_{\rm{sym},i}(\rho,\v{q})
=&\frac{1}{i!}\frac{\partial^i}{\partial\delta^i}\left[\sum_{J=\rm{n,p}}\frac{\tau_3^{J,i}
\Sigma^{J}_{\rm{S/V}}(\rho,\delta,
\v{q})}{2}\right]_{\delta=0}.
\end{align}
In the above definition, we use $\tau_3^{\rm{n}}=+1$ and $\tau_3^{\rm{p}}=-1$.

Through Eq.\,(\ref{sec1_Esym}), one can obtain the symmetry energy in msQCDSR as
\begin{align}\label{Esym-msQCDSR}
E_{\rm{sym}}(\rho)=&\frac{k_{\rm{F}}^2}{6{e}_{\rm{F}}^{\ast}}+{\frac{2\pi^2\sigma_{\rm{N}}}{\mathscr{M}^2m_{\rm{q}}}
\frac{M_0^{\ast}}{{e}_{\rm{F}}^{\ast}}\frac{\alpha}{\beta}\rho
+ \frac{4\pi^2\rho}{\mathscr{M}^2}}
,\end{align}
with its decomposition given by
\begin{align}
E_{\rm{sym}}^{\rm{kin}}(\rho)=&\frac{k_{\rm{F}}^2}{6{e}_{\rm{F}}^{\ast}},\label{ddd1}\\
E_{\rm{sym}}^{\rm{1st,S}}(\rho)=&\frac{1}{2}\frac{M_0^{\ast}}{{e}_{\rm{F}}^{\ast}}
\Sigma_{\rm{sym}}^{\rm{S}}(\rho)=\frac{2\pi^2\sigma_{\rm{N}}}{\mathscr{M}^2m_{\rm{q}}}
\frac{M_0^{\ast}}{{e}_{\rm{F}}^{\ast}}\frac{\alpha}{\beta}\rho,\label{ddd2}\\
E_{\rm{sym}}^{\rm{1st,V}}(\rho)=&\frac{1}{2}\Sigma_{\rm{sym}}^{\rm{V}}(\rho)=
\frac{4\pi^2\rho}{\mathscr{M}^2},\label{ddd3}
\end{align}
where $e_{\rm{F}}^{\ast}=(k_{\rm{F}}^2+M_0^{\ast,2})^{1/2}$,
and the sum of the last two terms is the potential part of the symmetry
energy in msQCDSR $E_{\rm{sym}}^{\rm{pot}}(\rho)=E_{\rm{sym}}^{\rm{1st,S}}(\rho)+E_{\rm{sym}}^{\rm{1st,V}}(\rho)$.
Several features of these expressions should be pointed out:

1. At low densities, the leading order term in the symmetry energy is the kinetic part, which roughly scales as $\rho^{2/3}$ since at low densities $e_{\rm{F}}^{\ast}\approx M$.
The potential part of the symmetry energy contributes starting from the linear terms in density $\rho$.
Moreover, both the scalar and vector self-energy contributions to the symmetry energy are positive in msQCDSR, i.e., $E_{\rm{sym}}^{\rm{1st,S}}(\rho)>0$ and $E_{\rm{sym}}^{\rm{1st,V}}(\rho)>0$, and this finding will be verified in the full QCDSR.

2. Perhaps much more important is that, through Eq.\,(\ref{ddd2}) and Eq.\,(\ref{ddd3}), one establishes
the connection between the symmetry energy and some other fundamental quantities, such as the quark mass $m_{\rm{q}}$ and the nucleon-sigma term $\sigma_{\rm{N}}$.
For instance, a larger $m_{\rm{q}}$ corresponds to a smaller $E_{\rm{sym}}^{\rm{1st,S}}(\rho)$ (dependence
of $E_{\rm{sym}}^{\rm{1st,S}}(\rho)$ on $\sigma_{\rm{N}}$ is non-trivial since $\sigma_{\rm{N}}$ also affects the ratio $\alpha/\beta$, and will be discussed in detail in Section~\ref{SEC_FullQSR},
see the fourth panel of Fig.\,\ref{fig_EsymDecomSign}).

3. Tracing back to the physical origin of the isospin factor $\alpha/\beta$, see Eq.\,(\ref{chiral_cond-1}), one can simply obtain the $E_{\rm{sym}}^{\rm{1st,S}}(\rho)$
if other approaches could give the density/isospin dependence of the chiral condensates. For instance, ref.\,\cite{Lac10} gave a different isospin effect of the chiral
condensate, i.e., $\alpha/\beta\leftrightarrow c/f_{\pi}^2$ with $c$ a low energy coefficient and $f_{\pi}$ the pion decay constant, then
\begin{equation}E_{\rm{sym}}^{\rm{1st,S}}(\rho)
\sim\frac{\sigma_{\rm{N}}\rho}{m_{\rm{q}}f_{\pi}^2}\sim-\left(\frac{m_{\pi}}{m_{\rm{q}}}\right)^2\frac{\sigma_{\rm{N}}\rho}{\langle\overline{q}q\rangle_{\rm{vac}}}
,\end{equation}
where the last relation is obtained via the Gell-Mann--Oakes--Renner relation $m_{\pi}^2f_{\pi}^2=-m_{\rm{q}}\langle\overline{q}q\rangle_{\rm{vac}}$.
The above relation shows the $E_{\rm{sym}}^{\rm{1st,S}}(\rho)$ scales as the square of the ratio $m_{\pi}/m_{\rm{q}}$.
It is fair to say that the $E_{\rm{sym}}^{\rm{1st,S}}(\rho)$ is the most non-trivial term in the decomposition of the symmetry energy and
its density dependence largely characterizes the change of the density behavior of the total symmetry energy.
More discussions on this issue will be given in the following sections.

4. On the other hand, the density dependence of $E_{\rm{sym}}^{\rm{1st,V}}(\rho)$
is much simpler and even when other contributions are included in the QCDSR equations, it only changes slightly.

An important problem related to the structure of the symmetry energy is the parabolic approximation of the EOS of ANM. Let us briefly discuss it in msQCDSR and let the more detailed investigations to the next section.
Parabolic approximation of EOS of ANM could be conventionally characterized as
\begin{equation}
E_{\rm{n}}^{\rm{para}}(\rho)\approx
E_0(\rho)+E_{\rm{sym}}(\rho),
\end{equation}
or equivalently the parabolic approximation of the symmetry energy,
\begin{equation}
E_{\rm{sym}}^{\rm{para}}(\rho)\approx E_{\rm{n}}(\rho)-E_0(\rho),
\end{equation}
i.e., the $E_{\rm{sym}}^{\rm{para}}(\rho)$ is the difference between the EOS of PNM and that of SNM.
Consequently, the high order effect of the EOS of ANM is given by,
\begin{equation}\label{def_EHO}
E_{\rm{HO}}(\rho)\equiv
E_{\rm{sym}}^{\rm{para}}(\rho)-E_{\rm{sym}}(\rho),\end{equation}
or,
\begin{equation}E_{\rm{HO}}(\rho)=E_{\rm{sym,4}}(\rho)+E_{\rm{sym,6}}(\rho)+\cdots
,
\end{equation}
since $E_{\rm{n}}(\rho)= E_0(\rho)+E_{\rm{sym}}(\rho)+E_{\rm{sym,4}}(\rho)+E_{\rm{sym,6}}(\rho)+\cdots$,
where $E_{\rm{sym,4}}(\rho)$, $E_{\rm{sym,6}}(\rho)$, $\cdots$ are the fourth-order symmetry energy, sixth-order symmetry energy, $\cdots$ \,\cite{Cai12xx}.
In Fig.\,\ref{fig_EHOMostSimple}, the high order EOS $E_{\rm{HO}}(\rho)$ as a function of density in
msQCDSR is shown. At the saturation density, the high order effect is found to be about
$E_{\rm{HO}}(\rho_0)\approx-2.4\,\rm{MeV}$.
Although many investigations on the high order symmetry energy both from microscopic calculations and phenomenological models\,\cite{LCK08,Cai12xx,Ste06,Kai15,Nan16,Agr17,PuJ17,ZhZ17,RWang17,ZWZhang19} indicates that the magnitude of the high order term, especially the fourth-order symmetry energy, is small, e.g., roughly $
|E_{\rm{HO}}(\rho\lesssim\rho_0)|\lesssim1\,\rm{MeV}$ at
$\rho_0\approx0.16\,\rm{fm}^{-3}$, there are still no fundamental
symmetries and/or principles guaranteeing its smallness.
Thus it is an interesting issue to explore the high order EOS of the ANM in the framework of QCDSR, and this is the main subject of the next section.

\begin{figure}[h!]
\centering
  \includegraphics[width=8.5cm]{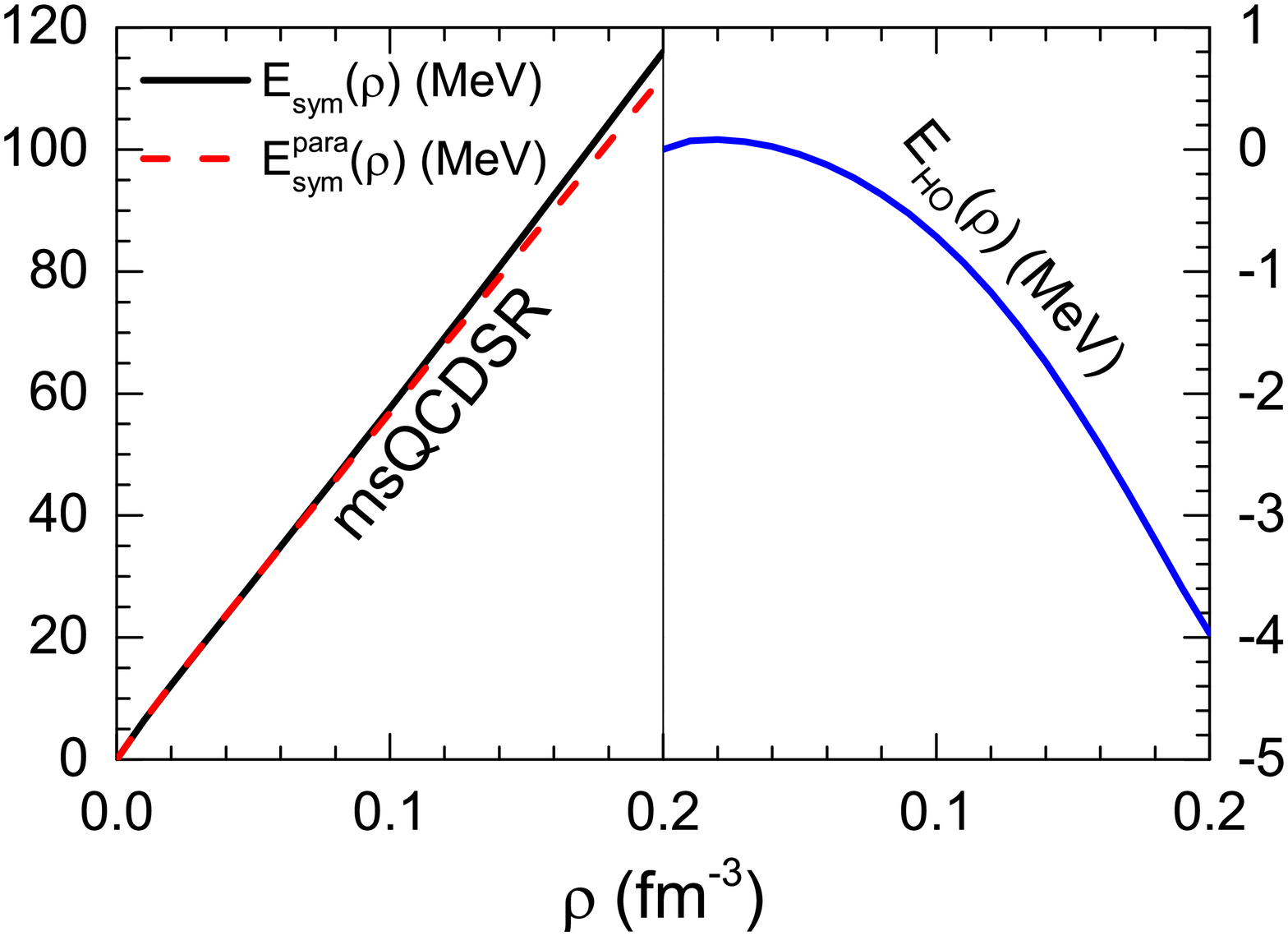}
  \caption{(Color Online) High order effect of the EOS as a function of density in msQCDSR, parameters are $\omega_0=1.4\,\rm{GeV},\sigma_{\rm{N}}=45\,\rm{MeV},m_{\rm{q}}=3.5\,\rm{MeV},m_{\rm{s}}=95\,\rm{MeV},\mathscr{M}^2=1.05\,\rm{GeV}^2$ (default set).}
  \label{fig_EHOMostSimple}
\end{figure}

Finally, let us qualitatively discuss the possible relation between the symmetry energy and the chiral condensates.
In order to make the discussion general, we assume that the chiral condensates has the following density structure,
\begin{align}
\langle\overline{\rm{u}}\rm{u}\rangle_{\rho,\delta}=&\langle\overline{q}q\rangle_{\rm{vac}}
+\mathscr{D}_{\rm{u}}^1\mathscr{F}_{\rm{u}}^1(\delta)\rho+\mathscr{D}_{\rm{u}}^{\theta}\mathscr{F}_{\rm{u}}^{\theta}(\delta)
\rho^{\theta}+\cdots,\\
\langle\overline{\rm{d}}\rm{d}\rangle_{\rho,\delta}=&\langle\overline{q}q\rangle_{\rm{vac}}
+\mathscr{D}_{\rm{d}}^1\mathscr{F}_{\rm{d}}^1(\delta)\rho+\mathscr{D}_{\rm{d}}^{\theta}\mathscr{F}_{\rm{d}}^{\theta}(\delta)
\rho^{\theta}+\cdots,
\end{align}
with $\theta>1$, and we also assume that
\begin{equation}
\mathscr{D}_{\rm{u}}^1=\mathscr{D}_{\rm{d}}^1\equiv\mathscr{D}^1,~~
\mathscr{D}_{\rm{u}}^{\theta}=\mathscr{D}_{\rm{d}}^{\theta}\equiv
\mathscr{D}^{\theta},\end{equation} then we have
\begin{align}
\langle\overline{\rm{u}}\rm{u}\rangle_{\rho,\delta}-\langle\overline{\rm{d}}\rm{d}\rangle_{\rho,\delta}
\approx&\mathscr{D}^1[\mathscr{F}_{\rm{u}}^1(\delta)-\mathscr{F}_{\rm{d}}^1(\delta)]\rho\notag\\
&+\mathscr{D}^{\theta}[\mathscr{F}_{\rm{u}}^{\theta}(\delta)-\mathscr{F}_{\rm{d}}^{\theta}(\delta)]\rho^{\theta}
.\end{align} Neglecting the high order term, i.e., the $\theta$ term
temporarily, we have
\begin{equation}
\langle\overline{\rm{u}}\rm{u}\rangle_{\rho,\delta}-\langle\overline{\rm{d}}\rm{d}\rangle_{\rho,\delta}
\approx\mathscr{D}^1[\mathscr{F}_{\rm{u}}^1(\delta)-\mathscr{F}_{\rm{d}}^1(\delta)]\rho,
\end{equation}
and consequently the first order symmetry scalar self-energy is given by
\begin{align}
\Sigma^{\rm{S}}_{\rm{sym}}(\rho)\approx&-\frac{4\pi^2\mathscr{D}^1}{\mathscr{M}^2}\cdot[\mathscr{F}_{\rm{u}}^1(1)-\mathscr{F}_{\rm{d}}^1(1)]\rho
,\end{align} and the corresponding contribution to the symmetry energy
is
\begin{equation}
E_{\rm{sym}}^{\rm{1st,S}}(\rho)\approx-\frac{2\pi^2\mathscr{D}^1}{\mathscr{M}^2}\frac{M_0^{\ast}}{{e}_{\rm{F}}^{\ast}}\cdot
[\mathscr{F}_{\rm{u}}^1(1)-\mathscr{F}_{\rm{d}}^1(1)]\rho
.\end{equation}
Due to the
empirical knowledge on the nucleon Dirac effective mass in SNM,
the scalar self-energy part of the symmetry energy strongly depends on the
structure of the u/d quark condensates, especially
the isospin part, i.e., the properties of function
$\mathscr{F}_{\rm{u/d}}^1$. The $\mathscr{D}^1$ is given by $ \mathscr{D}^1={\sigma_{\rm{N}}}/{2m_{\rm{q}}}
$, thus
\begin{align}
E_{\rm{sym}}^{\rm{1st,S}}(\rho)\approx-\frac{\pi^2\sigma_{\rm{N}}\langle\overline{q}q\rangle_{\rm{vac}}}{\mathscr{M}^2m_{\rm{q}}}
\cdot\frac{
[\mathscr{F}_{\rm{u}}^1(1)-\mathscr{F}_{\rm{d}}^1(1)]\rho}{\langle\overline{q}q\rangle_{\rm{vac}}}.
\end{align}
If the u quark in PNM restores its chiral symmetry first, i.e., $\langle
\overline{\rm{u}}\rm{u}\rangle_{\rho}$ approaches to zero earlier
than $\langle \overline{\rm{d}}\rm{d}\rangle_{\rho}$ as the density increases, then
$E_{\rm{sym}}^{\rm{1st,S}}(\rho)$ is positive; on the opposite side,
$E_{\rm{sym}}^{\rm{1st,S}}(\rho)$ is negative. In the msQCDSR, one has $E_{\rm{sym}}^{\rm{1st,S}}(\rho)>0$, i.e., the u quark in
PNM restores its chiral symmetry earlier than d quark.
When we consider higher order terms in density (characterized by $\mathscr{F}^{\theta}_{\rm{u/d}}$) in the chiral
condensates, different patterns may emerge. It indicates that one can constrain the density dependence of the chiral
condensates via the empirical knowledge of the symmetry energy within a reasonable density region. This is one of the motivations
to include the higher order density terms in the chiral condensates as in Eq.\,(\ref{chiral_cond}),
which will be further explored in detail in Section~\ref{SEC_HighCond}.

\setcounter{equation}{0}
\section{More Discussions on $E_{\textmd{HO}}(\rho)$}
\label{SEC_HOEsti}

In this section, we study the effects of the high order effects of
the EOS of ANM. In order to study the parabolic approximation of the EOS of ANM,
we add terms into the QCDSR equations eventually. For example, based on the msQCDSR, we add the
four-quark condensates of the form $\langle
\overline{q}q\rangle^2_{\rho,\delta}$, i.e.,
\begin{equation}
\langle\overline{q}q\rangle_{\rho,\delta}^2\longrightarrow(1-f)\langle\overline{q}q\rangle_{\rm{vac}}^2
+f\langle\overline{q}q\rangle_{\rho,\delta}^2\equiv
B_4^{q}(f)\equiv\widetilde{\langle\overline{q}q\rangle_{\rho,\delta}^2},
\end{equation}
into the QCDSR equation,
and call the corresponding QCDSR the simple QCDSR (abbreviated as sQCDSR), which is denoted by $\mathscr{C}_{\rm{B}}=1$.
Specifically, the relevant QCDSR equations are modified as (where $t=-1$ is adopted
in Eq.\,(\ref{sQCDSRp}) and Eq.\,(\ref{sQCDSRn}))
\begin{align}
\rm{proton (p)}:~~\frac{\mathscr{M}^6}{32\pi^4}\longrightarrow&\frac{\mathscr{M}^6}{32\pi^4}+\frac{2}{3}B_4^{\rm{d}}(f),\label{sQCDSRp}\\
\rm{neutron (n)}:~~\frac{\mathscr{M}^6}{32\pi^4}\longrightarrow&\frac{\mathscr{M}^6}{32\pi^4}+\frac{2}{3}B_4^{\rm{u}}(f),\label{sQCDSRn}
\end{align}
compared with Eq.\,(\ref{msQCDSRp2}) and Eq.\,(\ref{msQCDSRn2}) for the msQCDSR, respectively.
Similarly,
$\mathscr{C}_4=1$ means four-dimensional condensates are
included based on $\mathscr{C}_{\rm{B}}=\mathscr{C}_{\rm{A}}=1$,
etc. The whole order of adding different types of effects is as follows: ``msQCDSR'' (only three-dimensional condensates) $\to$
$\mathscr{C}_{\rm{B}}=1$ (sQCDSR with $B_4^{q}(f)$ included) $\to$
$\mathscr{C}_{\rm{A}}=1$ (anomalous effects $L^{-2\Gamma_{\eta}+\Gamma_{\mathcal{O}_n}}$) $\to$ $\mathscr{C}_{\rm{4}}=1$
(four-dimensional condensates)
$\to$
$\mathscr{C}_{\rm{5}}=1$ (five-dimensional condensates) $\to$ $\mathscr{C}_{\rm{H}}=1$ (quasi-hole effects) $\to$
$\mathscr{C}_{\rm{D}}=1$ (four-quark condensates of type $\langle q^{\dag}q\rangle_{\rho,\delta}^2$, which is absent in vacuum) $\to$ ``full QCDSR (fQCDSR)''
($\mathscr{C}_{\rm{C}}=1$).
\begin{figure}[h!]
\centering
  \includegraphics[width=8.5cm]{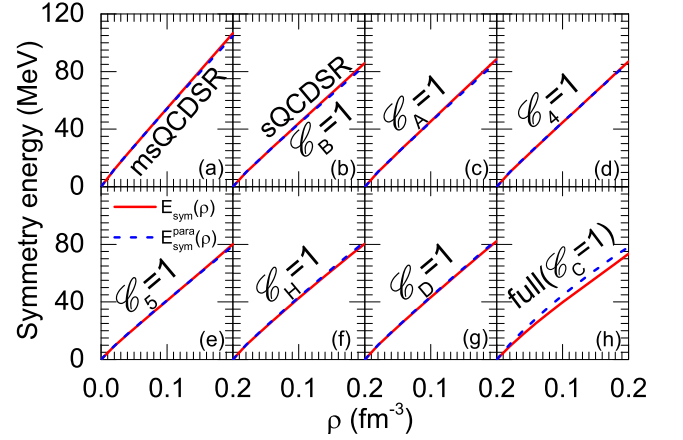}
  \caption{(Color Online) Symmetry energy as well as its parabolic approximation.
  The physical meaning is given in the main context. Parameters used are the
same as these used in Fig.\,\ref{fig_EHOMostSimple}.}
  \label{fig_EHO_order_1}
\end{figure}
In these calculations, the Ioffe parameter $t=-1$ is
adopted. In Fig.\,\ref{fig_EHO_order_1}, the symmetry energy and its parabolic approximation
are shown as functions of density in the above order. It is obvious from the figure that the
parabolic approximation well behaves until the continuum
excitations ($\mathscr{C}_{\rm{C}}=1$), characterized by the functions $E_0$, $E_1$ and $E_2$, i.e., Eqs.\,(\ref{def_E0}), (\ref{def_E1}), and (\ref{def_E2}), are included.
For example, in the full QCDSR (abbreviated as fQCDSR), the $E_{\rm{HO}}$ at $\rho_0$ is found to be about $E_{\rm{HO}}(\rho_0)\approx5.7\,\rm{MeV}$.
However, the $E_{\rm{HO}}(\rho_0)$ in the QCDSR to order $\mathscr{C}_{\rm{D}}=1$ is about 0.7\,MeV.

\begin{figure}[h!]
\centering
  \includegraphics[width=8.cm]{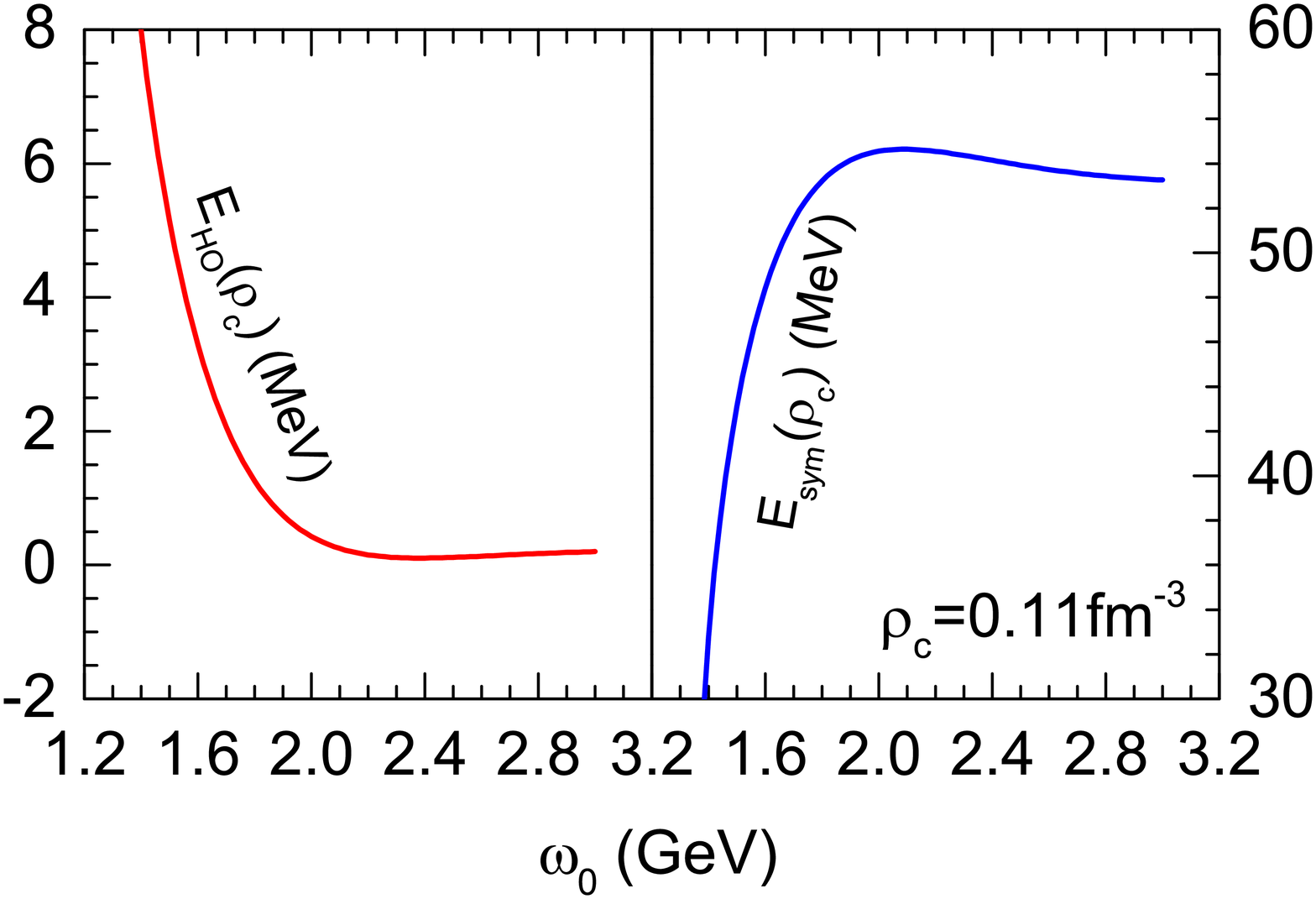}\\
  \includegraphics[width=8.cm]{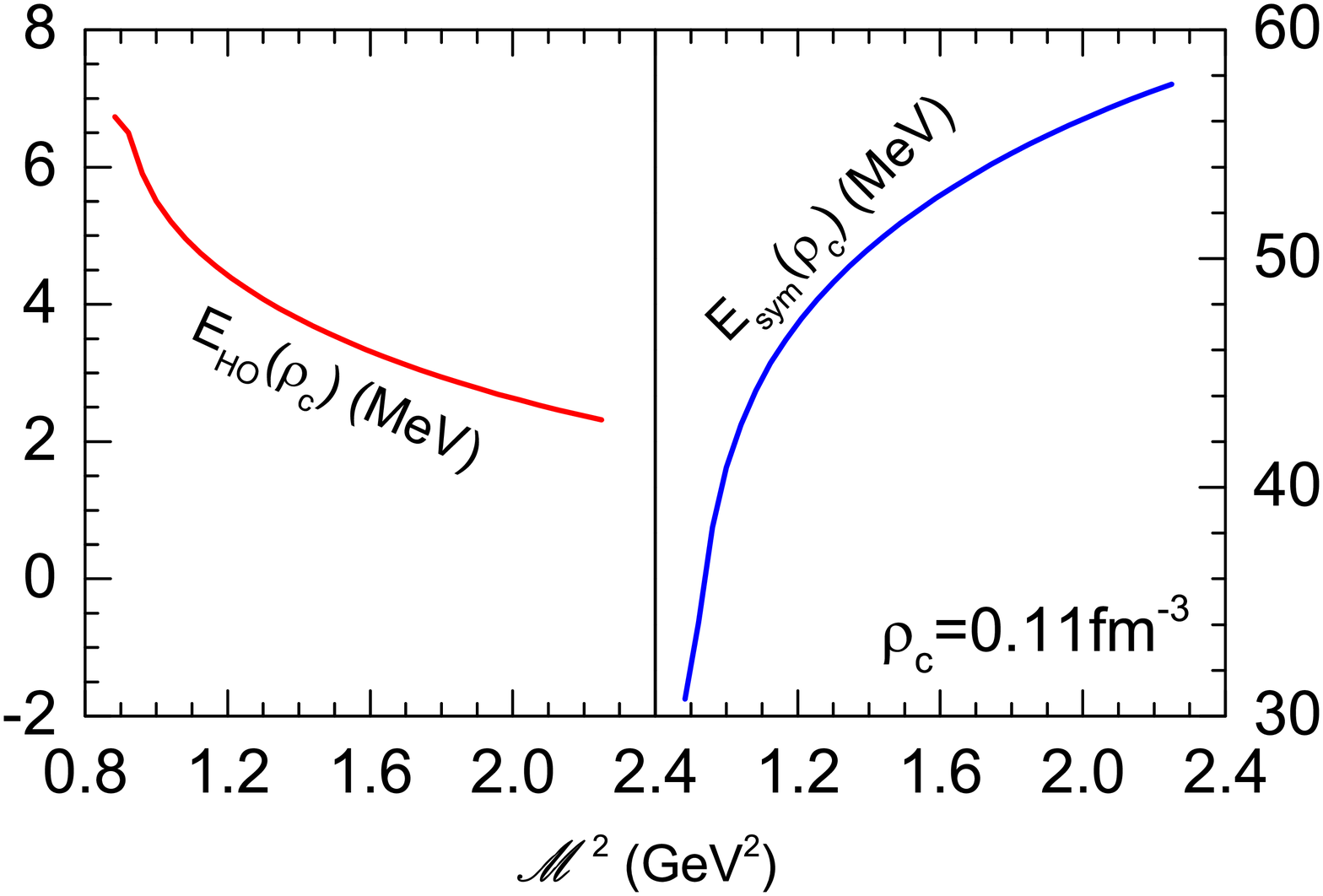}
  \caption{(Color Online) $E_{\rm{HO}}$ and $E_{\rm{sym}}$ at $\rho_{\rm{c}}=0.11\,\rm{fm}^{-3}
  $ as functions of continuum excitations factor $\omega_0$ (upper) and Borel mass squared $\mathscr{M}^2$ (lower) in fQCDSR.
 } \label{fig_w0BMEHOEsym}
\end{figure}

The investigations on the the dependence of the high order effect $E_{\rm{HO}}(\rho)$ on the intrinsic parameters of the QCDSR
will be useful for understanding the behavior of the parabolic approximation of the ANM EOS. In the following, we study
the dependence of $E_{\rm{HO}}(\rho)$ on the continuum excitation parameter $\omega_0$, the Borel mass $\mathscr{M}$, and the effective four-quark
condensates parameter $f$. In Fig.\,\ref{fig_w0BMEHOEsym}, we show the
$E_{\rm{HO}}(\rho)$ and $E_{\rm{sym}}(\rho)$ at (the cross density)
$\rho_{\rm{c}}=0.11\,\rm{fm}^{-3}$\,\cite{Zha13,Che12a,Li12,Dan14} as functions of the continuum
excitations factor $\omega_0$ (upper) and Borel mass squared $\mathscr{M}^2$ (lower) using the fQCDSR equations.
Since the symmetry energy $E_{\rm{sym}}(\rho)$ at the cross density $\rho_{\rm{c}}$ is better constrained\,\cite{Zha13,Dan14} than that at the saturation density,
we will also focus on the symmetry energy at $\rho_{\rm{c}}$ besides its value at $\rho_0$ in the following sections.
Two main features of Fig.\,\ref{fig_w0BMEHOEsym} are necessary to be pointed out:
1). With the value of $\omega_{\rm{0}}$ increasing, i.e., as the
effects of the continuum excitations reduces, the parabolic
approximation of the EOS of ANM becomes better, and this finding is consistent with the results shown in} panels (g) and (h) of Fig.\,\ref{fig_EHO_order_1} (if $\omega_0\to\infty$, there will be no continuum excitation effects).
2). As the Borel mass (squared) increases, the parabolic approximation becomes better,
however, the symmetry energy also becomes larger.
The symmetry energy and the $E_{\rm{HO}}$ display almost opposite variation tendency with the $\omega_0$ and $\mathscr{M}^2$.
For example, the symmetry energy at $\rho_{\rm{c}}$ with $\mathscr{M}^2=1.05\,\rm{GeV}^2$
is found to be about 43.1\,MeV, which is much larger than the empirical value about 26.6\,MeV\,\cite{Zha13},
however, the corresponding $E_{\rm{HO}}(\rho_{\rm{c}})$ is found to be about 5.1\,MeV.
These results show that the parabolic approximation of the EOS of ANM in QCDSR is heavily broken,
indicating the fourth-order symmetry energy even the higher order terms maybe large.

Now let us study the four-quark effective parameter $f$ on the EOS qualitatively. The effects of $f$ can be demonstrated semi-analytically in the sQCDSR (including the contributions from the
four-quark condensates of the type
$\langle\overline{q}q\rangle_{\rho,\delta}^2$ based on the msQCDSR).
and then according to Eq.\,(\ref{sec1_SNM}) and Eq.\,(\ref{sec1_PNM}), one can obtain the
$E_0(\rho)$ and $E_{\rm{n}}(\rho)$, respectively.
The $\langle\overline{q}q\rangle_{\rho,\delta}^2$ type of the four-quark condensates in SNM can be approximated as
\begin{align}
B_4(f)=&(1-f)\langle\overline{q}q\rangle_{\rm{vac}}^2+f(\langle\overline{q}q\rangle_{\rm{vac}}+a\rho)^2\notag\\
\approx&
\langle\overline{q}q\rangle_{\rm{vac}}^2+2fa\langle\overline{q}q\rangle_{\rm{vac}}\rho
\equiv A+Bf\rho,
\end{align}
where $B_4$ (without superscript) denotes the corresponding term
in SNM, $a=\sigma_{\rm{N}}/2m_{\rm{q}},
A=\langle\overline{q}q\rangle_{\rm{vac}}^2,B=2a\langle\overline{q}q\rangle_{\rm{vac}}$.
In the above approximation, the term proportional to $\rho^2$ is omitted, since we are only interested in low density behavior of the EOS.
The nucleon effective mass is given by (Ioffe parameter $t=-1$ in this estimate)
\begin{align}
M_0^{\ast}(\rho)=&-\frac{({\mathscr{M}^4}/{4\pi^2})\langle\overline{q}q\rangle_{\rho}}{
{\mathscr{M}^6}/{32\pi^4}+(2/3)B_4(f)}\notag\\
\approx&-\frac{\mathscr{M}^4}{4\pi^2D}\langle\overline{q}q\rangle_{\rho}\cdot\left(1-\frac{2B}{3D}f\rho\right)\label{zz_zz},
\end{align} where $\langle\overline{q}q\rangle_{\rho}$ is the chiral condensate in SNM, and
\begin{equation}
D=\frac{\mathscr{M}^6}{32\pi^4}+\frac{2A}{3}=\frac{\mathscr{M}^6}{32\pi^4}+\frac{2}{3}\langle\overline{q}q\rangle_{\rm{vac}}^2.
\end{equation}
Thus, $
M_0^{\ast}(\rho)\approx{I}_1+{I}_2\rho+\mathcal{O}(\rho^2)$, with
\begin{align}
{I}_1=&-\frac{\mathscr{M}^4\langle\overline{q}q\rangle_{\rm{vac}}}{4\pi^2D}
,\\
{I}_2=&-\frac{\mathscr{M}^4}{4\pi^2D}\left(a-\frac{2Bf\langle\overline{q}q\rangle_{\rm{vac}}}{
3D}\right).\end{align} In a very similar manner, one can obtain the
approximation for the vector self-energy
\begin{equation}
\Sigma_{\rm{V}}^0(\rho)\approx\frac{\mathscr{M}^4}{12\pi^2D}8\langle
q^{\dag}q\rangle_{\rho}\cdot\left(1-\frac{2B}{3D}f\rho\right),\end{equation}
using the expression for $\langle q^{\dag}q\rangle_{\rho}$, then $
\Sigma_{\rm{V}}^0(\rho)\approx {S}_2\rho+\mathcal{O}(\rho^2)$ with
${S}_2={\mathscr{M}^4}/{\pi^2D}$. Then according to Eq.\,(\ref{sec1_SNM}), one obtains
\begin{align}
E_0(\rho)\approx&{I}_1-M_{}+\frac{3}{10{I}_1}\left(\frac{3\pi^2}{2}\right)^{2/3}\rho^{2/3}\notag\\
&+\frac{1}{2}({I}_2+{S}_2)\rho,
\end{align}
to order $\rho$. The first term on the right hand side should be zero since $E_0(0)=0$, leading to
$M=-{\mathscr{M}^4\langle\overline{q}q\rangle_{\rm{vac}}}/{4\pi^2D}$\,\cite{Footnote3}, and one obtains the following approximation for the
EOS of SNM,
\begin{equation}\label{E0App}
E_0(\rho)\approx E_0^{\rm{FFG}}(\rho)+\frac{1}{2}({I}_2+{S}_2)\rho,
\end{equation}
where $E_0^{\rm{FFG}}(\rho)=3k_{\rm{F}}^2/10M$ is the free Fermi gas (FFG) prediction on the EOS of SNM,
using the expressions for $I_2$ and $S_2$, the EOS of SNM can finally be written as
\begin{align}\label{E0s}
E_0(\rho)
\approx&E_0^{\rm{FFG}}(\rho)+
\frac{1}{2}\frac{M\rho}{\langle\overline{q}q\rangle_{\rm{vac}}}\notag\\
&\times\left[\frac{\sigma_{\rm{N}}}{2m_{\rm{q}}}\left(1+\frac{16\pi^2f}{3}\frac{M\langle\overline{q}q\rangle_{\rm{vac}}}{\mathscr{M}^4}\right)-4\right]
.
\end{align}

Similarly, the nucleon effective mass in PNM could be approximated as $M_{\rm{n}}^{\ast}(\rho)\approx
{I}_1+W_2\rho$, with $ {W}_2=\left(1-{\alpha}/{\beta}\right){I}_2$, and $
\Sigma_{\rm{V}}^{\rm{n}}(\rho)\approx5S_2\rho/4+\mathcal{O}(\rho^2)$, then the EOS of PNM is given approximately by
\begin{align}
E_{\rm{n}}(\rho)
\approx&E_{\rm{n}}^{\rm{FFG}}(\rho)+\frac{1}{2}\left({W}_2+\frac{5}{4}{S}_2\right)\rho\notag\\
=&E_{\rm{n}}^{\rm{FFG}}(\rho)+\frac{1}{2}\frac{M\rho}{\langle\overline{q}q\rangle_{\rm{vac}}}\notag\\
&\times\left[\left(1-\frac{\alpha}{\beta}\right)\frac{\sigma_{\rm{N}}}{2m_{\rm{q}}}\left(1+\frac{16\pi^2f}{3}\frac{M\langle\overline{q}q\rangle_{\rm{vac}}}{\mathscr{M}^4}\right)-5\right]
,\label{Ens}
\end{align}
where $E_{\rm{n}}^{\rm{FFG}}(\rho)=3k_{\rm{F,n}}^2/10M\sim\rho^{2/3}$ is the FFG prediction on the EOS of PNM.
This expression is already very interesting. For example, since the EOS of PNM at very low densities (say densities smaller than $0.01\,\rm{fm}^{-3}$) could be
determined very accurate by simulations or microscopic calculations, there exists a relation between several fundamental quantities and the model parameters in QCDSR,
such as the four-quark effective parameter $f$, the nucleon-sigma term $\sigma_{\rm{N}}$ (the factor $\alpha/\beta$ also depends on $\sigma_{\rm{N}}$),
the chiral condensate in vacuum, the light quark mass, and the Borel mass.
Specifically, a positive $f$ parameter leads to a reduction on the $E_{\rm{n}}(\rho)$, which will be verified numerically in
the following sections (e.g., see Fig.\,\ref{fig_R1PNM}).
Moreover, as discussed in Subsection~\ref{sb_FS}, the parameter $f$ would be essentially determined by the EOS of PNM at
a very low density $\rho_{\rm{vl}}\approx0.02\,\rm{fm}^{-3}$, and it is also indicated in Eq.\,(\ref{Ens}).
Based on the approximations for $E_0(\rho)$ and $E_{\rm{n}}(\rho)$, the symmetry energy obtained in the parabolic approximation is roughly given by
\begin{align}\label{Esyms}
E_{\rm{sym}}^{\rm{para}}(\rho)
\approx&E_{\rm{sym,para}}^{\rm{FFG}}(\rho)
-\frac{1}{2}\frac{M\rho}{\langle\overline{q}q\rangle_{\rm{vac}}}\notag\\
&\times\left[\frac{\alpha}{\beta}\frac{\sigma_{\rm{N}}}{2m_{\rm{q}}}\left(1+\frac{16\pi^2f}{3}\frac{M\langle\overline{q}q\rangle_{\rm{vac}}}{\mathscr{M}^4}\right)+1\right],
\end{align}
where the kinetic symmetry energy
in the parabolic approximation is given by
\begin{align}E_{\rm{sym,para}}^{\rm{FFG}}(\rho)\equiv& E_{\rm{n}}^{\rm{FFG}}(\rho)
-E_0^{\rm{FFG}}(\rho)\notag\\
=&(2^{2/3}-1)E_0^{\rm{FFG}}(\rho)\approx0.59E_0^{\rm{FFG}}(\rho).
\end{align}

On the other hand, when using the exact nucleon self-energy decomposition formula Eq.\,(\ref{sec1_Esym}) for calculating the symmetry energy,
one obtains $E_{\rm{sym}}^{\rm{kin}}(\rho)=(3/5)E_0^{\rm{FFG}}(\rho)$, $E_{\rm{sym}}^{\rm{1st,S}}(\rho)
\approx-2^{-1}(\alpha/\beta)\rho I_2$, and $E_{\rm{sym}}^{\rm{1st,V}}(\rho)\approx8^{-1}\rho S_2$, in
the same approximation level (i.e., in sQCDSR).
Interestingly, the potential part of the symmetry energy in the two approaches is found to be same at this order (i.e., order $\rho$), and only the kinetic part introduces
the corresponding high order effects, which is about $-0.01E_0^{\rm{FFG}}(\rho)$.
Going beyond the linear order in density, the $f$ parameter will come into play in the high order effects of the EOS $E_{\rm{HO}}(\rho)$.
Moreover, the above qualitative analysis (approximation) on the $E_{\rm{n}}(\rho)$, $E_0(\rho)$, and $E_{\rm{sym}}(\rho)$
is useful for further investigations on, e.g., the correlation between the role played by the four-quark condensates and the EOS of ANM, as mentioned just above.
Furthermore, if one takes $f=0$ in Eq.\,(\ref{Ens}), then the $E_{\rm{n}}(\rho)$ could be written as\,\cite{Cai17QCDSR}
\begin{align}
E_{\rm{n}}(\rho)
\approx&E_{\rm{n}}^{\rm{FFG}}(\rho)+\frac{1}{2}\frac{M\rho}{\langle\overline{q}q\rangle_{\rm{vac}}}\left[\left(1-\frac{\alpha}{\beta}\right)\frac{\sigma_{\rm{N}}}{2m_{\rm{q}}}-5\right]
,\label{Ens-1}
\end{align}
which depends only on several fundamental quantities, such as $m_{\rm{q}}$, $\langle\overline{q}q\rangle_{\rm{vac}}$,
and $\sigma_{\rm{N}}$, and not on the effective parameters $f$ and $\mathscr{M}^2$.
Despite its simplicity, Eq.\,(\ref{Ens-1}) already has the power of quantitative predictions at very low densities.
We will discuss more on this point in the following sections.

\begin{figure}[h!]
\centering
  \includegraphics[width=8.5cm]{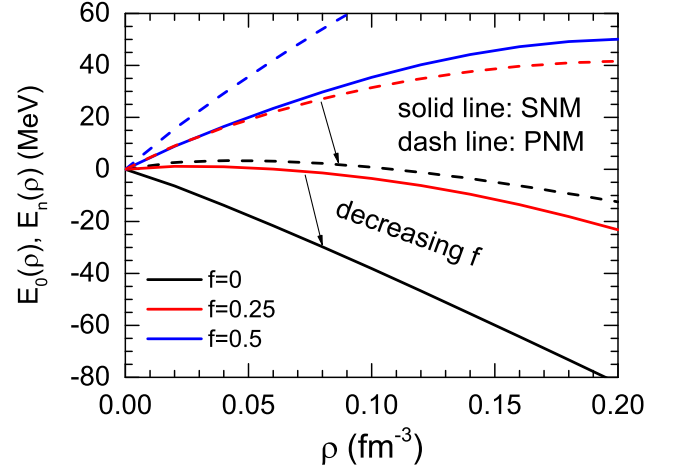}
  \caption{(Color Online) EOS of SNM and that of PNM as functions of density with different $f$ using in fQCDSR.
Parameters are the same as these used in Fig.\,\ref{fig_EHOMostSimple}.}
  \label{fig_fSNMPNM}
\end{figure}

\begin{figure}[h!]
\centering
  \includegraphics[width=8.cm]{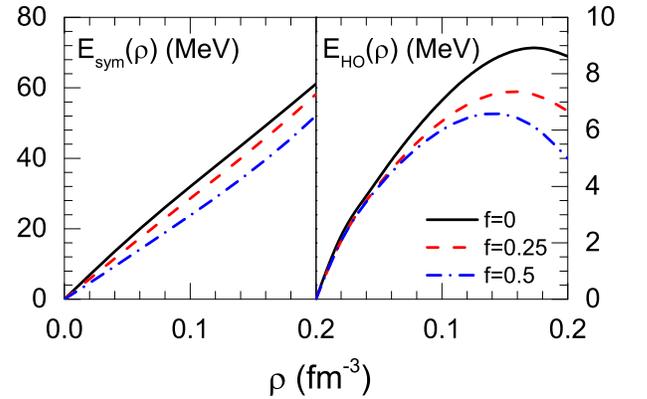}
  \caption{(Color Online) $E_{\rm{sym}}$ (left) and $E_{\rm{HO}}
  $ (right) as functions of density with different $f$ in fQCDSR.
Other parameters are the same as these used in Fig.\,\ref{fig_EHOMostSimple}.}
  \label{fig_fEHOEsym}
\end{figure}

In Fig.\,\ref{fig_fSNMPNM}, the effects of the parameter $f$ on
the EOS of SNM and that of PNM in the fQCDSR are shown.
Although the predictions on $E_0(\rho)$ and $E_{\rm{n}}(\rho)$ are quantitatively incorrect compared to their empirical constraints,
the effects of $f$ are obvious, which are consistent with the estimate given in Eq.\,(\ref{E0s}) and Eq.\,(\ref{Ens}),
i.e., as $f$ increases, the $E_0(\rho)$ and $E_{\rm{n}}(\rho)$ are both enhanced correspondingly.
However, the enhancement on $E_{\rm{n}}(\rho)$ due to $f$ is less than that on $E_0(\rho)$ as shown from Eq.\,(\ref{E0s})
and Eq.\,(\ref{Ens}) by a factor $\alpha/\beta$, leading to the result that as $f$ increases, the the symmetry energy is reduced, see Eq.\,(\ref{Esyms}).
The numerical results on the $E_{\rm{sym}}(\rho)$ and $E_{\rm{HO}}(\rho)$ using fQCDSR equations are shown in Fig.\,\ref{fig_fEHOEsym}.
For example, the high order effects in the EOS at $\rho_{\rm{c}}=0.11\,\rm{fm}^{-3}$ with $f=0$, $0.25$ and $0.5$ are found to be about
7.5\,MeV, 6.6\,MeV, and 6.3\,MeV, respectively, and those values at $\rho_0=0.16\,\rm{fm}^{-3}$ are found to be about 8.8\,MeV, 7.4\,MeV, and 6.4\,MeV, respectively.
These results again, when combining the features shown in Fig.\,\ref{fig_w0BMEHOEsym} and that the Borel mass squared was
constrained to fall within about $0.8\,\rm{GeV}^2\lesssim\mathscr{M}^2\lesssim1.4\,\rm{GeV}^2$\,\cite{Iof84} together with
the continuum excitation factor about $2\,\rm{GeV}^2\lesssim s_0^{\ast}\sim\omega_0^2\lesssim3\,\rm{GeV}^2$\,\cite{Coh95}, indicate
that the high order term $E_{\rm{HO}}(\rho)$ at the cross density $\rho_{\rm{c}}$ and at the saturation density $\rho_0$
are generally not small (e.g., $\lesssim1\,\rm{MeV}$).

Finally, it is useful to generally analyze the origin of the possible breakdown of the parabolic approximation of the EOS of ANM in QCDSR framework.
In the msQCDSR, the density dependence of the chiral condensates has the
following structure,
\begin{align}
\langle\overline{q}q\rangle_{\rho,\delta}^{\rm{u,d}}
=&\langle\overline{q}q\rangle_{\rm{vac}}+a(1\mp\xi\delta)\rho,
\end{align}see Eq.\,(\ref{chiral_cond-1}), where $a=\sigma_{\rm{N}}/2m_{\rm{q}}$ and
$\xi=\alpha/\beta$.
Moreover, the nucleon Dirac effective mass and the vector self-energy are given as $M_J^{\ast}\sim\langle\overline{q}q\rangle_{\rho,\delta}^{\rm{u,d}}$
and $\Sigma_{\rm{V}}^J\sim\langle q^{\dag}q\rangle_{\rho,\delta}^{\rm{u,d}}$, see Eq.\,(\ref{zz_MJ}) and Eq.\,(\ref{zz_SVJ}), respectively.
Then, the corresponding contribution originated from the scalar self-energy to the EOS of SNM via Eq.\,(\ref{sec1_SNM}) can be obtained from
the following estimate,
\[
E_0(\rho)\sim\frac{1}{\rho}\int_0^{\rho}\d\rho\left(\Sigma_{\rm{S}}^0+\Sigma_{\rm{V}}^0+\frac{k_{\rm{F}}^2}{2(M+\Sigma_{\rm{S}}^0)}+\cdots\right)
\]
or, roughly $\sim a\rho/2$, where ``$\cdots$'' in the above expression denotes relativistic corrections. Moreover, the relevant contribution to EOS of PNM is roughly
given by $a(1-\xi)\rho/2$ via Eq.\,(\ref{sec1_PNM}). Consequently, the symmetry energy obtained using
the parabolic approximation, i.e., $E_{\rm{sym}}^{\rm{para}}(\rho)$, is roughly $-a\xi\rho/2$.
It can be easily shown that the contribution of the vector self-energy to the symmetry energy exactly cancels in the parabolic approximation and
in the self-energy decomposition for the $E_{\rm{sym}}(\rho)$ in msQCDSR, both are $4\pi^2\rho/\mathscr{M}^2$, see Eq.\,(\ref{ddd3}).
Consequently, the scalar self-energy contribution to the symmetry energy via the decomposition
is roughly given by $-({M_0^{\ast}}/{{e}_{\rm{F}}^{\ast}})a\xi\rho/2$ via simply subtracting $a(1-\xi)\rho/2$ and $a\rho/2$ and multiplying a factor $M_0^{\ast}/e_{\rm{F}}^{\ast}$.
Thus $E_{\rm{HO}}(\rho)= E_{\rm{sym}}^{\rm{para}}(\rho)
-E_{\rm{sym}}(\rho)$ is found to be
\begin{equation}\label{zz_EHOes}
E_{\rm{HO}}(\rho)\sim-\frac{1}{2}a\xi\rho\left(1-\frac{M_0^{\ast}}{{e}_{\rm{F}}^{\ast}}\right),
\end{equation}
where the overall factor is unimportant for the qualitative demonstration.
At low densities, $e_{\rm{F}}^{\ast}=(M_0^{\ast,2}+k_{\rm{F}}^2)^{1/2}\approx k_{\rm{F}}^2/2M_0^{\ast}+M_0^{\ast}
\approx M_0^{\ast}$, indicating that the expression in the parentheses in Eq.\,(\ref{zz_EHOes}) is small.
As density increases, the $E_{\rm{HO}}(\rho)$ by Eq.\,(\ref{zz_EHOes}) increases eventually, leading to the increasing of the
high order effects eventually. For example, keeping only the leading order term in $(k_{\rm{F}}/M_0^{\ast})^2$, one obtains the high order effect
as
\begin{equation}
E_{\rm{HO}}(\rho)\sim-a\xi k_{\rm{F}}^2\rho/4M_0^{\ast,2},
\end{equation}
in msQCDSR. It shows that the high order effect $E_{\rm{HO}}(\rho)$ is negative (see the left panel of Fig.\,\ref{fig_EHOMostSimple}).

Furthermore, when considering higher order terms in density in the
chiral condensates, e.g.,
\begin{equation}
\langle\overline{q}q\rangle_{\rho,\delta}^{\rm{u,d}}
=\langle\overline{q}q\rangle_{\rm{vac}}+a(1\mp\xi\delta)\rho+b(1\mp\zeta\delta)\rho^{\theta},~~\theta>1
,\end{equation} and assuming other quark/gluon condensates remain
unchanged, then the integration of EOS from Eq.\,(\ref{sec1_SNM}) and Eq.\,(\ref{sec1_PNM})
gives roughly $-({\theta+1})^{-1}b\zeta\rho^{\theta}$ to the symmetry energy
from the $\rho^{\theta}$ term. And the corresponding term using the self-energy
decomposition of the symmetry energy is given by $-(M_0^{\ast}/e_{\rm{F}}^{\ast})b\zeta\rho^{\theta}/2$,
consequently the difference becomes now
\begin{align}
E_{\rm{HO}}^{\theta\rm{\,term}}(\rho)=&E_{\rm{sym}}^{\rm{para, $\theta$ term}}(\rho)-E_{\rm{sym}}^{\theta\rm{ term}}(\rho)\notag\\
\sim&-b\zeta\rho^{\theta}\left(\frac{1}{\theta+1}-\frac{1}{2}\frac{M_0^{\ast}}{e_{\rm{F}}^{\ast}}\right)\notag\\
\approx&\frac{1}{2}b\zeta\rho^{\theta}\frac{\theta-1}{\theta+1}\label{EHOXX}
,\end{align}
the last line is valid at low densities where $M_0^{\ast}\approx e_{\rm{F}}^{\ast}$ is a good approximation.
It is obvious from this expression the $E_{\rm{HO}}(\rho)$ will not be small even at low densities, and
the high order effects become more and more important as density increases since when keeping term in $(k_{\rm{F}}/M_0^{\ast})^2$, one has
\begin{equation}
E_{\rm{HO}}^{\theta\rm{\,term}}(\rho)\approx \frac{1}{2}b\zeta\rho^{\theta}\frac{\theta-1}{\theta+1}
-\frac{1}{4}b\zeta\left(\frac{k_{\rm{F}}}{M_0^{\ast}}\right)^2\rho^{\theta}.
\end{equation}
The first term is absent if $\theta=1$ and in this case the second term is relatively small owing to the small factor $(k_{\rm{F}}/M_0^{\ast})^2$.
On the other hand, the higher order terms in density in the chiral condensates
are naturally essential as density increases, e.g., the linear approximation (\ref{chiral_cond-1})
breaks down eventually.
Combining these analyses, it is intuitive to conclude that one of the main reasons for the breakdown of the parabolic approximation for the EOS of ANM in QCDSR maybe the higher order density terms in the chiral condensates, and this finding will be justified in the numerical studies in the following sections.
The other effects, such as the continuum excitations, may also lead to the breakdown of the parabolic approximation.

\setcounter{equation}{0}
\section{Full Calculations on EOS of ANM}\label{SEC_FullQSR}

After three sections on the nucleon mass in vacuum, the self-energy structure of the symmetry energy and the high order effects of the EOS of ANM
(mainly qualitatively) given above, we now systematically investigate the EOS of ANM through the full QCDSR calculations.
It is necessary to mention again that the default parameters used are the
same as these used in Fig.\,\ref{fig_EHOMostSimple} except $\omega_0$, i.e.,
$m_{\rm{q}}=3.5\,\rm{MeV},m_{\rm{s}}=95\,\rm{MeV},f=0,\mathscr{M}^2=1.05\,\rm{GeV}^2$,
$\sigma_{\rm{N}}=45\,\rm{MeV}$, and $\omega_0=1.5\,\rm{GeV}$.
In this section, the dependence of the physical
quantities on $\omega_0,\mathscr{M}^2,f$ and $\sigma_{\rm{N}}$ will be
studied carefully.
Moreover, the Ioffe parameter $t\approx-1.22$ is determined in this section self-consistently via the scheme for the nucleon mass in vacuum given in Section~\ref{SEC_MassVacuum}.
This section is organized as follows: The nucleon self-energies in SNM will be
explored first in Subsection~\ref{ss_1}; then the symmetry energy with its self-energy decomposition will be studied
in Subsection~\ref{ss_2}; Subsection~\ref{ss_3} studies the correlation between the symmetry energy and the
quark/gluon condensates,
and finally the first QCDSR parameter set in this work, i.e., QCDSR-1 (or the naive QCDSR in the sense only the linear
density terms in the chiral condensates (i.e., Eq.~\ref{chiral_cond}) are included), will finally be given in Subsection~\ref{ss_4}.
Shortcoming of the QCDSR-1 together with the possible improvements will also be given.

\subsection{Nucleon Self-energies in SNM}\label{ss_1}

\begin{figure}[h!]
\centering
  \includegraphics[width=8.6cm]{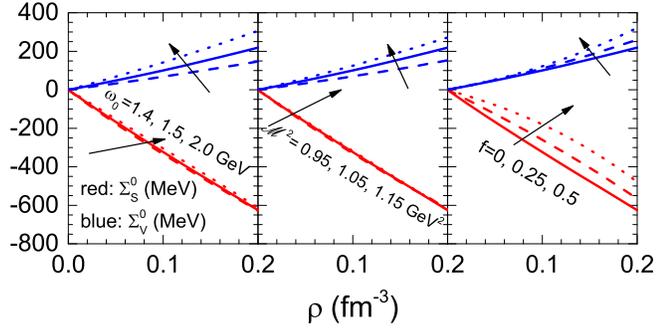}
  \caption{(Color Online) Nucleon self-energies in SNM as functions of
  density with different values of $\omega_0$ (left), $\mathscr{M}^2$ (central) or $f$ (right),
  respectively.}
  \label{fig_SNMSE_w0BMf}
\end{figure}

In Fig.\,\ref{fig_SNMSE_w0BMf}, the nucleon self-energies in SNM as functions of density with different values of
$\omega_0,\mathscr{M}^2$ or $f$, respectively, are shown. It is obvious that
these three parameters all have obvious effects on the vector
self-energy $\Sigma_{\rm{V}}^0$,  while $\omega_0$ and $\mathscr{M}^2$ have little
impacts on the scalar self-energy $\Sigma_{\rm{S}}^0$, and the effects of the parameter
$f$ on both the $\Sigma_{\rm{S}}^0$ and $\Sigma_{\rm{V}}^0$ are
shown to be sizable (similar phenomena are also displayed in Fig.\,\ref{fig_fSNMPNM}).

\begin{figure}[h!]
\centering
  \includegraphics[width=8.5cm]{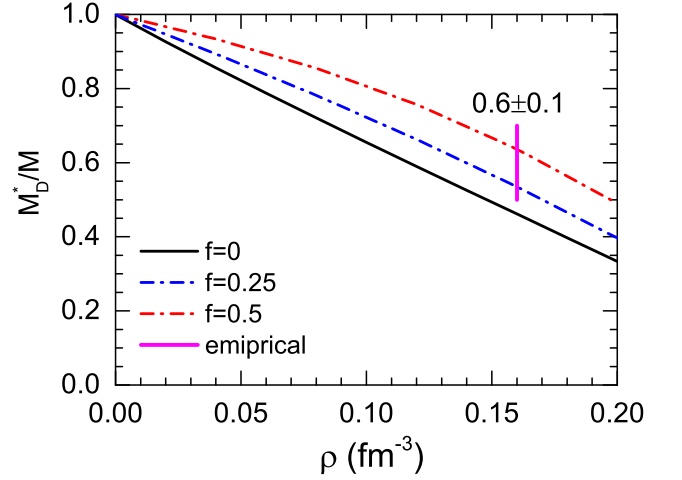}
  \caption{(Color Online) Nucleon Dirac effective mass in SNM as
 a function of
  density with different $f$, $M_{\rm{D}}^{\ast}\equiv M_0^{\ast}$. }
  \label{fig_fDirac}
\end{figure}
\begin{figure}[h!]
\centering
  \includegraphics[width=8.5cm]{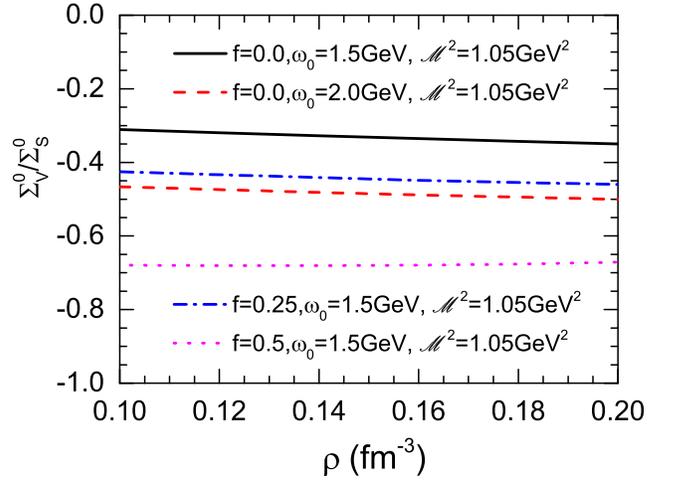}
  \caption{(Color Online) Ratio of $\Sigma_{\rm{V}}^0$ and $\Sigma_{\rm{S}}^0$ as
 a function of density with different $f$ or $\omega_0$.}
  \label{fig_RatioVS}
\end{figure}
\begin{figure}[h!]
\centering
  \includegraphics[width=8.5cm]{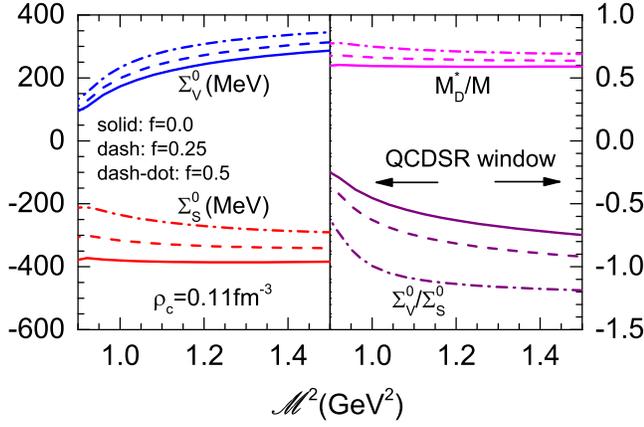}
  \caption{(Color Online) $\Sigma_{\rm{V}}^0$ and $\Sigma_{\rm{S}}^0
  $ as
  functions of Borel mass squared with different $f$ at
  $\rho_{\rm{c}}=0.11\,\rm{fm}^{-3}$.
}
  \label{fig_BMSV}
\end{figure}
\begin{figure}[h!]
\centering
  \includegraphics[width=8.5cm]{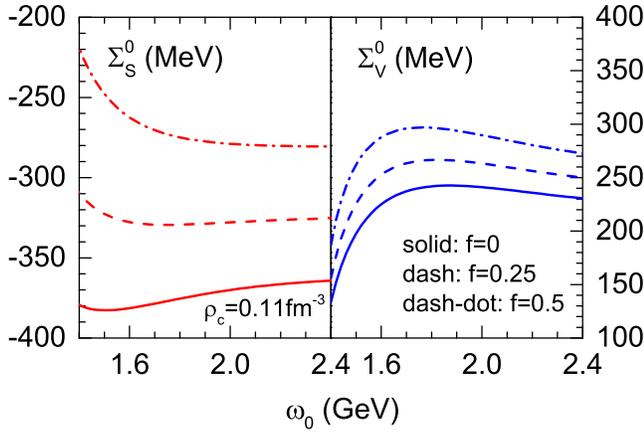}
  \caption{(Color Online) $\Sigma_{\rm{S}}^0$ (left) and $\Sigma_{\rm{V}}^0
  $ (right) as
  functions of $\omega_0$ with different $f$ at
  $\rho_{\rm{c}}=0.11\,\rm{fm}^{-3}$.
 }
  \label{fig_w0SV}
\end{figure}

\begin{figure}[h!]
\centering
  \includegraphics[width=8.cm]{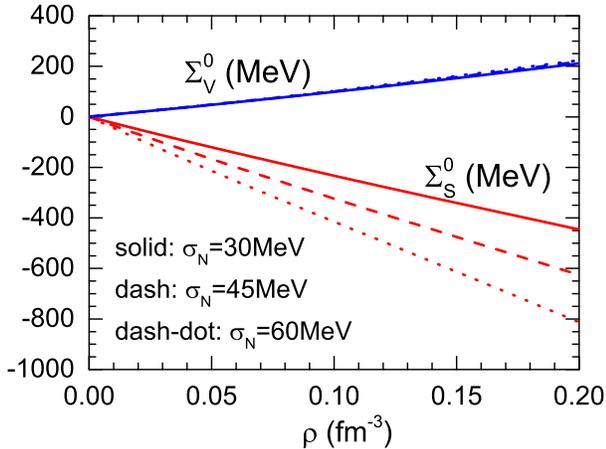}
  \caption{(Color Online) $\Sigma_{\rm{V}}^0$ and $\Sigma_{\rm{S}}^0
  $ as functions of density with different values of $\sigma_{\rm{N}}$.}
  \label{fig_SignSE}
\end{figure}

Fig.\,\ref{fig_fDirac} shows the nucleon Dirac effective mass in
SNM $M_{\rm{D}}^{\ast}\equiv M_0^{\ast}=M+\Sigma_{\rm{S}}^0(\rho)$, as a function of density with different $f$, where the empirical constraint
on the nucleon Dirac effective mass in SNM about $ M_{\rm{D}}^{\ast}(\rho_0)/M\approx0.6\pm0.1$, e.g., by
analyzing the energy level splitting data in several typical finite nuclei (e.g., see ref.\,\cite{Men06}), is also shown.
QCDSR gives a reasonable
result for $M_{\rm{D}}^{\ast}$, for instance, the $M_{\rm{D}}^{\ast}(\rho_0)/M$ is found to be about 0.45
for $f=0$, 0.53 for $f=0.25$, and 0.64 for $f=0.5$. In Fig.\,\ref{fig_RatioVS}, the ratio of $\Sigma_{\rm{V}}^0$ and
$\Sigma_{\rm{S}}^0 $ as a function of density with different
$f$ or $\omega_0$ is shown. Specifically, $
{\Sigma_{\rm{V}}^0(\rho)}/{\Sigma_{\rm{S}}^0(\rho)}\approx-0.7\sim-0.3$, or ${\Sigma_{\rm{V}}^0(\rho)}/{\Sigma_{\rm{S}}^0(\rho)}
\approx-0.5\pm0.2$,
within a wide range of densities, is obtained. The nearly constant ratio
${\Sigma_{\rm{V}}^0(\rho)}/{\Sigma_{\rm{S}}^0(\rho)}$ indicates the density dependence of the
self-energies is almost linear.
In Fig.\,\ref{fig_BMSV} and Fig.\,\ref{fig_w0SV}, the
$\Sigma_{\rm{V}}^0$ and $\Sigma_{\rm{S}}^0$ as functions of
the Borel mass squared or the continuum excitation factor $\omega_0$ with different $f$ at
the cross density
$\rho_{\rm{c}}=0.11\,\rm{fm}^{-3}$ are shown. The $\mathscr{M}^2$--dependence of $\Sigma_{\rm{V}}^0$ and $\Sigma_{\rm{S}}^0$
at other densities is also studied, which shows very similar behaivor. Interestingly, the QCDSR window for the
Borel mass squared $\mathscr{M}^2$, i.e., the region where the dependence of the quantities on $\mathscr{M}^2$ is weak even to be insensitive, from Fig.\,\ref{fig_BMSV} is found to be roughly about
$1\,\rm{GeV}^2\lesssim\mathscr{M}^2\lesssim1.5\,\rm{GeV}^2$, which is consistent with the early studies on it,
e.g., ref.\,\cite{Iof84} gave about $0.8\,\rm{GeV}^2\lesssim\mathscr{M}^2\lesssim1.4\,\rm{GeV}^2$.
Furthermore, the effects of $f$ on both the scalar and the vector self-energies are found to be obvious
from Fig.\,\ref{fig_BMSV}, see also Fig.\,\ref{fig_SNMSE_w0BMf}.
Figs.\,\ref{fig_SNMSE_w0BMf}, \ref{fig_fDirac}, \ref{fig_RatioVS}, \ref{fig_BMSV}, and \ref{fig_w0SV} together
demonstrate that the four-quark condensates have sizable effects on the self-energies.
Investigations on the four-quark condensates from more fundamental approaches (instead of only using the effective parameter $f$)
thus will be extremely important for making further progress in QCDSR method.
In Fig.\,\ref{fig_SignSE}, the density dependence
of $\Sigma_{\rm{V}}^0$ and $\Sigma_{\rm{S}}^0$ with different
$\sigma_{\rm{N}}$ is shown.
It is clearly shown that the sigma term $\sigma_{\rm{N}}$ strongly
affects the scalar self-energy. The reason is that the density dependence of $\Sigma_{\rm{S}}^0$ is directly determined by the $\sigma_{\rm{N}}$.
For instance, in the msQCDSR, one has,
\begin{equation}\Sigma_{\rm{S}}^0(\rho)=-\frac{8\pi^2}{\mathscr{M}^2}\left(\langle\overline{q}q\rangle_{\rm{vac}}+\frac{\sigma_{\rm{N}}\rho}{2m_{\rm{q}}}
\right)-M,\label{zz_ES0}
\end{equation}
i.e., $\Sigma_{\rm{S}}^0(\rho)$ linearly decreases as the density $\rho$ increases.
On the other hand, the effects of $\sigma_{\rm{N}}$ on the vector
self-energy $\Sigma_{\rm{V}}^0$ are found to be essentially small, through the full QCDSR calculations.
Actually, the $\Sigma_{\rm{V}}^0$ is independent of the $\sigma_{\rm{N}}$ in the msQCDSR.

\begin{figure}[h!]
\centering
  \includegraphics[width=8.5cm]{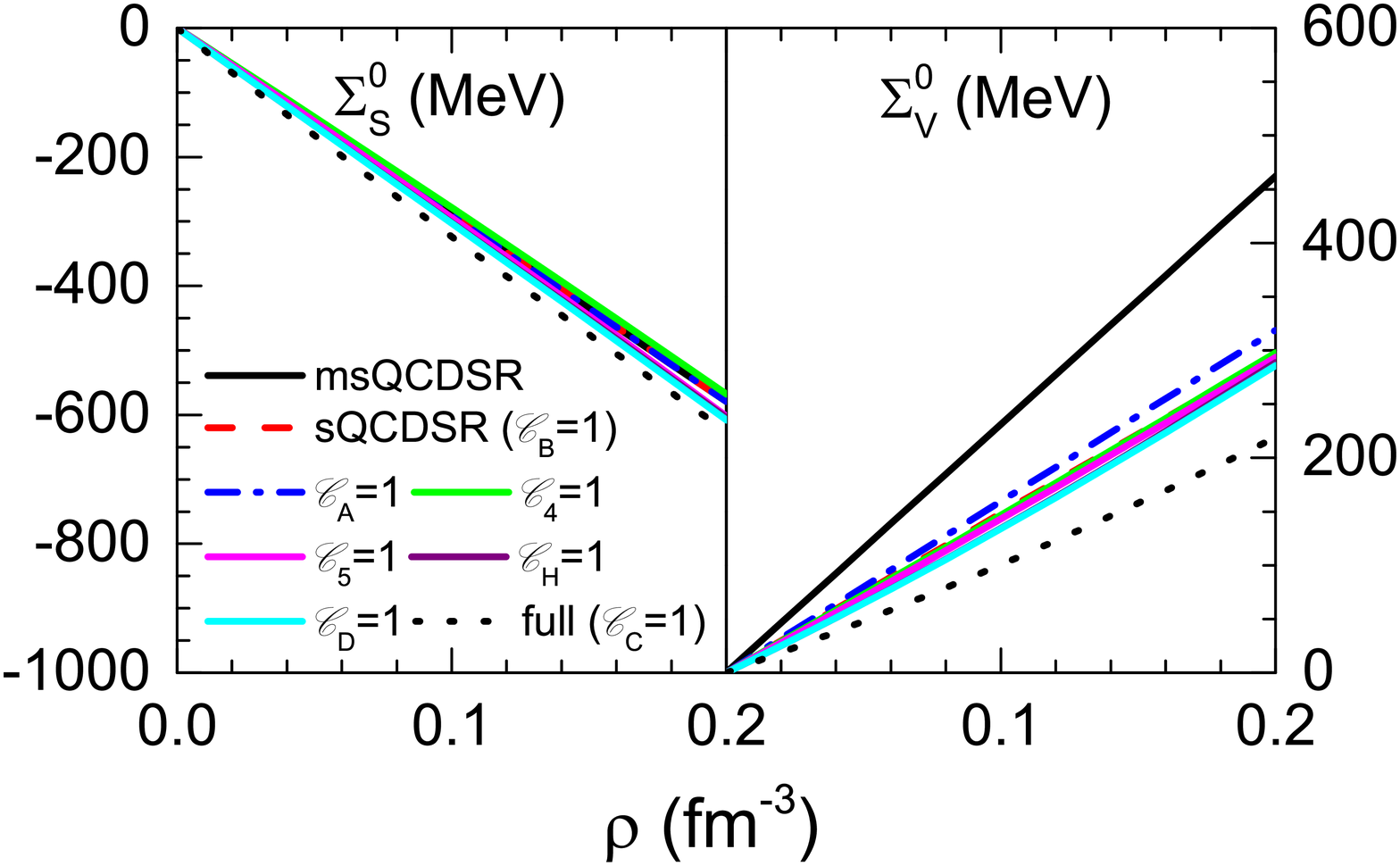}
  \caption{(Color Online) Density dependence of $\Sigma_{\rm{S}}^0$ (left) and $\Sigma_{\rm{V}}^0
  $ (right) obtained order by order as
 the same meaning in Fig.\,\ref{fig_EHO_order_1}.}
  \label{fig_SNMSE_order}
\end{figure}

In Fig.\,\ref{fig_SNMSE_order}, the density dependence of $\Sigma_{\rm{V}}^0$ and
$\Sigma_{\rm{S}}^0$ obtained order by order according to the same
scheme used in Fig.\,\ref{fig_EHO_order_1} is shown.
Since the density dependence of the quark/gluon condensates adopted in this section is linear (without the higher order density terms), the predictions on the dependence of the
self-energies are also found to be roughly linear.
The change on the scalar self-energy $\Sigma_{\rm{S}}^0$ from msQCDSR to fQCDSR is relatively smaller
than that on the vector self-energy $\Sigma_{\rm{V}}^0$. For example, at $\rho_0=0.16\,\rm{fm}^{-3}$, $\Sigma_{\rm{S}}(\rho_0)
$ changes from $-464$\,MeV in msQCDSR to $-506$\,MeV in fQCDSR. However, for the $\Sigma_{\rm{V}}(\rho_0)$, it changes from 370\,MeV in msQCDSR
to 170\,MeV in fQCDSR, inducing a $-54\%$ relative change.

Another feature in Fig.\,\ref{fig_SNMSE_order} is that the predictions on the scalar self-energy
in the msQCDSR (black solid line) and in the sQCDSR (red dash line) are exactly the same. This can be proved as follows. In the msQCDSR, the
expression for the effective mass is given by (in order to make a general proof here the Ioffe parameter can take any value instead of
being set to $-1$, $\langle\overline{q}q\rangle_{\rho}$ is the chiral condensate at finite densities in SNM)
\begin{equation}
M_0^{\ast}(\rho)=-\frac{16\pi^2c_1}{c_4}\frac{\langle\overline{q}q\rangle_{\rho}}{
\mathscr{M}^2},\end{equation} or equivalently $
{c_1}/{c_4}=-{\mathscr{M}^2{M}_0^{\ast}}/{16\pi^2\langle\overline{q}q\rangle_{\rho}}
$, taking it at zero density leads to $c_1/c_4=-{\mathscr{M}^2{M}}/{16\pi^2\langle\overline{q}q\rangle_{\rm{vac}}}$, thus one obtains the following formula for the nucleon scalar self-energy as a function of density,
\begin{align}\label{1defe}
\Sigma_{\rm{S}}^{0,\rm{msQCDSR}}(\rho)=&M_0^{\ast}(\rho)-M\notag\\
=&M
\left[\frac{\langle\overline{q}q\rangle_{\rho}}{\langle\overline{q}q\rangle_{\rm{vac}}}-1\right],\end{align}
which is independent of the parameters $c_1$ and $c_4$.
Similarly, in the sQCDSR, one has an extra term, i.e.,
$c_1\langle\overline{q}q\rangle_{\rm{vac}}^2/6\equiv c_1\mathcal{I}$ in the QCDSR equations, and the corresponding effective mass is
given by
\begin{equation}
M_0^{\ast}(\rho)=-\frac{({c_1}/{16\pi^2})\mathscr{M}^4\langle\overline{q}q\rangle_{\rho}
}{({c_4}/{256\pi^4})\mathscr{M}^6+c_1\mathcal{I}},\end{equation}
solving it gives
\begin{equation}
\frac{c_1}{c_4}=-\frac{({{M}}/{256\pi^4})\mathscr{M}^6
}{(1/{16\pi^2})\mathscr{M}^4\langle\overline{q}q\rangle_{\rm{vac}}
+M\mathcal{I}},\end{equation}
or equivalently,
\begin{equation}
M=-\frac{({c_1}/{16\pi^2})\mathscr{M}^4\langle\overline{q}q\rangle_{\rm{vac}}
}{({c_4}/{256\pi^4})\mathscr{M}^6+c_1\mathcal{I}},
\end{equation}
thus
\begin{align}
\Sigma_{\rm{S}}^{0,\rm{sQCDSR}}(\rho)=&M_0^{\ast}(\rho)-M\notag\\
=&-\frac{\displaystyle\frac{c_1}{c_4}\frac{\mathscr{M}^4\langle\overline{q}q\rangle_{\rm{vac}}}{
16\pi^2}\left[\frac{\langle\overline{q}q\rangle_{\rho}}{\langle\overline{q}q\rangle_{\rm{vac}}}-1\right]}{\displaystyle
\frac{\mathscr{M}^6}{256\pi^4}+\frac{c_1\mathcal{I}}{c_4}}\notag\\
=&M\left[\frac{\langle\overline{q}q\rangle_{\rho}}{\langle\overline{q}q\rangle_{\rm{vac}}}-1\right],
\end{align}
i.e., the scalar self-energy $\Sigma_{\rm{S}}^0(\rho)$
obtained in
sQCDSR is still given by Eq.\,(\ref{1defe}), furnishing the proof.
When other condensates are included in the QCDSR equations, however, there will be no close expression for the scalar self-energy.

\subsection{The Symmetry Energy}\label{ss_2}

\begin{figure}[h!]
\centering
  \includegraphics[width=8.5cm]{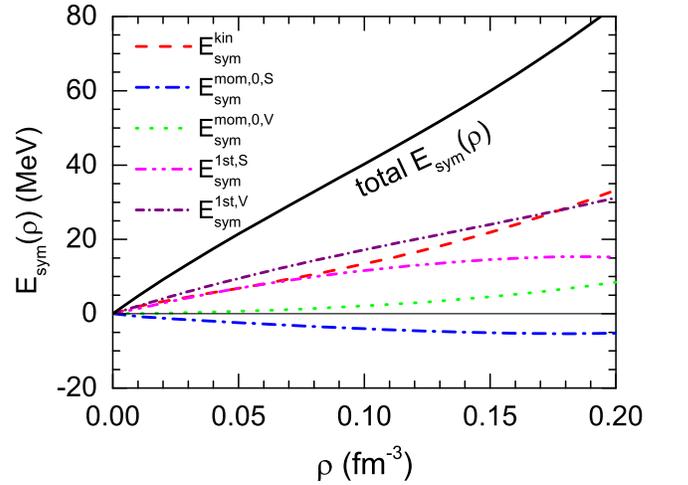}
  \caption{(Color Online) Symmetry energy with its self-energy
  decompositions as functions of density in fQCDSR.}
  \label{fig_EsymDecomp}
\end{figure}

In this subsection, we generally study the self-energy decomposition of the symmetry energy
through the full QCDSR equations without any fitting scheme.  The five terms in Eq.\,(\ref{sec1_Esym}) are the kinetic symmetry energy $E_{\rm{sym}}^{\rm{kin}}(\rho)$, contributions from momentum dependence of the
scalar self-energy $E_{\rm{sym}}^{\rm{mom,0,S}}(\rho)$ and that of the vector self-energy $E_{\rm{sym}}^{\rm{mom,0,V}}(\rho)$, first order symmetry scalar self-energy $E_{\rm{sym}}^{\rm{1st,S}}(\rho)$ and the first order symmetry vector self-energy $E_{\rm{sym}}^{\rm{1st,V}}(\rho)$,
respectively.
In Fig.\,\ref{fig_EsymDecomp}, each term
as a function of density is shown. For example,  at the cross density
$\rho_{\rm{c}}=0.11\,\rm{fm}^{-3}$, one has $
E_{\rm{sym}}^{\rm{kin}}(\rho_{\rm{c}})\approx14.9\,\rm{MeV}$, $
E_{\rm{sym}}^{\rm{mom,0,S}}(\rho_{\rm{c}})\approx-4.3\,\rm{MeV}$, $
E_{\rm{sym}}^{\rm{mom,0,V}}(\rho_{\rm{c}})\approx2.5\,\rm{MeV}$, $
E_{\rm{sym}}^{\rm{1st,S}}(\rho_{\rm{c}})\approx12.3\,\rm{MeV}$, and similarly $
E_{\rm{sym}}^{\rm{1st,V}}(\rho_{\rm{c}})\approx18.6\,\rm{MeV}$, leading to the total symmetry energy about
$ E_{\rm{sym}}(\rho_{\rm{c}})\approx44.1\,\rm{MeV}$. Obviously, the value of
the total symmetry energy at $\rho_{\rm{c}}$ is much larger than
the empirical constraints on it, e.g., the one obtained by analyzing the binding energy difference between a heavy isotope pair gives $
E_{\rm{sym}}(\rho_{\rm{c}})\approx26.65\pm0.2\,\rm{MeV}$\,\cite{Zha13}.
As a reference, we also give these symmetry energy terms at the saturation density $\rho_0$, they are about
23.9\,MeV, $-$5.0\,MeV, 5.2\,MeV, 14.7\,MeV, and 25.4\,MeV, leading to $E_{\rm{sym}}(\rho_0)\approx64.2\,\rm{MeV}$.
Despite the discrepancy between the symmetry energy from the fQCDSR and the empirical constraints is large,
QCDSR predicts the sign of the momentum-dependence of the nucleon self-energies, i.e.,
\begin{align}
E_{\rm{sym}}^{\rm{mom,0,S}}(\rho)=&\frac{k_{\rm{F}}}{6}\frac{M_0^{\ast}}{{e}_{\rm{F}}^{\ast}}\left.
\frac{\d\Sigma_{\rm{S}}^0}{\d|\v{k}|}\right|_{|\v{k}|=k_{\rm{F}}}<0,\label{f1g1_1}\\
E_{\rm{sym}}^{\rm{mom,0,V}}(\rho)=&\frac{k_{\rm{F}}}{6}\left.\frac{\d\Sigma_{\rm{S}}^0}{\d|\v{k}|}\right|_{|\v{k}|=k_{\rm{F}}}>0,\label{f1g1_2}
\end{align}
and these relations will be further studied in Subsection~\ref{ss_4} and the following sections when the fitting scheme (Subsection \ref{sb_FS}) is adopted.

\begin{figure}[h!]
\centering
  \includegraphics[width=8.5cm]{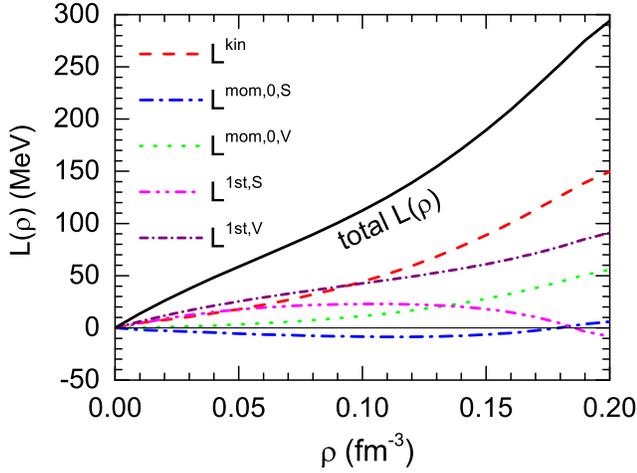}
  \caption{(Color Online) The same as Fig.\,\ref{fig_EsymDecomp} but for the slope parameter of the symmetry energy.}
  \label{fig_LDecomp}
\end{figure}

The slope parameter of the symmetry energy, i.e., $
L(\rho)=3\rho{\text{d}E_{\rm{sym}}(\rho)}/{\text{d}\rho}$ (it is easy to recognize meaning of $L$ from the context
since we use the same letter for the slope parameter of the symmetry energy and the anomalous dimension of the interpolation
field defined in Eq.\,(\ref{def_L_a})), can similarly be decomposed as different nucleon self-energy terms, see ref.\,\cite{Cai12} for more details.
However, the most straightforward manner is to calculate each symmetry energy decomposition term, i.e., $L^{j}(\rho)=3\rho\text{d} E_{\rm{sym}}^j(\rho)/\text{d}\rho$
with $j=$``kin'', ``mom,0,S'', ``mom,0,V'', ``1st,S'', and ``1st,V''.
In Fig.\,\ref{fig_LDecomp}, the
density-dependence of these terms is shown, and for instance, at the cross density $\rho_{\rm{c}}$,
one obtains $ L^{\rm{kin}}(\rho_{\rm{c}})\approx51.4\,\rm{MeV}$,
$L^{\rm{mom,0,S}}(\rho_{\rm{c}})\approx-8.7\,\rm{MeV}$,
$L^{\rm{mom,0,V}}(\rho_{\rm{c}})\approx13.7\,\rm{MeV}$,
$L^{\rm{1st,S}}(\rho_{\rm{c}})\approx23.0\,\rm{MeV}$, and $
L^{\rm{1st,V}}(\rho_{\rm{c}})\approx45.8\,\rm{MeV}$, leading to $
L(\rho_{\rm{c}})\approx125.1\,\rm{MeV}$. The value of
$L(\rho_{\rm{c}})$ is similarly found to be much larger than the constraint on it, e.g., $
L(\rho_{\rm{c}})\approx46\pm4.5\,\rm{MeV}$ from ref.\,\cite{Zha13} by analysing the correlation
between the neutron skin thickness of the neutron-rich heavy nuclei and the $L$ parameter.
As a reference, we also list these terms at the saturation density, they are 100.6\,MeV, $-$4.9\,MeV, 32.6\,MeV,
14.9\,MeV, and 65.9\,MeV, respectively, and $L\equiv L(\rho_0)\approx209.2\,\rm{MeV}$.
Thus, the prediction on the $L$ parameter from the QCDSR without any fitting scheme is much larger than the
empirical one both at the cross density and at the saturation density, for example, the $L$ parameter at $\rho_0$
was nowadays better constrained to be around about $60\pm30$\,MeV (see, e.g., refs.\,\cite{Che12a,LiBA13,Oer17,Che17xx}).

\begin{figure}[h!]
\centering
  \includegraphics[width=8.5cm]{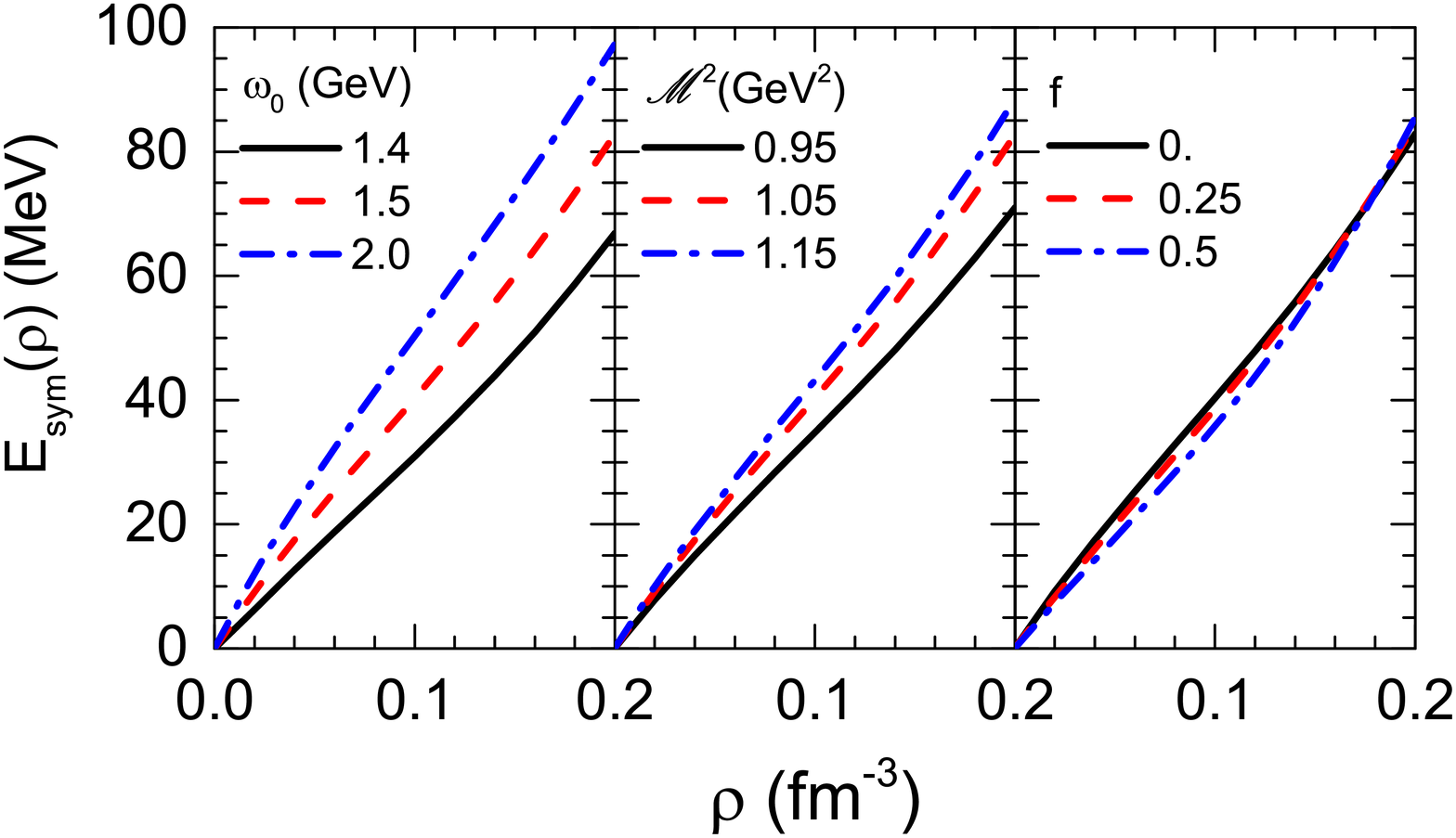}
  \caption{(Color Online) Symmetry energy as a function of density with different $\omega_0$ (left),
$\mathscr{M}^2$ (central) or $f$ (right).}
  \label{fig_Esym_w0BMf}
\end{figure}

In Fig.\,\ref{fig_Esym_w0BMf}, the symmetry energy as a
function of density with different $\omega_0,\mathscr{M}^2$ or $f$ is shown.
It is clearly shown from the figure that the effects of $\omega_0$ and $\mathscr{M}^2$ are
comparatively larger than those of $f$.
This could be easily understood, e.g., according to Eq.\,(\ref{Esyms}), that the effects of the $f$--term on the symmetry energy is mainly characterized by
the factor $\xi\equiv(\alpha/\beta)f\approx0.1f$. Thus there is no surprise that
the effects of the parameter $f$ on
the symmetry energy is smaller than these on the EOS of
SNM or that of PNM since the relevant term in $E_0(\rho)$ or $E_{\rm{n}}(\rho)$ is directly proportional to $f$
or $(1-\alpha/\beta)f$, see the relevant analyses in Section~\ref{SEC_HOEsti}, i.e., Eq.\,(\ref{E0s}) and Eq.\,(\ref{Ens}).
It means that the effects of $f$ on the symmetry energy is roughly cancelled, see the left panel of Fig.\,\ref{fig_fEHOEsym} for
a similar calculation.

\begin{figure}[h!]
\centering  
  \includegraphics[width=8.7cm]{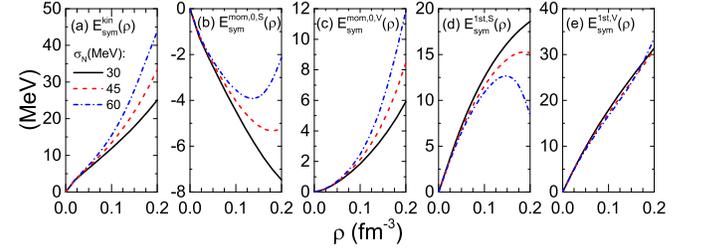}
  \caption{(Color Online) Density dependence of the self-energy decomposition of the symmetry energy with different $\sigma_{\rm{N}}$.}
  \label{fig_EsymDecomSign}
\end{figure}

\begin{figure}[h!]
\centering
  \includegraphics[width=8.cm]{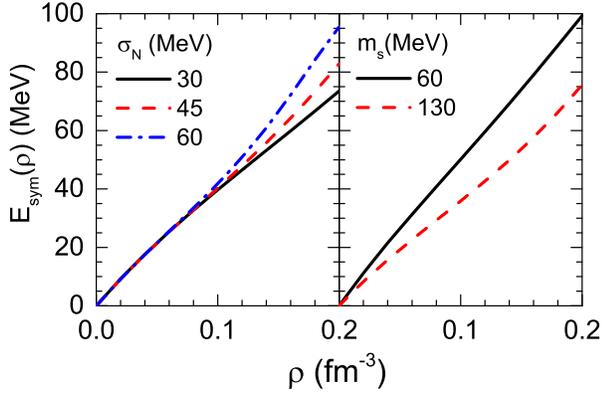}
  \caption{(Color Online) Symmetry energy as a function of density with different $\sigma_{\rm{N}}$ (left)
  or $m_{\rm{s}}$ (right).}
  \label{fig_EsymSignMs}
\end{figure}

In Fig.\,\ref{fig_EsymDecomSign}, the density dependence of the self-energy decomposition
of the symmetry energy is shown with different $\sigma_{\rm{N}}$.
The effects of $\sigma_{\rm{N}}$ on the kinetic symmetry energy $E_{\rm{sym}}^{\rm{kin}}(\rho)$ is found to be large and it
could be understood as follows: as $\sigma_{\rm{N}}$ increases, then according to, e.g., Eq.\,(\ref{zz_M0ast}) from
the msQCDSR, one finds that the effective mass $M_0^{\ast}(\rho)$ is reduced.
Consequently, $e_{\rm{F}}^{\ast}$ is also reduced, leading to the enhancement on the kinetic symmetry energy $k_{\rm{F}}^2/6e_{\rm{F}}^{\ast}$.
For instance, the $E_{\rm{sym}}^{\rm{kin}}(\rho_0)$ changes from about 19.2\,MeV at $\sigma_{\rm{N}}=30\,\rm{MeV}$ to 30.9\,MeV at $\sigma_{\rm{N}}=60\,\rm{MeV}$,
roughly a 61\% relative increase.
However, on the other hand, the factor $y\equiv M_0^{\ast}/e_{\rm{F}}^{\ast}\approx1-x^2/2$ decreases as $x\equiv k_{\rm{F}}/M_0^{\ast}$ increases.
Then according to Eq.\,(\ref{ddd2}), whether $E_{\rm{sym}}^{\rm{1st,S}}(\rho)
\sim y\sigma_{\rm{N}}$ is enhanced or reduced depends on the competition between the enhancement factor $\sigma_{\rm{N}}$
and the reduction factor $y$, and the final effect is found to reduce the $E_{\rm{sym}}^{\rm{1st,S}}(\rho)$
at a fixed density, as shown in the panel (d) of Fig.\,\ref{fig_EsymDecomSign}, i.e., the factor $y$ wins the competition over $\sigma_{\rm{N}}$.
Besides $E_{\rm{sym}}^{\rm{kin}}(\rho)$ and $E_{\rm{sym}}^{\rm{1st,S}}(\rho)$, the effects of $\sigma_{\rm{N}}$
on the other three terms are very non-trivial and could not be analyzed in a semi-analytical manner. The $E_{\rm{sym}}^{\rm{mom,0,S}}(\rho)$ and $E_{\rm{sym}}^{\rm{mom,0,V}}(\rho)$,
for instance, are found to be enhanced from about $-$4.8\,MeV to $-$3.6\,MeV and 2.2\,MeV to 3.0\,MeV as $\sigma_{\rm{N}}$ changes from 30\,MeV to 60\,MeV,
respectively, and the $E_{\rm{sym}}^{\rm{1st,V}}(\rho)$ is almost unaffected.
Moreover, as shown in Fig.\,\ref{fig_EsymDecomSign}, the enhancement due to the increasing of $\sigma_{\rm{N}}$ on the kinetic symmetry energy
is much larger than the reduction on $E_{\rm{sym}}^{\rm{1st,S}}(\rho)$, leading to
the enhancement on the total symmetry energy, as shown in the left panel in Fig.\,\ref{fig_EsymSignMs}.
For example, if one takes $\sigma_{\rm{N}}=45\pm15\,\rm{MeV}$ as shown in the left panel in
Fig.\,\ref{fig_EsymSignMs}, then the uncertainty on the symmetry energy is found to be about 3.5\,MeV (12.3\,MeV) at $\rho_{\rm{c}}$ ($\rho_0$).
Furthermore, the $\sigma_{\rm{N}}$ affects the density dependence of the symmetry energy in the sense that a smaller $\sigma_{\rm{N}}$
induces a softer symmetry energy, i.e., a smaller slope parameter $L$. Specifically, the relative change generated by varying $\sigma_{\rm{N}}$
from 30\,MeV to 60\,MeV on the $L$ parameter shown in the left panel of Fig.\,\ref{fig_EsymSignMs} is about 73\%.
The connection between the nucleon-sigma term $\sigma_{\rm{N}}$ and the parameter $L$ itself
provides a possible mechanism to investigate the density dependence of the symmetry energy.

\begin{figure}[h!]
\centering
  \includegraphics[width=8.cm]{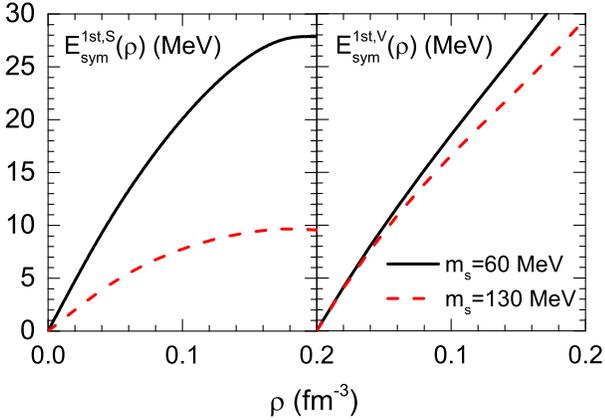}
  \caption{(Color Online) $E_{\rm{sym}}^{\rm{1st,S}}$ (left) and $E_{\rm{sym}}^{\rm{1st,V}}$ (right) as functions of the density with different $m_{\rm{s}}$.}
  \label{fig_EsymDecomMs}
\end{figure}

Interestingly, as shown in the right panel in Fig.\,\ref{fig_EsymSignMs}, a
larger mass of the s quark will induce a smaller symmetry energy.
In fact, one can find that the effects of s quark mass $m_{\rm{s}}$ come into play in the symmetry energy through the parameters $\alpha$ and $\beta$
defined in Eq.\,(\ref{zz_apm}). For instance, taking a smaller (larger) value of the s quark mass, e.g., $m_{\rm{s}}=60\,\rm{MeV}$ ($m_{\rm{s}}=130\,\rm{MeV}$),
one then obtains
$\alpha\approx0.29$ ($0.14$) and $\beta\approx1.71$ ($1.86$), leading to $\alpha/\beta\approx0.17$ ($0.08$).
Then according to the expression for $\Sigma_{\rm{sym}}^{\rm{S}}(\rho)$ from the msQCDSR (see Eq.\,(\ref{zz_ESsym})), one can easily find that a smaller (larger)
$m_{\rm{s}}$ induces a larger (smaller) $E_{\rm{sym}}^{\rm{1st,S}}(\rho)$,
shown in the left panel in Fig.\,\ref{fig_EsymDecomMs} (e.g.,
the $\Sigma_{\rm{sym}}^{\rm{1st,S}}(\rho_{\rm{c}})$ changes from about 8.2\,MeV to 21.5\,MeV as $m_{\rm{s}}$ changes from 130\,MeV
to 60\,MeV, inducing an effect about 13.3\,MeV).
On other hand, the s quark mass only affects the $E_{\rm{sym}}^{\rm{1st,V}}(\rho)$ in a more minor manner
through the total QCDSR equations, shown in the right panel in Fig.\,\ref{fig_EsymDecomMs} (the corresponding effect is found to be about 2.4\,MeV at $\rho_{\rm{c}}$),
i.e., the change on the $E_{\rm{sym}}^{\rm{1st,V}}(\rho)$ due to the $m_{\rm{s}}$ is relatively small compared to its effects on $E_{\rm{sym}}^{\rm{1st,S}}(\rho)$.
From the decomposition (\ref{sec1_Esym}), $m_{\rm{s}}$ obviously has no effects on the other three terms, i.e., $E_{\rm{sym}}^{\rm{kin}}(\rho)$,
$E_{\rm{sym}}^{\rm{mom,0,S}}(\rho)$ and $E_{\rm{sym}}^{\rm{mom,0,V}}(\rho)$ (since these terms only involve the symmetric quantities).
These analyses finally lead to the conclusion that considering the $m_{\rm{s}}$ uncertainties will induce sizable effects on the total symmetry energy.
To our best knowledge,
since there exists no similar analyses on the relation between the symmetry energy and the s quark mass,
our studies on it may provide new insights into the physical origin of as well as the uncertainties on the nuclear symmetry energy with respect to the s quark mass.

\begin{figure}[h!]
\centering
  \includegraphics[width=8.5cm]{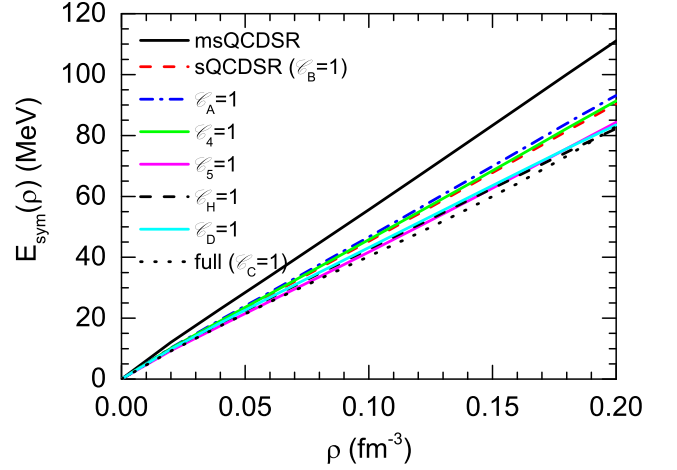}
  \caption{(Color Online) Density dependence of the symmetry energy
 obtained order by order as
 the same scheme in Fig.\,\ref{fig_EHO_order_1}.}
  \label{fig_Esym_order}
\end{figure}

In Fig.\,\ref{fig_Esym_order}, we show the density dependence of the
symmetry energy obtained order
by order. One can find that the four-quark condensates of the type (\ref{f4}) essentially
have large effects on the symmetry energy (from the msQCDSR with black solid line to the sQCDSR with red dash line denoted by ``$\mathscr{C}_{\rm{B}}=1$'').
For instance, the symmetry energy at $\rho_{\rm{c}}$
($\rho_0$) in the msQCDSR is found be about 61.2\,MeV (88.9\,MeV). On the other hand, the $E_{\rm{sym}}$
in the sQCDSR at $\rho_{\rm{c}}$ ($\rho_0$) is found to be about 49.8\,MeV (72.3\,MeV), generating a reduction about
11.4\,MeV (16.6\,MeV), respectively.
Once Eq.~(\ref{f4}) is included in the QCDSR equations, the following effects are found to be much smaller
compared to those from Eq.~(\ref{f4}).
These estimates demonstrate again the importance of the four-quark condensates (\ref{f4}).

\begin{figure}[h!]
\centering  
  \includegraphics[width=8.5cm]{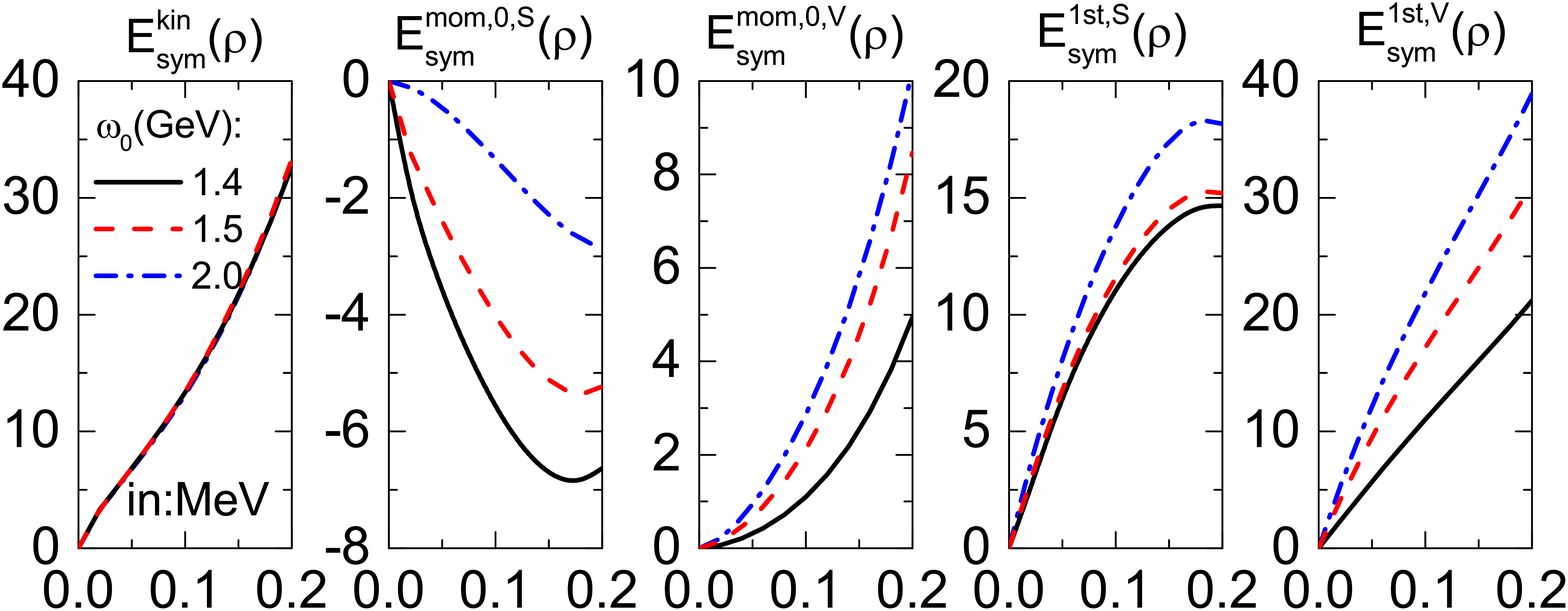}\\[0.cm]
  \includegraphics[width=8.5cm]{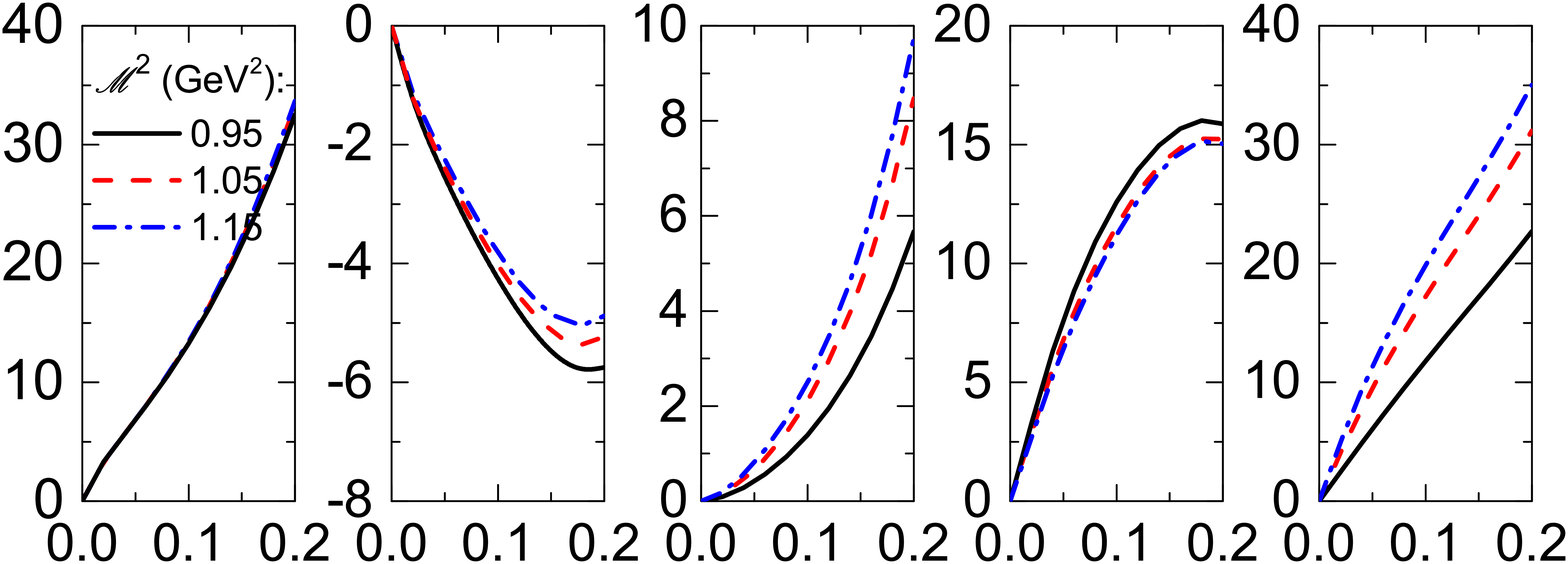}\\[0.cm]
  \includegraphics[width=8.5cm]{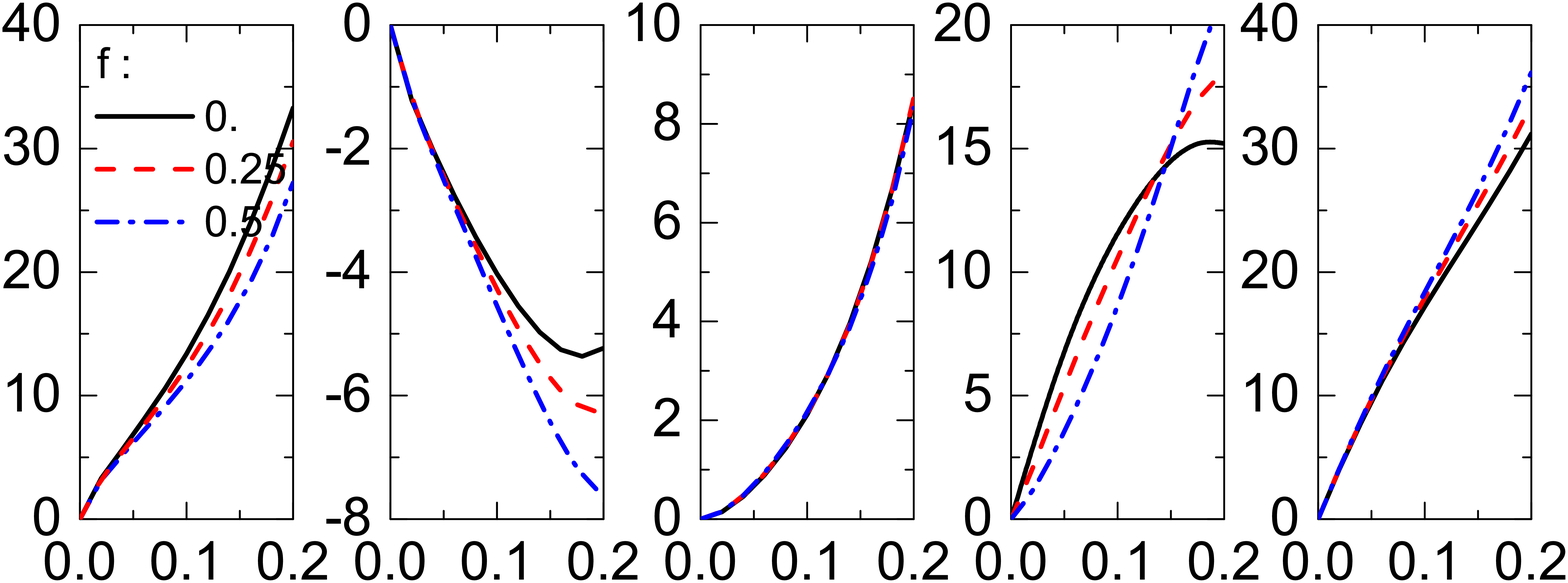}\\[0.cm]
  \includegraphics[width=8.5cm]{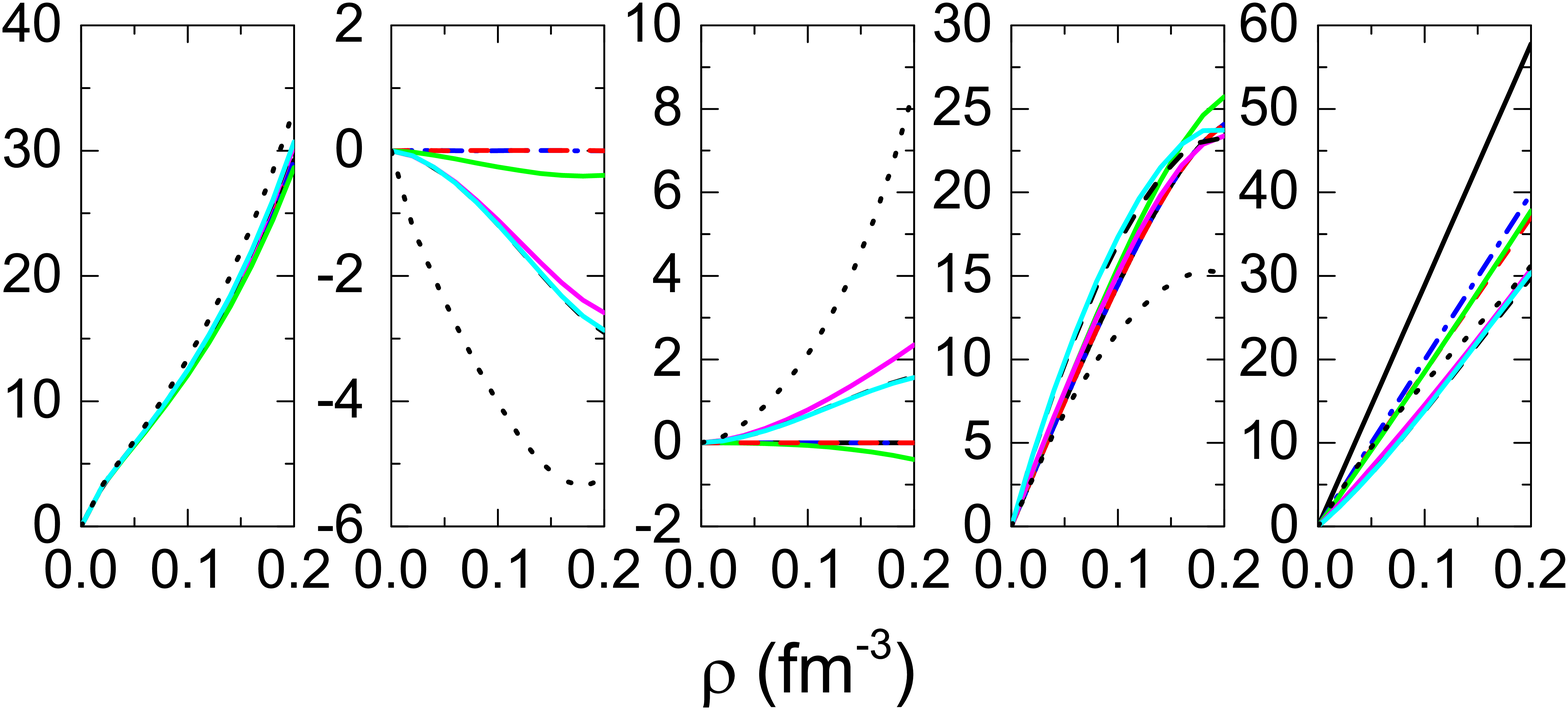}
  \caption{(Color Online) Self-energy decomposition of the symmetry energy with different $\omega_0
  $ (upper), $\mathscr{M}^2$ (second line), and $f$ (third line).
  The order by order calculations on these decompositions are also shown (fourth line), and the meaning of
  different curves is the same as those in Fig.\,\ref{fig_Esym_order}.}
  \label{fig_EsymDecomw0BMfSign}
\end{figure}

Finally, the self-energy decomposition of the symmetry energy with different
$\omega_0$/$\mathscr{M}^2$/$f$ is shown in
Fig.\,\ref{fig_EsymDecomw0BMfSign}. The order by order results on these terms
are also shown in the fourth row. Several features are necessary to be pointed from these figures:
1). At a fixed density, the effects of the continuum excitation factor $\omega_0$ on each decomposition term are positive correlated, i.e.,
a smaller (larger) $\omega_0$ corresponds to a smaller (larger) $E_{\rm{sym}}^{\rm{mom,0,S}}(\rho)$ or $E_{\rm{sym}}^{\rm{mom,0,V}}(\rho)$
or $E_{\rm{sym}}^{\rm{1st,S}}(\rho)$ or $E_{\rm{sym}}^{\rm{1st,V}}(\rho)$, and $\omega_0$ almost does not affect $E_{\rm{sym}}^{\rm{kin}}(\rho)$.
2). The effects of the Borel mass squared $\mathscr{M}^2$ on the $E_{\rm{sym}}^{\rm{mom,0,S}}(\rho)$ or $E_{\rm{sym}}^{\rm{mom,0,V}}(\rho)$
or $E_{\rm{sym}}^{\rm{1st,S}}(\rho)$ are similar like $\omega_0$. However, as $\mathscr{M}^2$ increases, $E_{\rm{sym}}^{\rm{1st,S}}(\rho)$
is shown to be reduced, and the overall effects of $\mathscr{M}^2$ are to enhance the symmetry energy (see also Fig.\,\ref{fig_Esym_w0BMf}).
3). As the four-quark condensates parameter $f$ increase, the effective mass $M_0^{\ast}$ (and consequently $e_{\rm{F}}^{\ast}$) increases according to Eq.\,(\ref{zz_zz}),
leading to the reduction on the kinetic symmetry energy at a fixed density, as shown in the left panel of row 3 in Fig.\,\ref{fig_EsymDecomw0BMfSign}.
4). For the order by order calculations, the continuum excitation effects are found to be large on $E_{\rm{sym}}^{\rm{mom,0,S}}(\rho)$, $E_{\rm{sym}}^{\rm{mom,0,V}}(\rho)$,
and $E_{\rm{sym}}^{\rm{1st,S}}(\rho)$, see the black dot lines shown in the second, third and fourth panels (from left) in the fourth row of
Fig.\,\ref{fig_EsymDecomw0BMfSign}.
Features shown in these figures clearly establish the close relationships between the self-energy decomposition terms
of the symmetry energy and the effective parameters appearing in the QCDSR, providing important
guidelines to understand the physical origin of the symmetry energy as well as the corresponding uncertainties.

\subsection{Correlation between $E_{\textmd{sym}}(\rho)$
and Condensates}\label{ss_3}

In this subsection, we study the correlation of the nucleon self-energies and the
symmetry energy with the quark/gluon condensates.
These explorations are useful for generally establishing the connection between the EOS of ANM in nuclear physics,
and the condensates encapsulating the degrees of freedom of quarks and gluons in hadronic physics (and related issues in QCD).
The condensates and their
uncertainties are listed as follows (see Subsection~\ref{sb_QG}),
\begin{align}
\langle\overline{q}q\rangle_{\rm{vac}}&:~~-(255\sim220\,\rm{MeV})^3,~-(252\,\rm{MeV})^3;\\
\left\langle\frac{\alpha_{\rm{s}}}{\pi}G^2\right\rangle_{\rm{vac}}&:~(330\pm30\,\rm{MeV})^4,~
(330\,\rm{MeV})^3;\\
G_{\rm{a}}&:~~325\pm75\,\rm{MeV},~
325\,\rm{MeV};\\
G_{\rm{s}}&:~~100\pm10\,\rm{MeV},~
100\,\rm{MeV};\\
\vartheta_1&:~~0.1\leq\vartheta_1\leq0.6,~0.35;\\
\vartheta_3&:~~0.0\leq\vartheta_3\leq1.0,~0.51;\\
\varphi_1&:~~0.2\leq\varphi_1\leq0.8,~0.55;\\
\varphi_2&:~~0.1\leq\varphi_2\leq0.6,~0.34;\\
\varphi_3&:~~0.0\leq\varphi_3\leq0.3,~0.145;\\
\langle
g_{\rm{s}}\overline{q}\sigma\mathcal{G}q\rangle_{\rm{sym}}^{\rm{p}}&:~~
0.62\,\rm{GeV}^2\sim3\,\rm{GeV}^2,~
0.62\,\rm{GeV}^2;\\
\langle
g_{\rm{s}}{q}^{\dag}\sigma\mathcal{G}q\rangle_{\rm{sym}}^{\rm{p}}&:~
-0.33\,\rm{GeV}^2\sim0.66\,\rm{GeV}^2,~
0.66\,\rm{GeV}^2;\\
\Lambda_{\rm{QCD}}&:~~0.17\pm0.5\,\rm{GeV},~0.17\,\rm{GeV},
\end{align}
where the last term in each line represents the central value of the
quantity, and
\begin{align}
G_{\rm{a}}=&\left\langle\frac{\alpha_{\rm{s}}}{\pi}(\v{E}^2-\v{B}^2)\right\rangle,~~
G_{\rm{s}}=\left\langle\frac{\alpha_{\rm{s}}}{\pi}(\v{E}^2+\v{B}^2)\right\rangle.
\end{align}
Other parameters introduced in QCDSR are $
\alpha_{\rm{s}}=0.5,f=0$, as well as $ \mu=0.5\,\rm{GeV},
\mathscr{M}^2=1.05\,\rm{GeV}^2, \omega_0=1.5\,\rm{GeV},
\sigma_{\rm{N}}=45\,\rm{MeV}, m_{\rm{q}}=3.5\,\rm{MeV}$ and $
m_{\rm{s}}=95\,\rm{MeV}$.

\begin{figure}[h!]
\centering
  \includegraphics[width=8.8cm]{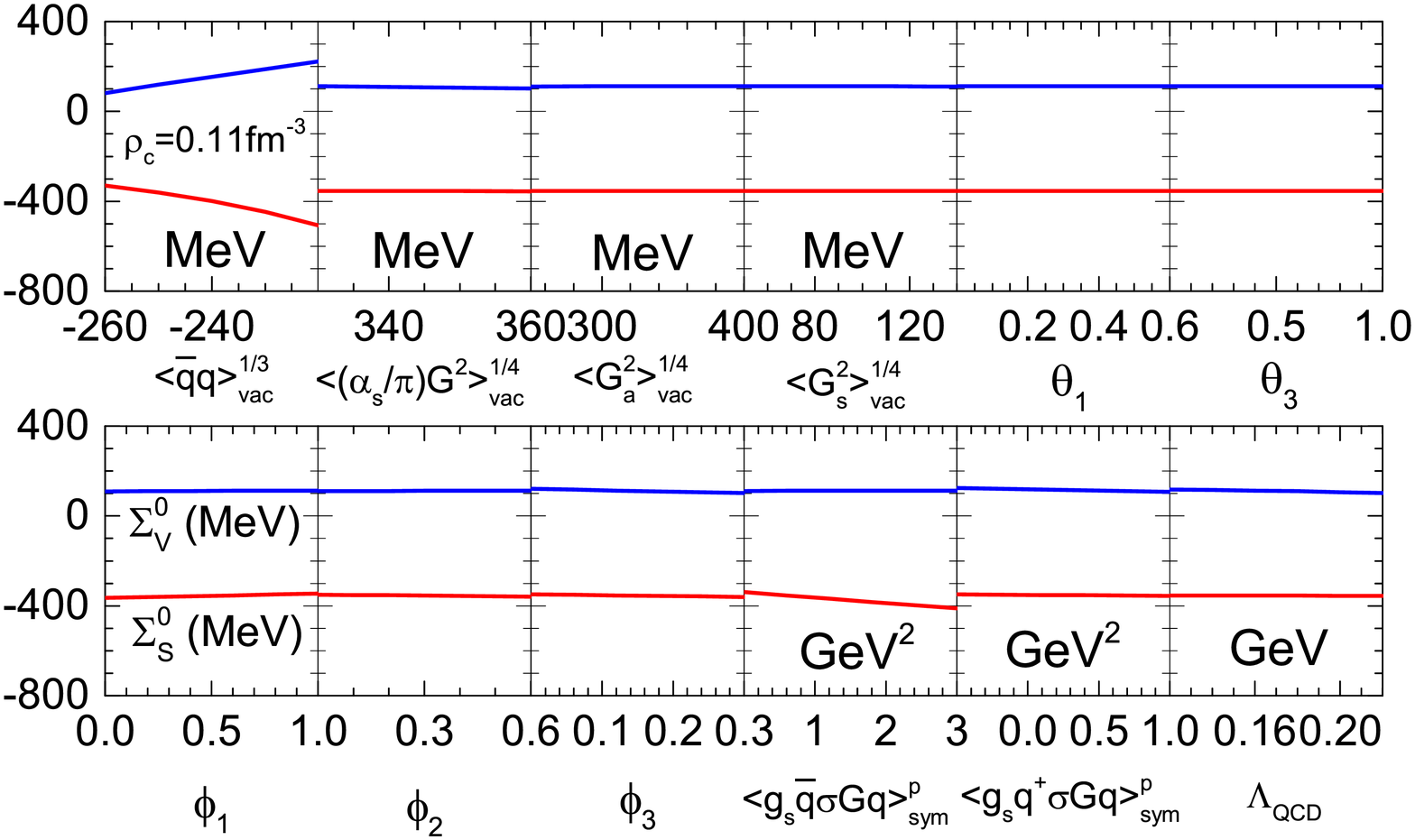}\\[0.5cm]
  \includegraphics[width=8.8cm]{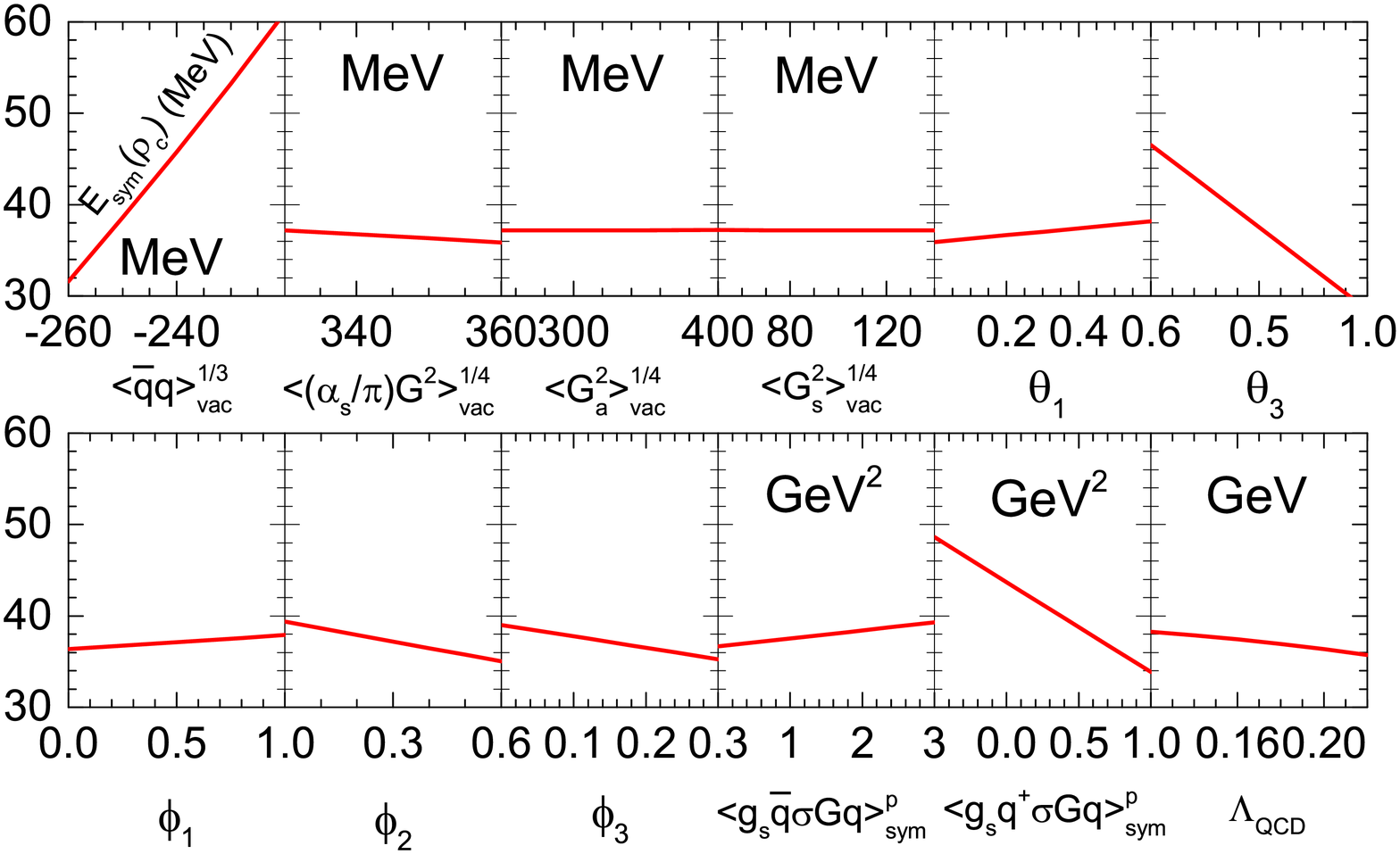}
  \caption{(Color Online) Correlations of the nucleon self-energies in SNM (upper) and the symmetry energy (lower) with the
eleven condensate parameters as well as $\Lambda_{\rm{QCD}}$,
  at $\rho_{\rm{c}}=0.11\,\rm{fm}^{-3}$.}
  \label{fig_CA_SV0}
\end{figure}

In general, one can in the framework of QCDSR study different correlations between the
condensates and other physical quantities, through the QCDSR equations.
For example, in Fig.\,\ref{fig_CA_SV0}, the correlation between the nucleon self-energies in SNM $\Sigma_{\rm{S/V}}^0$(upper panel)/nuclear symmetry energy (low panel)
at the cross density $\rho_{\rm{c}}$ and the condensate properties is shown.
Similarly, a very similar correlation pattern will be found if one studies the correlation at other densities, e.g., at $\rho_0$.
For the scalar self-energy $\Sigma_{\rm{S}}^0$, since, e.g.,
$\Sigma_{\rm{S}}^0\propto-\langle\overline{q}q\rangle_{\rm{vac}}$ in the msQCDSR (see Eq.\,(\ref{zz_ES0})),
a larger chiral condensate in vacuum naturally leads to a smaller $\Sigma_{\rm{S}}^0$.
Besides the strong dependence on $\langle\overline{q}q\rangle_{\rm{vac}}$, both $\Sigma_{\rm{S}}^0$
and $\Sigma_{\rm{V}}^0$ are independent of the other condensate properties except that $\Sigma_{\rm{S}}^0$
weakly depends on the five-dimensional parameter $\langle g_{\rm{s}}\overline{q}\sigma\mathcal{G}q\rangle_{\rm{sym}}^{\rm{p}}$.

On the other hand, the dependence of the symmetry energy on the quark/gluon condensates
show very fruitful patterns (see the lower panel of Fig.\,\ref{fig_CA_SV0}).
For instance, the positive correlation between $E_{\rm{sym}}(\rho)$ and $\langle\overline{q}q\rangle_{\rm{vac}}$.
This could be understood as follows: as the $\langle\overline{q}q\rangle_{\rm{vac}}$ increases (e.g., moving from left to right on the horizon axes),
the nucleon Dirac effective mass $M_0^{\ast}=M+\Sigma_{\rm{S}}^0$ is reduced correspondingly (since $\Sigma_{\rm{S}}^0$ increases),
leading to an enhancement on the factor $e_{\rm{F}}^{\ast,-1}=[k_{\rm{F}}^2+(M+\Sigma_{\rm{S}}^0)^2]^{-1/2}$
at a fix density, thus consequently the kinetic symmetry energy $E_{\rm{sym}}^{\rm{kin}}(\rho)
=k_{\rm{F}}^2/6e_{\rm{F}}^{\ast}$ is enhanced. While on the other hand, other symmetry energy decomposition terms (defined in Eq.\,(\ref{sec1_Esym})) depend weakly on the chiral
condensate in vacuum. More specifically, the symmetry energy at $\rho_{\rm{c}}$ is changed from 35.1\,MeV at $\langle\overline{q}q\rangle_{\rm{vac}}=-(255\,\rm{MeV})^3$
to 61.0\,MeV at $\langle\overline{q}q\rangle_{\rm{vac}}=-(220\,\rm{MeV})^3$,
generating an enhancement about 25.9\,MeV.
More interestingly, the symmetry energy is found to depend on several other quantities characterizing
the high mass dimensional condensates. And among which the symmetry energy displays the strongest correlation with the $\vartheta_3$
and the $\langle g_{\rm{s}}q^{\dag}\sigma \mathcal{G}q\rangle_{\rm{sym}}^{\rm{p}}$ parameters.
It is not intuitive to understand this strong correlation, however,
these two parameters together determine the density dependence of the mixing condensates of quarks and gluons, i.e., $\langle
g_{\rm{s}}q^{\dag}\sigma \mathcal{G}q\rangle_{\rho,\delta}\approx(1\mp\vartheta_3\delta)\langle g_{\rm{s}}q^{\dag}\sigma \mathcal{G}q\rangle_{\rm{sym}}^{\rm{p}}$,
see Eq.\,(\ref{yy1}), and to our best knowledge, it is the first time to relate the nuclear symmetry energy
to the quark and gluon mixing condensates, with the latter the very fundamental quantities in hadronic physics.
Thus the strong connection between $E_{\rm{sym}}(\rho)$ and $\langle
g_{\rm{s}}q^{\dag}\sigma \mathcal{G}q\rangle_{\rho,\delta}$ provides a useful bridge to explore
the properties of EOS of ANM using the knowledge from other physics branches (e.g., hadronic physics here).

\subsection{QCDSR-1 (naive QCDSR)}\label{ss_4}

Based on the qualitative analyses given in the above sections and the full QCDSR calculations
given in this section, we now give the first set of the QCDSR parameter to study the EOS of ANM.
In this subsection and the following two sections, the main attentions will be given to the density dependence of the
EOS of SNM, the symmetry energy and the EOS of PNM, while the other physical quantities such as dependence of the self-energies
on the Borel mass squared, etc., will not be given (which actually could be obtained very similarly as the analyses in the above subsections).
According to the fitting scheme given in Subsection~\ref{sb_FS}, the four-quark condensates parameter $f$
is found to be about $f\approx0.50$. Interestingly, with only the $f$ parameter, the PNM EOS $E_{\rm{n}}(\rho)$
at the density $\rho_{\rm{vl}}\approx0.02\,\rm{fm}^{-3}$ and the symmetry energy $E_{\rm{sym}}(\rho)$
at $\rho_{\rm{c}}\approx0.11\,\rm{fm}^{-3}$ could be fitted reasonably within their empirical ranges (Subsection~\ref{sb_FS}),
i.e., $E_{\rm{n}}(\rho_{\rm{vl}})\approx 4.2\,\rm{MeV}$\,\cite{Tew13,Kru13} and $E_{\rm{sym}}(\rho_{\rm{c}})\approx26.65\pm0.2\,\rm{MeV}$\,\cite{Zha13}.
In the following, we call this QCDSR parameter set the QCDSR-1, or the naive QCDSR, in the sense that only the linear approximation
of the chiral condensates is used (i.e., without the $\Phi$-term in Eq.\,(\ref{chiral_cond})). Moreover, the twist-four four-quark condensates
will not be included in this section and in the next section, and we
explore this type of condensates on the EOS of ANM in some detail in Section~\ref{SEC_TWIST4}.

\begin{figure}[h!]
\centering
  \includegraphics[width=8.5cm]{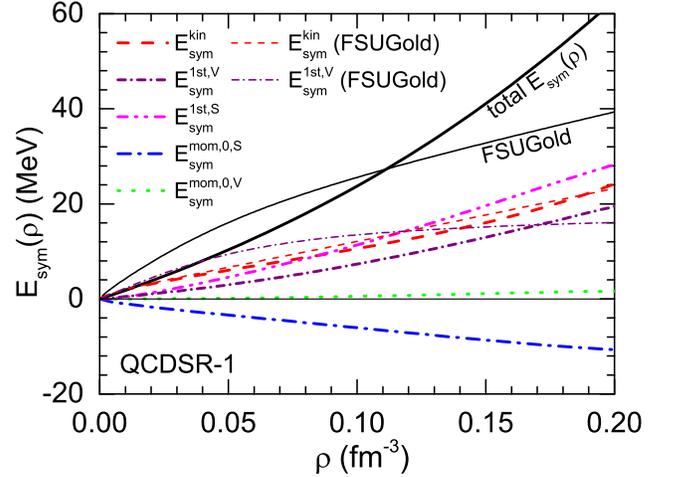}
  \caption{(Color Online) Density dependence of the symmetry energy and its self-energy decomposition in QCDSR-1. The corresponding results (shown by thin lines) from the RMF model calculations with the FUSGold interaction are also included for comparison (Note: The $E_{\rm{sym}}^{\rm{1st,S}}$, $E_{\rm{sym}}^{\rm{mom,0,S}}$ and $E_{\rm{sym}}^{\rm{mom,0,V}}$ are identically zero for FSUGold).}
  \label{fig_R1EsymDecom}
\end{figure}

In Fig.\,\ref{fig_R1EsymDecom}, the symmetry energy as well
as its self-energy decomposition (see Eq.\,(\ref{sec1_Esym})) are shown. Specifically, the symmetry energy
at cross density $\rho_{\rm{c}}=0.11\,\rm{fm}^{-3}$ is found to be about
$E_{\rm{sym}}(\rho_{\rm{c}})\approx26.9\,\rm{MeV}$ (slightly larger than the one adopted in the fitting scheme), while the
corresponding decomposition terms are found to be about $
E_{\rm{sym}}^{\rm{kin}}(\rho_{\rm{c}})\approx11.6\,\rm{MeV}$, $
E_{\rm{sym}}^{\rm{mom,0,S}}(\rho_{\rm{c}})\approx-6.6\,\rm{MeV}$, $
E_{\rm{sym}}^{\rm{mom,0,V}}(\rho_{\rm{c}})\approx0.6\,\rm{MeV}$, $
E_{\rm{sym}}^{\rm{1st,S}}(\rho_{\rm{c}})\approx12.9\,\rm{MeV}$, and
$ E_{\rm{sym}}^{\rm{1st,V}}(\rho_{\rm{c}})\approx8.4\,\rm{MeV}$,
respectively. As a reference, the symmetry energy as well as the corresponding decomposition terms at $\rho_0$ are found to be about 45.1\,MeV and 17.4\,MeV, $-$9.1\,MeV, 1.2\,MeV, 21.4\,MeV, and 14.2\,MeV, respectively.
It is obvious that the $E_{\rm{sym}}(\rho_0)$ from the naive QCDSR is still some larger than its empirical constraints (e.g., about $32\pm3\,\rm{MeV}$).
As is shown in Fig.\,\ref{fig_R1EsymDecom}, the momentum dependence of the nucleon scalar (vector) self-energy
is negative (positive) within the density region, which is consistent with the results
obtained in the last section without adopting the fitting scheme. However, as mentioned in the above sections, the magnitude of the momentum dependence
of the nucleon self-energies is essentially weak (both the blue and green lines), leading to a relative smaller contribution to the symmetry
energy.
For comparison, we also include in Fig.\,\ref{fig_R1EsymDecom} the corresponding predictions from the phenomenological nonlinear RMF model with the celebrated parameter set FSUGold \,\cite{Tod05}. Since there is no scalar and isovector channel (characterized by the so-called $\delta$ meson) in the FSUGold parameter set, the $E_{\rm{sym}}^{\rm{1st,S}}$ is identically zero, and since the conventional RMF approaches lack momentum-dependent interactions, the corresponding momentum-dependent terms $E_{\rm{sym}}^{\rm{mom,0,S/V}}$ are zero too. More specifically, the total symmetry energy in the FSUGold parameter set at the density $\rho_{\rm{c}}=0.11\,\rm{fm}^{-3}$/the density $\rho_0=0.16\,\rm{fm}^{-3}$ is about 27.1\,MeV /34.2\,MeV. In addition, it is seen from Fig.\,\ref{fig_R1EsymDecom} that the $E_{\rm{sym}}^{\rm{1st,V}}$ from FSUGold is significantly different from the prediction of QCDSR-1.
Furthermore, the symmetry energy (\ref{sec1_Esym}) could also be rewritten as\,\cite{Cai12,LCCX18}
\begin{equation}
E_{\rm{sym}}(\rho)=\frac{k_{\rm{F}}^2}{6M_{\rm{L}}^{\ast}}+\frac{1}{2}\left[\frac{M_0^{\ast}}{e_{\rm{F}}^{\ast}}\Sigma^{\rm{S}}_{\rm{sym}}+
\Sigma^{\rm{V}}_{\rm{sym}}\right],\label{sec1_Esymxx}
\end{equation}
where $M_{\rm{L}}^{\ast}$ is the nucleon Landau effective mass in SNM, defined as\,\cite{LCCX18}
\begin{equation}M_{\rm{L}}^{\ast}(\rho)
\equiv k_{\rm{F}}[\d e_{\rm{F}}/\d|\v{k}|]_{|\v{k}|=k_{\rm{F}}}
\end{equation} with $e_{\rm{F}}(\rho,|\v{k}|)$
the total single nucleon energy in SNM.
Consequently, from the sum of the kinetic symmetry energy and the two terms related to the momentum dependence
of the self-energies, one can easily obtain the Landau mass. In QCDSR-1, the nucleon Landau mass in SNM
is given by $M_{\rm{L}}^{\ast}/M\approx1.29$ at the saturation density (the Dirac mass will be given shortly).
It is also interesting to notice
that the overall density dependence of the symmetry energy is very different for the QCDSR-1 and FSUGold, reflecting that the slope parameter $L$ are different in these predictions.
Specifically, the slope parameter of the symmetry
energy could be obtained directly through its definition, and at $\rho_{\rm{c}}$ the $L$ parameter $L(\rho_{\rm{c}})$ in QCDSR-1 is found to be $105.9$\,MeV, which is much larger than
the empirical constraints, see, e.g., refs.\,\cite{Zha13,Dan14}.
On the other hand, the $L(\rho_{\rm{c}})$ in FSUGold is about $50.0$\,MeV.
While at $\rho_0=0.16\,\rm{fm}^{-3}$, the $L$ parameter in QCDSR-1 (FSUGold) is found to be about $196.8$\,MeV/ ($63.9$\,MeV), once again showing the one obtained from QCDSR-1 is much larger than the phenomenological prediction.
The large symmetry energy
at the saturation density as well as the $L$
parameter both at the cross density and the saturation density indicate that the linear
approximation of the condensates, especially the chiral condensates (\ref{chiral_cond}),
already breaks down at densities significantly less than the saturation density $\rho_0$.
It is necessary to consider the effective higher order terms in density in the chiral condensate, and
this is the main task of the next section.

\begin{figure}[h!]
\centering
  \includegraphics[width=8.5cm]{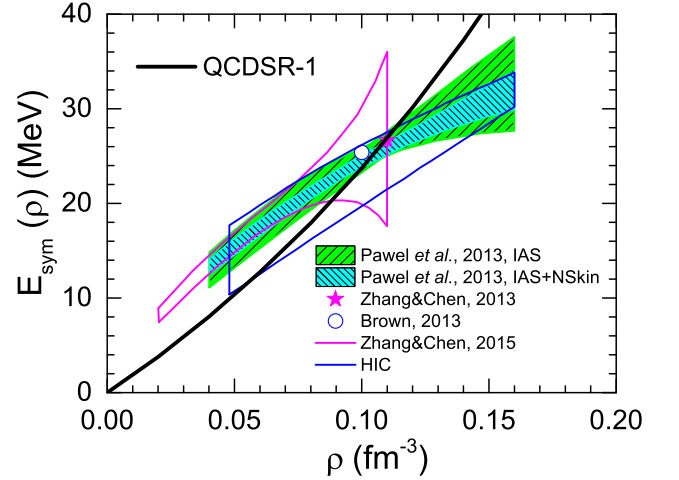}
  \caption{(Color Online) Symmetry energy obtained by QCDSR-1.
   Also shown include the results from the analysis of
isobaric analog states (IAS)\,\cite{Dan14} while the shaded area
enclosed by the solid cyan line and labeled by ``IAS+NSkin'' results
when the IAS analysis is supplemented with additional constraints
from neutron skin data, the constraint on the symmetry energy from
heavy ion collisions (HIC)\,\cite{Tsa12} (blue band). A recent study
on the symmetry energy at around $\rho_0/3$ using the electric
dipole polarizability in $^{208}$Pb is also shown for
comparison\,\cite{Zha15} (magenta band). Two studies on the symmetry
energy at the cross density\,\cite{Zha13} are shown, they are the
constrain on the symmetry energy at
$\rho_{\rm{c}}\approx0.11\,\rm{fm}^{-3}$ using
isotope binding energy difference\,\cite{Zha13} which is labeled by
a magenta star and that from fit of ground state properties of
double magic nucleus using Skyrme CSkp functionals\,\cite{Bro14} (blue
circle).}
  \label{fig_R1Esym}
\end{figure}

In Fig.\,\ref{fig_R1Esym}, the total symmetry energy obtained in QCDSR-1 as
a function of density is shown. Although the symmetry energy roughly passes through the constraint at
$\rho_{\rm{c}}\approx0.11\,\rm{fm}^{-3}$\,\cite{Zha13,Dan14,Zha15,Bro14}, its density behavior both at low densities $\lesssim0.05\,\rm{fm}^{-3}$ and at densities $\gtrsim0.12\,\rm{fm}^{-3}$ is shown to be  inconsistent with the
predictions from other approaches in the sense that the symmetry energy obtained in QCDSR-1 is too stiff.
Several other typical constraints on the symmetry energy are also shown in Fig.\,\ref{fig_R1Esym} include the results from the analysis of
isobaric analog states (IAS)\,\cite{Dan14} while the shaded area
enclosed by the solid cyan line and labeled by ``IAS+NSkin'' results
when the IAS analysis is supplemented with additional constraints
from neutron skin data, the constraint on the symmetry energy from
heavy ion collisions (HIC) which is labeled by the blue
band\,\cite{Tsa12}. A recent study on the symmetry energy at low
densities (around $\rho_0/3$) using the electric dipole
polarizability in $^{208}$Pb is also shown for
comparison\,\cite{Zha15} by the magenta band. Two studies on the
symmetry energy at the cross density\,\cite{Zha13} are shown, they
are the constrain on the symmetry energy at
$\rho_{\rm{c}}\approx0.11\,\rm{fm}^{-3}$ to be
$E_{\rm{sym}}(\rho_{\rm{c}})\approx26.65\pm0.20\,\rm{MeV}$ using
isotope binding energy difference\,\cite{Zha13} which is labeled by
a magenta star and that from fit of ground state properties of
double magic nucleus using Skyrme CSkp functionals which found a
value of
$E_{\text{sym}}(\rho_{\text{c}})\approx25.4\pm0.8\,\text{MeV}$ at
$\rho_{\text{c}}\approx0.10\,\text{fm}^{-3}$\,\cite{Bro14} (blue
circle).
Moreover, since we are mainly interested in the symmetry energy from QCDSR method at densities $\lesssim\rho_0$,
some other constraints on $E_{\rm{sym}}(\rho)$ at densities $\gtrsim\rho_0$ will not be compared here,
e.g., ref.\,\cite{Rus16} constraints the symmetry energy between $\rho_0$ to about $2\rho_0$.

\begin{figure}[h!]
\centering
  \includegraphics[width=8.5cm]{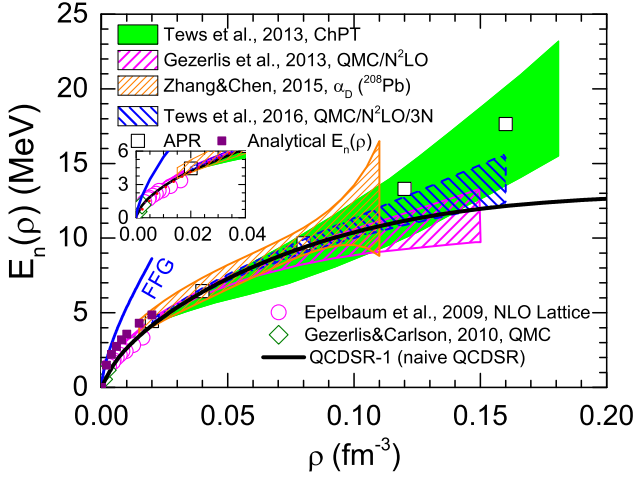}
  \caption{(Color Online) EOS of PNM obtained by QCDSR-1.
Also shown are the results from ChPT\,\cite{Tew13,Kru13} (green band), QMC
simulations combing with chiral force to next-to-next-to-leading
order (N$^2$LO) with\,\cite{Tew16} (blue band) and
without\,\cite{Gez13} (magenta band) leading-order chiral
three-nucleon interactions forces, next-to-leading order (NLO)
lattice calculation\,\cite{Epe09} (magenta circle), QMC simulations for PNM at very low densities\,\cite{Gez10} (green
diamond), the APR EOS\,\cite{APR} (black open squares), the free Fermi gas (FFG) prediction (blue line), and the analytical approximation of $E_n(\rho)$ (i.e., Eq.~(\ref{Ens-1})) (purple solid squares). The results from analyzing the electric dipole
polarizability in $^{208}$Pb \,\cite{Zha15} is
also shown for comparison.}
  \label{fig_R1PNM}
\end{figure}

In Fig.\,\ref{fig_R1PNM}, we show the EOS of PNM as a function of density.
Also included in Fig.\,\ref{fig_R1PNM} are the results from
ChPT\,\cite{Tew13,Kru13} (green band), QMC
simulations {combined} with chiral force to next-to-next-to-leading
order (N$^2$LO) with\,\cite{Tew16} (blue band) and
without\,\cite{Gez13} (magenta band) leading-order chiral
three-nucleon interactions forces, next-to-leading order (NLO)
lattice calculation\,\cite{Epe09a} (magenta circle), {and} QMC simulations
for PNM at very low densities\,\cite{Gez10} (green diamond). The
result from analyzing experimental data on the electric dipole polarizability $\alpha_{\text D}$ in
$^{208}$Pb\,\cite{Zha15} is also shown for comparison.
The inset in Fig.\,\ref{fig_R1PNM} shows the EOS of PNM at very low
densities.
Interestingly, by artificially neglecting the contributions from dimension-four and higher order terms (and only keeping
the four-quark condensates of the type (\ref{f4})), one obtains an effective approximation for EOS of PNM, see Eq.\,(\ref{Ens})
and Eq.\,(\ref{Ens-1}),
and the latter (Eq.\,(\ref{Ens-1})) clearly
demonstrates how the chiral condensate goes into play in the EOS
of PNM, i.e., the second term characterized by several constants
($\xi,\sigma_{\rm{N}},m_{\rm{q}}$ and
$\langle\overline{q}q\rangle_{\rm{vac}}$) is negative, leading to a
reduction on the $E_{\rm{n}}(\rho)$ compared to the FFG prediction.
In Fig.\,\ref{fig_R1PNM}, we also plot the results from Eq.\,(\ref{Ens-1}) at densities $\lesssim0.02\,\rm{fm}^{-3}$ (violet solid square).
One can see that the approximation Eq.\,(\ref{Ens-1}) can already produce reasonably the $E_{\rm{n}}(\rho)$ at low densities.
The FFG prediction on the $E_{\rm{n}}^{\rm{FFG}}(\rho)$, however, already becomes larger at, e.g., $\rho_{\rm{vl}}=0.02\,\rm{fm}^{-3}$,
and even a relativistic correction $-k_{\rm{F,n}}^4/56M^3$ to the $E_{\rm{n}}^{\rm{FFG}}(\rho)$
will not make the situation much better, strongly indicating that the non-interacting Fermi model lacks the fundamental information
between nucleons and nucleons to produce the correct EOS.
Furthermore, it is seen from Fig.\,\ref{fig_R1PNM} that the prediction on the
$E_{\rm{n}}(\rho)$ from QCDSR-1 is consistent with several QMC
simulations and lattice computation at densities $\lesssim0.02\,\rm{fm}^{-3}$,
showing that QCDSR is a reliable approach in the study of EOS of PNM, especially at
lower densities, where the naive QCDSR is good enough.

Although the only adjustable parameter $f$ in the naive QCDSR
is fixed by the $E_{\rm{n}}(\rho)$ at the very low density of $0.02\,\rm{fm}^{-3}$
(thus the $E_{\rm{n}}(\rho)$ at densities greater than $0.02\,\rm{fm}^{-3}$ has no fitting requirements),
the prediction on the EOS of PNM at densities $\lesssim0.1\,\rm{fm}^{-3}$ in QCDSR-1
is found to be well-behaved compared with the APR EOS, demonstrating that the QCDSR with the linear density
approximation for the
chiral condensates can be quantitatively applied to study the EOS of PNM within these density region.
However, as density even increases, the systematic deviation between the $E_{\rm{n}}(\rho)$
obtained by QCDSR-1 and that predicted by APR EOS becomes large and this
can not be improved by simply adjusting the parameter $f$, indicating on the other hand that
the leading-order linear density approximation for the chiral condensates
does not work well enough and the higher order density terms in the chiral
condensates are needed for the PNM calculations at these densities.
For example, at $\rho_0\approx0.16\,\rm{fm}^{-3}$, the difference between the ARP EOS
and the $E_{\rm{n}}(\rho)$ from QCDSR-1 is found to be about 5.4\,MeV.
Once one considers the term $\Phi
g\rho^2$ in Eq.\,(\ref{chiral_cond}) for PNM, and recalculate the $E_{\rm{n}}(\rho)$ under the fitting scheme,
we find that compared with the case of the naive QCDSR, the obtained prediction can be largely improved
to fit the APR EOS. This feature suggests that the QCDSR with
effective higher order density terms in quark condensates can be
used to study the EOS of dense nucleonic matter at higher densities.
The relevant investigations will be given in the next section.

\begin{figure}[h!]
\centering
  \includegraphics[width=8.5cm]{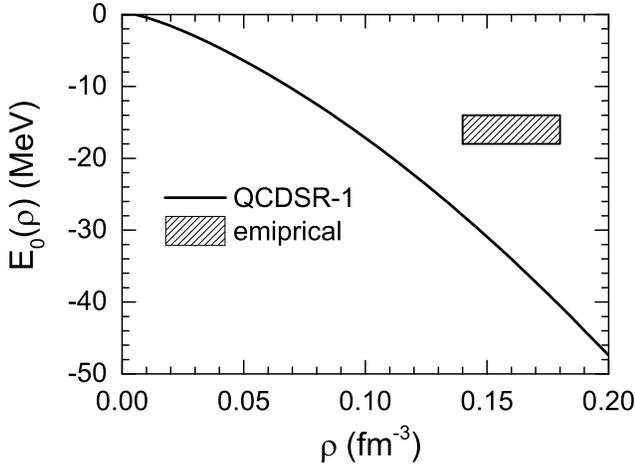}
  \caption{EOS of SNM obtained by QCDSR-1. Empirical constraints on the saturation density
  and the binding energy, i.e., $(\rho_0,E_0(\rho_0))=(0.16\pm0.02\,\rm{fm}^{-3},-16\pm2\,\rm{MeV})$, is also shown.}
  \label{fig_R1SNM}
\end{figure}

Since in our fitting scheme (see Subsection~\ref{sb_FS}), the constraints
on the $E_{\rm{n}}(\rho)$ at a very low density $\rho_{\rm{vl}}$
and on the symmetry energy at $\rho_{\rm{c}}$
are used without using any empirical constraints on the EOS of SNM (even the symmetry energy constraint is not used in QCDSR-1), it is interesting to explore the consequent prediction
on the $E_0(\rho)$ from the QCDSR-1.
In Fig.\,\ref{fig_R1SNM}, we show the density dependence of the EOS of SNM obtained in QCDSR-1.
The empirical constraints on the saturation density about
$\rho_0\approx0.16\pm0.02\,\rm{fm}^{-3}$ and the corresponding
binding energy $E_0(\rho_0)\approx-16\pm2\,\rm{MeV}$ are also shown in Fig.\,\ref{fig_R1SNM}.
For instance, the $E_0(\rho)$ at $0.16\,\rm{fm}^{-3}$ ($0.11\,\rm{fm}^{-3}$) in QCDSR-1 is found to be about $-$34.0\,MeV ($-$19.7\,MeV).
And in the meanwhile, the saturation
density of QCDSR-1 itself is found to be about
$\rho_0^{\rm{QCDSR-1}}\approx0.6\,\rm{fm}^{-3}$, with the corresponding binding energy about
$-$100.0\,MeV, showing that the symmetric matter in QCDSR-1 is very deep bounded.
It actually is another cue that the effective $\Phi$-term in Eq.\,(\ref{chiral_cond}) is important, indicating the breakdown of the chiral condensates at linear order at densities even smaller than the saturation density.
Moreover, it is really a very difficult problem on how to obtain the correct (even reasonable) saturation properties of the SNM
in the microscopic theories (see, e.g., ref.\,\cite{Dri19} in the framework of ChPT).
Based on the symmetry energy and the EOS of SNM and the EOS PNM,
one can estimate the high order effects in the EOS $E_{\rm{HO}}(\rho)$, and the detailed results on $E_{\rm{HO}}(\rho)$ will
not be given here (see Fig.\,\ref{fig_R2EHO}).
However as a reference, for example, the $E_{\rm{HO}}(\rho)$ is found to be about 3.6\,MeV (1.1\,MeV) at $\rho_{\rm{c}}$ ($\rho_0$),
demonstrating again that the $E_{\rm{HO}}(\rho)$ is generally non-negligible in QCDSR.

\begin{figure}[h!]
\centering
  \includegraphics[width=8.5cm]{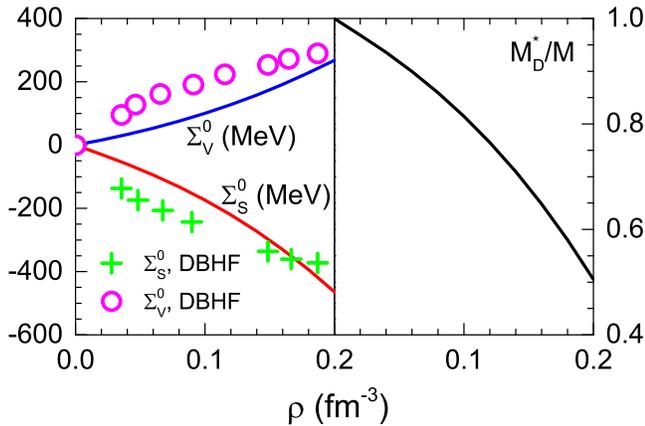}
  \caption{(Color Online) Density dependence of the nucleon self-energies
 in SNM and the nucleon Dirac effective mass obtained in QCDSR-1.
  The predictions on the self-energies from the Dirac--Brueckner--Hartree--Fock (DBHF) approach\,\cite{Bro90} is also shown for comparison.}
  \label{fig_R1SEDirac}
\end{figure}

In Fig.\,\ref{fig_R1SEDirac}, the nucleon self-energies
in SNM and the nucleon Dirac effective mass as functions
of density are shown. The predictions on the self-energies by the
Dirac--Brueckner--Hartree--Fock (DBHF) calculations\,\cite{Bro90} are also shown for comparison.
It is obvious from the figure that although the overall tendency of the density dependence of the self-energies
is consistent compared to the DBHF predictions, the density behavior roughly
characterized, e.g., by the index $\sigma$ in $\rho^{\sigma}$, is very different.
It demonstrates onces again that the higher density terms in the chiral condensates
are important, since $\Sigma_{\rm{S}}^0(\rho)$ is directly related to the chiral condensates $\langle\overline{q}q\rangle_{\rho}$.
Moreover, the nucleon Dirac effective mass at $\rho_0=0.16\,\rm{fm}^{-3}$ is found to
be $M_{\rm{D}}^{\ast}/M\approx0.65$, which is consistent with the
empirical constraints\,\cite{LCK08} (since from Fig.\,\ref{fig_R1SEDirac} it is clearly shown that around $0.16\,\rm{fm}^{-3}
$ the prediction on $\Sigma_{\rm{S}}^0$ is consistent with the DBHF prediction).

As a short summary of this section, we determine the effective parameter $f$ in the QCDSR to
fix the $E_{\rm{n}}(\rho)$ at $\rho_{\rm{vl}}\approx0.02\,\rm{fm}^{-3}$
with the prediction from ChPT, and the symmetry energy at the cross density $\rho_{\rm{c}}$
is found to be consistent with the empirical constraints. However, the symmetry energy
at the saturation density and the slope parameter $L$ at both the saturation and the cross
density are found to be too large, clearly indicating the breakdown of the linear
approximation for the chiral condensates. Besides, the EOS of PNM at densities larger around $0.1\,\rm{fm}^{-3}$
also shows systematic deviation from the APR EOS. Improving the EOS of PNM and the density
behavior of the symmetry energy is one of the main motivations to include higher order terms in density in the chiral
condensates, which is the main issue in the next section.

\setcounter{equation}{0}
\section{Higher Order Density Terms in Chiral Condensates, Symmetry Energy and QCDSR-2}\label{SEC_HighCond}

As studied in the last section,
the symmetry energy and its slope parameter in QCDSR-1 are
found to be too large at the saturation density, indicating the linear approximation
for the chiral condensates (i.e., Eq.~(\ref{chiral_cond-1})) breaks down already at densities less than the
saturation point since the density dependence of the chiral condensates is directly
related to the symmetry energy, e.g., see the expression for the symmetry energy obtained in the msQCDSR, i.e., Eq.\,(\ref{Esym-msQCDSR})
and Eq.\,(\ref{ddd2}). In this section, we study the possible higher order terms
in density in the chiral condensates introduced by the $\Phi$-term in Eq.\,(\ref{chiral_cond}).
Once the $\Phi$-term is included in the chiral condensates, two other effective parameters, i.e.,
$\Phi$ and $g$ are introduced, and $\Phi$, $g$ and $f$ will be determined by
the $E_{\rm{n}}(\rho)$ at a very low density $\rho_{\rm{vl}}=0.02\,\rm{fm}^{-3}$,
the symmetry energy at $\rho_{\rm{c}}=0.11\,\rm{fm}^{-3}$, and adjusting the PNM EOS to fit
the APR EOS as much as possible, see the fitting scheme given in Subsection~\ref{sb_FS}. Consequently, these parameters will be re-adjusted.
Besides, other parameters in the QCDSR are still taken as $\mathscr{M}^2=1.05\,\rm{GeV}^2,\omega_0=1.5\,\rm{GeV},\sigma_{\rm{N}}=45\,\rm{MeV},m_{\rm{q}}=3.5\,\rm{MeV}
$ and $m_{\rm{s}}=95\,\rm{MeV}$. One then obtains $f\approx0.43$, $\Phi'\equiv\Phi\times\langle\overline{q}q\rangle_{\rm{vac}}\approx3.45$\,\cite{Cai17QCDSR},
and $g=-0.64$. Moreover, the Ioffe parameter $t\approx-1.22$ is independent of the $\Phi$-term introduced,
since when using the nucleon mass in vacuum to fix the parameter $t$ it only depends on the vacuum properties of the condensates.

\begin{figure}[h!]
\centering
  \includegraphics[width=8.5cm]{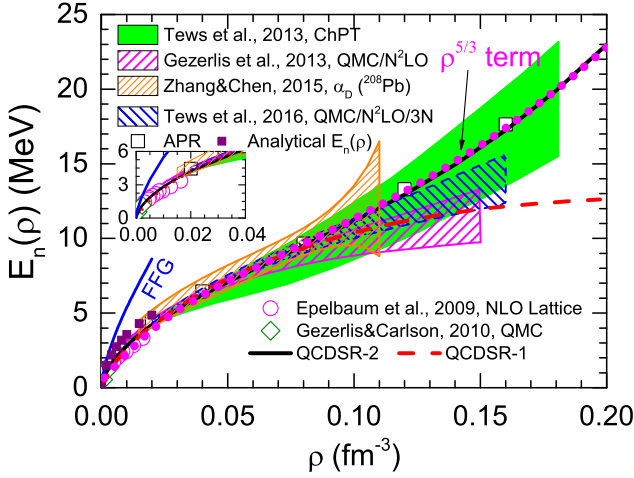}
  \caption{(Color Online) Same as Fig.\,\ref{fig_R1PNM} but for QCDSR-2\,\cite{Cai17QCDSR}. The $E_{\rm{n}}(\rho)$ from QCDSR-1 is also shown for comparison. See the text for details.}
  \label{fig_R2PNM}
\end{figure}

We abbreviate the corresponding QCDSR parameter set the QCDSR-2.
Based on the obtained $\Phi$ and $g$,
we can estimate the density below which the $\Phi$-term
has minor contribution to the quark condensates.
This density can be estimated as $|\Phi(1-g)\rho^2|\ll|(\sigma_{\rm{N}}/2m_{\rm{q}})(1-\xi)\rho|$,
i.e., the last term in Eq.\,(\ref{chiral_cond}) is significantly less than the second term in Eq.\,(\ref{chiral_cond}), and we thus obtain
$\rho\ll\rho_{\rm{es}}\approx2.13\,\rm{fm}^{-3}$.
Therefore, the effects of $\Phi$ and $g$ on the $E_{\rm{n}}(\rho)$ are trivial at substantial densities,
e.g., when one artificially takes $\Phi'=0$ and keeping $f$ fixed, the $E_{\rm{n}}(\rho_{\rm{vl}})$ $(E_{\rm{n}}(0.1\,\rm{fm}^{-3})$ changes
from 4.20\,MeV to 4.22\,MeV (from 11.15\,MeV to 10.01\,MeV).
Thus it is reasonable to expect that effects of $\Phi$ and
$g$ on the $E_{\rm{n}}(\rho)$ at low densities $\lesssim0.1\,\rm{fm}^{-3}$ are small.
However, as the densities even increase, there is no guarantee that the $\Phi$-term
still has small effects on the EOS of PNM since the $E_{\rm{n}}(\rho)$ is obtained by integrating over the density, see Eq.\,(\ref{sec1_PNM})\,\cite{Cai17QCDSR}.
In Fig.\,\ref{fig_R2PNM}, the
$E_{\rm{n}}(\rho)$ obtained in QCDSR-2 is shown, and the prediction
on the EOS of PNM from QCDSR-1 is also shown for comparison. Other constraints on $E_{\rm{n}}(\rho)$ shown in the figure are the
same as those in Fig.\,\ref{fig_R1PNM}.
Compared with the prediction on the $E_{\rm{n}}(\rho)$ by the naive QCDSR, once one considers the term $\Phi
g\rho^2$ in Eq.\,(\ref{chiral_cond}) for PNM, and recalculates the corresponding EOS, one finds that the obtained prediction is
largely improved to fit the APR EOS. For instance, the EOS of
PNM at $0.12\,\rm{fm}^{-3}$ is now found to be 12.9\,MeV, which is very close to
the APR prediction 13.3\,MeV. Moreover, the overall fitting
between the EOS of PNM from QCDSR-2 and the APR EOS is much better compared with the prediction
by QCDSR-1, e.g.,
at densities $\lesssim0.16\,\rm{fm}^{-3}$.
These features suggest that the QCDSR with
effective higher order density terms in quark condensates can be
used to study the EOS of dense nucleonic matter at higher densities.
It is also necessary to point out that using a different higher order density term in Eq.\,(\ref{chiral_cond})
and re-fix the parameters $f$, $\Phi$ and $g$ by the same fitting scheme, the density behavior of the $E_{\rm{n}}(\rho)$
is almost unchanged. For example, when adopting a $\rho^{5/3}$ term, i.e., $\Phi(1\mp g\delta)\rho^{5/3}$, then $f\approx0.46$,
$\Phi'\equiv \Phi\times\langle\overline{q}q\rangle_{\rm{vac}}^{2/3}\approx1.61$ and $g\approx-0.34$ could be obtained,
and the corresponding $E_{\rm{n}}(\rho)$ is shown in Fig.\,\ref{fig_R2PNM} by the magenta dot line.
It is obvious seen that using a different higher order density term in the chiral condensates
will not change our conclusions on the EOS of PNM\,\cite{Cai17QCDSR}.
In the following, we will not consider other higher order density terms in Eq.~(\ref{chiral_cond}) except the one has the form $\Phi (1\mp g\delta)\rho^2$.

\begin{figure}[h!]
\centering
  \includegraphics[width=8.5cm]{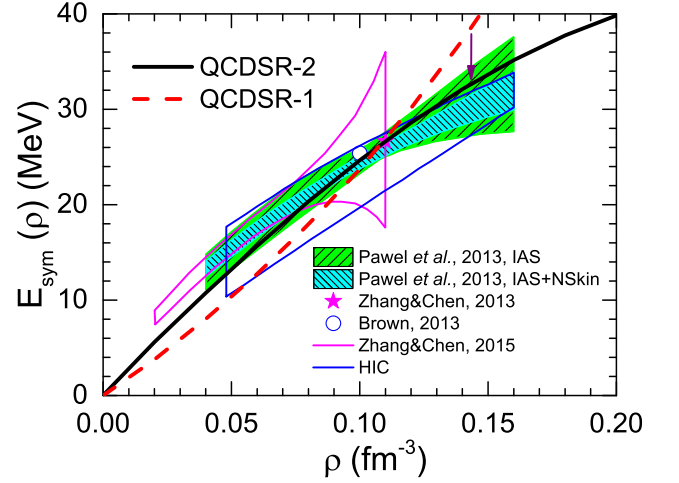}
  \caption{(Color Online) Symmetry energy obtained in QCDSR-2. The $E_{\rm{sym}}(\rho)$ from QCDSR-1 is also shown for comparison. See the text for details.}
  \label{fig_R2Esym}
\end{figure}

In Fig.\,\ref{fig_R2Esym}, the density dependence of the
symmetry energy obtained in QCDSR-2 is shown, and the $E_{\rm{sym}}(\rho)$ from QCDSR-1 is also shown here for comparison.
Compared to the prediction on the symmetry energy from QCDSR-1, the $E_{\rm{sym}}(\rho)$
from QCDSR-2 is now largely improved, e.g., from about $0.04\,\rm{fm}^{-3}$ to about the saturation density $\rho_0$.
For example, the $E_{\rm{sym}}(\rho)$ now can roughly
pass through the constraints obtained from the IAS studies (green band).
More specifically, when evaluating at $\rho\approx0.04\,\rm{fm}^{-3}$, the symmetry energy
changes from 8.0\,MeV in the naive QCDSR to about 10.8\,MeV in QCDSR-2, introducing a relative 35\% change,
and more interestingly now $E_{\rm{sym}}(0.04\,\rm{fm}^{-3})$ is very close to the lower limit
predicted by the IAS studies about 11.1\,MeV.
Similarly, the symmetry energy at $\rho_0=0.16\,\rm{fm}^{-3}$ changes from 45.1\,MeV in QCDSR-1 to
about 35.3\,\rm{MeV} in QCDSR-2, and the relative change is about $-$22\%.
All these features demonstrate that the higher order density terms in the chiral condensates~(\ref{chiral_cond})
are essentially needed to describe a reasonable density behavior of the symmetry energy.

\begin{figure}[h!]
\centering
  \includegraphics[width=8.5cm]{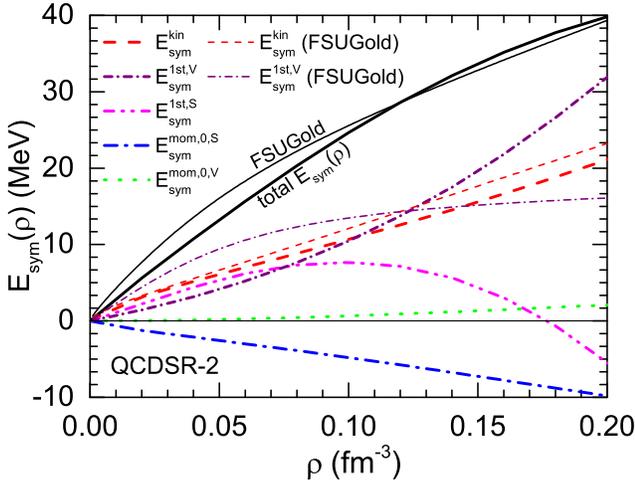}
  \caption{(Color Online) Same as Fig.\,\ref{fig_R1EsymDecom}
  but for QCDSR-2.}
  \label{fig_R2EsymDecom}
\end{figure}

Similarly, the nucleon self-energy decomposition (Eq.\,(\ref{sec1_Esym})) of the
symmetry energy obtained in QCDSR-2 is shown in Fig.\,\ref{fig_R2EsymDecom}.
The corresponding results from the RMF model with FSUGold \,\cite{Tod05} are also
included for comparison.
At the
cross density $\rho_{\rm{c}}=0.11\,\rm{fm}^{-3}$, for instance, one now obtains $
E_{\rm{sym}}^{\rm{kin}}(\rho_{\rm{c}})\approx11.7\,\rm{MeV}$,
$E_{\rm{sym}}^{\rm{mom,0,S}}(\rho_{\rm{c}})\approx-5.3\,\rm{MeV}$,
$E_{\rm{sym}}^{\rm{mom,0,V}}(\rho_{\rm{c}})\approx0.8\,\rm{MeV}$,
$E_{\rm{sym}}^{\rm{1st,S}}(\rho_{\rm{c}})\approx7.4\,\rm{MeV}$, and
$E_{\rm{sym}}^{\rm{1st,V}}(\rho_{\rm{c}})\approx12.1\,\rm{MeV}$, and
consequently $E_{\rm{sym}}(\rho_{\rm{c}})\approx26.7\,\rm{MeV}$ (the fitting scheme).
Moreover, the symmetry energy at the saturation density is found to be about 35.2\,MeV
with its self-energy decomposition terms about 16.6\,MeV, $-$7.8\,MeV,
1.5\,MeV, 3.0\,MeV, and 21.8\,MeV, respectively.
Besides the weak momentum dependence of the self-energies in SNM revealed several times in
the above sections, one can find interestingly that the $E_{\rm{sym}}^{\rm{1st,S}}(\rho)$ starts to decrease at
a critical density about $0.09\,\rm{fm}^{-3}$ and to even be negative around the saturation density.
This change could be traced back to the effective $\Phi$-term introduced in the chiral condensates (\ref{chiral_cond}).
These features demonstrate again the importance of the higher order terms in density in the
chiral condensates.
It is necessary to point out that using a different higher order terms in density (such as a term proportional to $\rho^{4/3}$
or $\rho^{5/3}$ as discussed in Subsection~\ref{sb_FS}) and
refits the parameters of the model, the changes in the density behavior of the symmetry energy
will be almost the same as the one shown in Fig.\,\ref{fig_R2EsymDecom}, i.e., the $E_{\rm{sym}}^{\rm{1st,S}}(\rho)$
starts to decreases at a critical density and even to become negative.
Moreover, the $E_{\rm{sym}}^{\rm{1st,V}}(\rho)$
becomes dominant at densities $\gtrsim0.1\,\rm{fm}^{-3}$ compared with the kinetic symmetry energy,
e.g., $E_{\rm{sym}}^{\rm{1st,V}}(\rho_0)\approx22.0\,\rm{MeV}$ and $E_{\rm{sym}}^{\rm{kin}}(\rho_0)\approx16.7\,\rm{MeV}$,
leading to $E_{\rm{sym}}^{\rm{1st,V}}(\rho_0)/E_{\rm{sym}}^{\rm{kin}}(\rho_0)\approx1.32$.
Furthermore, as the effective $\Phi$-term is included, the density dependence of the symmetry energy obtained from QCDSR-2 is found to be closer to the one from the phenomenological FSUGold parameter set. However, as in the case of QCDSR-1, the $E_{\rm{sym}}^{\rm{1st,V}}$ from FSUGold is again significantly different from the prediction of QCDSR-2. Compared with the QCDSR approach, therefore, the phenomenological nonlinear RMF model with FSUGold exhibits very different self-energy decomposition of the symmetry energy.

\begin{figure}[h!]
\centering
  \includegraphics[width=8.2cm]{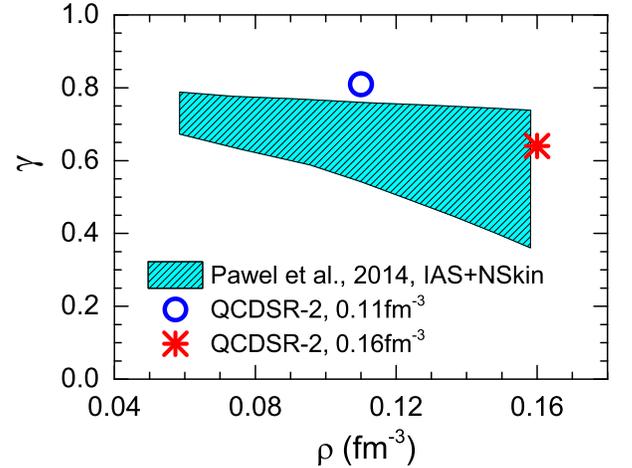}
  \caption{(Color Online) The parameter $\gamma(\rho)\equiv L(\rho)/3E_{\rm{sym}}(\rho)$ obtained in QCDSR-2
  at $\rho_{\rm{c}}$ and at $\rho_0$. See the text for details.}
  \label{fig_R2gamma}
\end{figure}

The $L$ parameter could be obtained through the density behavior of the symmetry energy.
Specifically, one finds that at $\rho_{\rm{c}}$ ($\rho_0$) the $L$ parameter is about 64.7\,MeV (67.5\,MeV).
Compared with the predictions on the $L$ parameter by the QCDSR-1, i.e., 105.9\,MeV (at $\rho_{\rm{c}}$)
and 196.8\,MeV (at $\rho_0$), the $\Phi$-term introduces a relative amount about $-$39\% ($-$66\%) on the $L$ at $\rho_{\rm{c}}$ ($\rho_0$).
Interestingly, although the $L(\rho_{\rm{c}})$ is slightly larger than nowadays best empirical
constraints (e.g., $L(\rho_{\rm{c}})$ about $46\pm4.5\,$MeV from ref.\,\cite{Zha13}),
the $L$ parameter at the saturation density is found to be consistent with its
empirical value about $60\pm30\,\rm{MeV}$ (e.g., see refs.\,\cite{Che12a,LiBA13,Oer17}).
Moreover, a relevant quantity for the discussion on the density behavior of the symmetry energy is given by ref.\,\cite{Dan14}
\begin{equation}\label{def_gamma}
\gamma(\rho)= \frac{\d\log
E_{\rm{sym}}(\rho)}{\d\log\rho}=\frac{L(\rho)}{3E_{\rm{sym}}(\rho)}.\end{equation}
The $\gamma$ parameter introduced in the parameterized form of the symmetry energy,
e.g., $E_{\rm{sym}}(\rho)\sim(\rho/\rho_0)^{\gamma}$, is often used in heavy-ion collisions simulations\,\cite{Tsa12}.
For instance, the $\gamma$ parameter of the potential part was constrained to be about $\gamma\approx0.72\pm0.19$ in ref.\,\cite{Rus16} from the comparison of the elliptic flow ratio of neutrons with respect to charged particles based on UrQMD predictions.
In our calculation in QCDSR-2, we find that the $\gamma$ parameter is about 0.81 (0.64)
at $\rho_{\rm{c}}$ ($\rho_0$). Ref.\,\cite{Dan14} gives the constraint
on $\gamma$ obtained by the IAS analyses, shown in Fig.\,\ref{fig_R2gamma}.
It is interesting to see that the $\gamma$ parameter obtained in QCDSR-2 at $\rho_{\rm{c}}$
is shown to be slightly larger than the blue band, however, the $\gamma$ parameter at $\rho_0$
is well consistent with the one given by ref.\,\cite{Dan14}.

\begin{figure}[h!]
\centering
  \includegraphics[width=8.5cm]{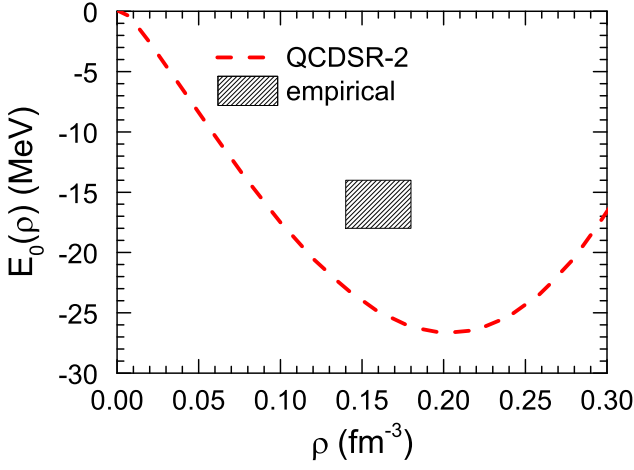}
  \caption{(Color Online) EOS of SNM obtained in QCDSR-2.}
  \label{fig_R2SNM}
\end{figure}

In Fig.\,\ref{fig_R2SNM}, we show the EOS of SNM obtained in QCDSR-2.
Due to the effective $\Phi$-term introduced in the chiral condensate, the
saturation density of QCDSR-2 is improved to be about $0.2\,\rm{fm}^{-3}$,
with the corresponding binding energy about $-$26.7\,MeV.
\begin{figure}[h!]
\centering
  \includegraphics[width=8.cm]{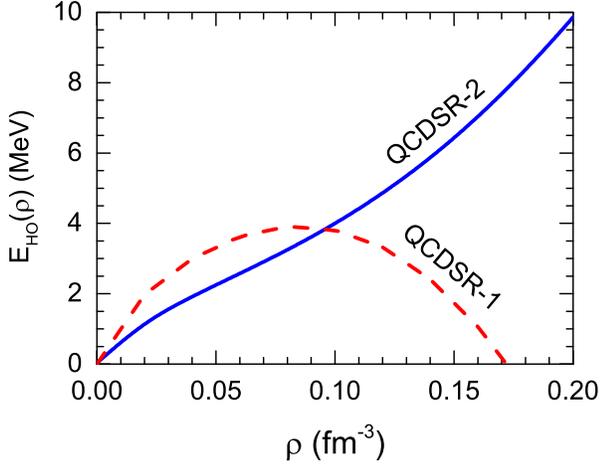}
  \caption{(Color Online) High order effects of the EOS of ANM obtained in QCDSR-1 and QCDSR-2.}
  \label{fig_R2EHO}
\end{figure}
The improvements on the EOS of PNM, the EOS of SNM
and the symmetry energy together verify again that the higher order terms in density in the
chiral condensates are important.
Similarly in Fig.\,\ref{fig_R2EHO}, the $E_{\rm{HO}}(\rho)\approx E_{\rm{sym,4}}(\rho)
+E_{\rm{sym,6}}(\rho)+\cdot\cdot\cdot$ as a function of density is shown
in both the QCDSR-1 and QCDSR-2. As discussed in Section~\ref{SEC_HOEsti},
the higher order terms in density in the chiral condensates may eventually break the parabolic approximation
of the EOS of ANM, see the discussion given around Eq.\,(\ref{EHOXX}), we find in QCDSR-2 the $E_{\rm{HO}}(\rho)$
is about 7.1\,MeV (4.4\,MeV) at $\rho_0$ ($\rho_{\rm{c}}$).
Although the high order term $E_{\rm{HO}}(\rho)$ is generally believed to have very little effects on, e.g., the nuclear structure
problems, it may induce sizable influence on the quantities in neutron stars\,\cite{Cai12xx,PuJ17}, such as the
core-crust transition density and the transition pressure.
Detailed investigation on the $E_{\rm{HO}}(\rho)$ in QCDSR will be extremely useful for further studies
on the relevant issues, which however is beyond the main scope of the present work.

\begin{figure}[h!]
\centering
  \includegraphics[width=8.2cm]{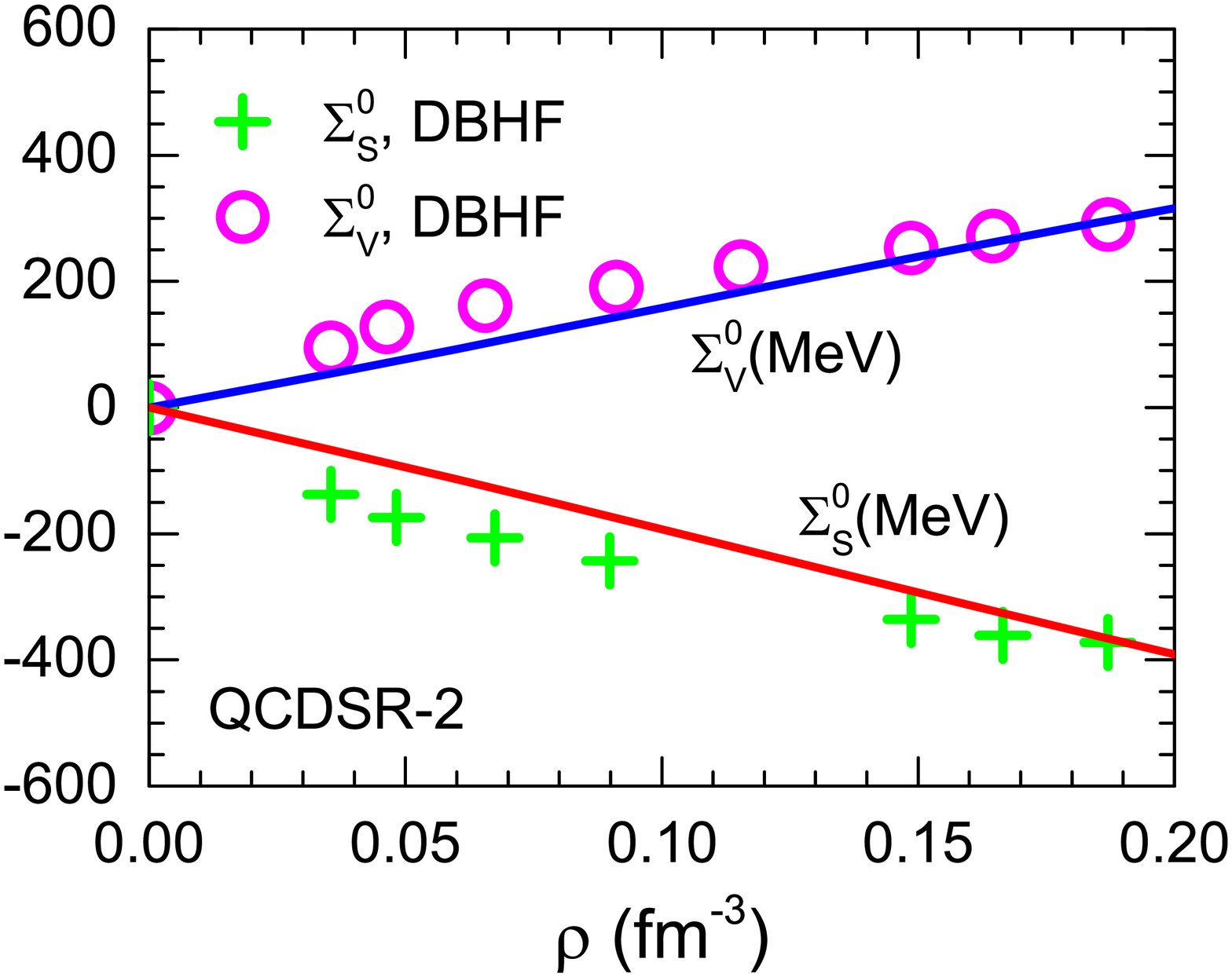}
  \caption{(Color Online) Density dependence of the nucleon self-energies in SNM obtained in QCDSR-2.}
  \label{fig_R2SESNM}
\end{figure}

In Fig.\,\ref{fig_R2SESNM}, the density dependence of the
nucleon self-energies in SNM is shown. Compared with the density behavior of the self-energies
obtained in QCDSR-1 shown in the left panel of Fig.\,\ref{fig_R1SEDirac}, the overall density behavior is largely
improved in QCDSR-2.
For instance, the $\Sigma_{\rm{S}}^0$ ($\Sigma_{\rm{V}}^0$) at $\rho_0\approx0.16\,\rm{fm}^{-3}$
is now shown to be about $-$312.3\,MeV (258.5\,MeV), which is close to the prediction
by the DBFH theories about $-$339.7\,MeV (267.7\,MeV).
Moreover, the nucleon Dirac effective mass in SNM is found to be about $M_{\rm{D}}^{\ast}(\rho_0)/M\approx0.68$,
via the $\Sigma_{\rm{S}}^0(\rho)$
since $M_{\rm{D}}^{\ast}=M+\Sigma_{\rm{S}}^0$.
Furthermore, the nucleon Landau effective mass could be obtained by the sum of $E_{\rm{sym}}^{\rm{kin}}(\rho)$,
$E_{\rm{sym}}^{\rm{mom,0,S}}(\rho)$, and $E_{\rm{sym}}^{\rm{mom,0,V}}(\rho)$, see Eq.\,(\ref{sec1_Esymxx}), consequently
$M_{\rm{L}}^{\ast}(\rho_0)/M\approx1.19$ through the decomposition of the symmetry energy (see Fig.\,\ref{fig_R2EsymDecom}).

\begin{figure}[h!]
\centering
  \includegraphics[width=8.3cm]{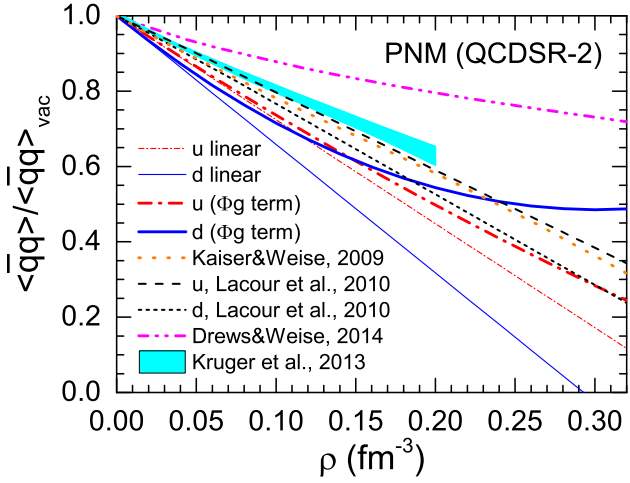}
  \caption{(Color Online) Density dependence of quark condensates in PNM from
QCDSR. Also shown are the results from ChPT\,\cite{Kru13a,Lac10,Kai09} and FRG
approach\,\cite{Dre14}. Taken from ref.\,\cite{Cai17QCDSR}.}
  \label{fig_R2QCondComparePNM}
\end{figure}

Since in QCDSR-2, the parameters $\Phi$ and $g$ are fixed, one can inversely study their effects on the chiral condensates~(\ref{chiral_cond}).
In Fig.\,\ref{fig_R2QCondComparePNM}, the density dependence of the
chiral condensates in QCDSR-2 as well as the corresponding predictions from
ChPT\,\cite{Kru13a,Lac10,Kai09} and the FRG method\,\cite{Dre14} is shown\,\cite{Cai17QCDSR}.
As demonstrated in Eq.\,(\ref{chiral_cond}) and shown in the figure, the chiral
condensate at low densities is dominated by the linear density term. More specifically, one has
$(\langle\overline{\rm{u}}\rm{u}\rangle_{\rho}
-\langle\overline{\rm{d}}\rm{d}\rangle_{\rho})/\langle\overline{q}q\rangle_{\rm{vac}}\approx
-\rho\sigma_{\rm{N}}\xi/
m_{\rm{q}}\langle\overline{q}q\rangle_{\rm{vac}}>0$ at low densities, since $\langle\overline{q}q\rangle_{\rm{vac}}$ is
negative. As density increases, the $\Phi$-term in
Eq.\,(\ref{chiral_cond}) starts to dominate and even to flip the
relative relation of the magnitude between $\langle\overline{\rm{u}}\rm{u}\rangle_{\rho}$ and
$\langle\overline{\rm{d}}\rm{d}\rangle_{\rho}$, leading to
$\langle\overline{\rm{u}}\rm{u}\rangle_{\rho}/\langle\overline{q}q\rangle_{\rm{vac}}<
\langle\overline{\rm{d}}\rm{d}\rangle_{\rho}/\langle\overline{q}q\rangle_{\rm{vac}}$
when the density is larger than about $0.15\,\rm{fm}^{-3}$. For instance,
the $\langle\overline{\rm{d}}\rm{d}\rangle_{\rho_0}/\langle\overline{q}q\rangle_{\rm{vac}}\,(\langle\overline{\rm{u}}\rm{u}\rangle_{\rho_0}/\langle\overline{q}q\rangle_{\rm{vac}})$
in PNM is found to change from 0.45 (0.56) in the linear density approximation
to 0.60 (0.59) with the inclusion of the $\Phi$-term,
leading to an enhancement of about 33\% (5\%). It is necessary to
point out that this flip is a direct consequence of the
inclusion of the higher order $\Phi$-term in Eq.\,(\ref{chiral_cond}).

More interestingly, one can find that the higher order $\Phi$-term in
Eq.\,(\ref{chiral_cond}) stabilizes the chiral condensate both for u and d
quarks at larger densities, while the leading-order linear density
approximation Eq.\,(\ref{chiral_cond}) leads to chiral symmetry restoration
at a density about $2\rho_0$\,\cite{Cai17QCDSR}.
The hindrance of the chiral symmetry restoration due to the higher order density
terms in quark condensates has important implications on the physical degrees
of freedom in the core of compact stars such as the neutron stars/quark stars where the matter is very close to the PNM.
These features are consistent with a recent analysis on the same issue based on the
FRG method\,\cite{Dre14}.

Furthermore, if one uses a different $\sigma_{\rm{N}}$, then by readjusting the values of the parameters
$f$, $\Phi$ and $g$ based on the fitting scheme (see Subsection~\ref{sb_FS}), it can
be demonstrated that the $\sigma_{\rm{N}}$ has very little influence
on the EOS of ANM.
Different $\sigma_{\rm{N}}$ leads to different
$\Phi$ and $g$, but the density dependence of the chiral condensates will change only
quantitatively, instead of qualitatively since the $\sigma_{\rm{N}}$ term (linear order)
is a perturbation to the vacuum chiral condensates, and similarly, the $\Phi$-term is a perturbation to the linear term.
It should be pointed out that the exploration on the $\sigma_{\rm{N}}$ itself is an important issue
in nuclear physics, and the exact knowledge on the $\sigma_{\rm{N}}$
will certainly help improving our understanding on the relevant aspects of the in-medium strong interaction.

\begin{figure}[h!]
\centering
  \includegraphics[width=8.5cm]{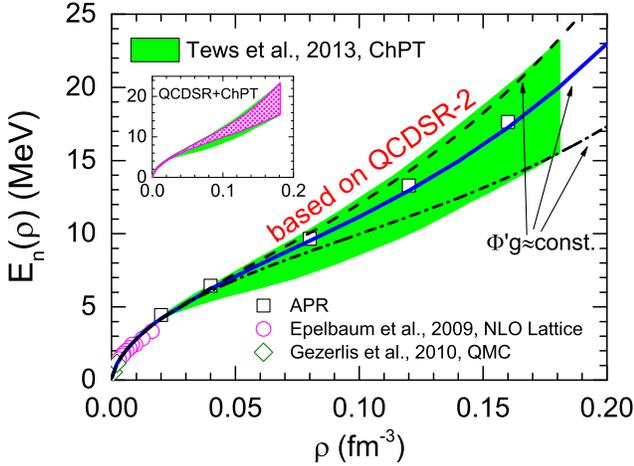}
  \caption{(Color Online) EOS of PNM with different $(\Phi',g)$
  to maximally span the green band (ChPT) shown
in Fig.\,\ref{fig_R2PNM} for the density dependence of $E_{\rm{n}}(\rho)$.
The black dash (dash-dot) line
corresponds to $ \Phi'\approx4.00,g\approx-0.552$ ($\Phi'\approx2.70,g\approx-0.817$).
See the text for details.}
  \label{fig_R2PNMcom}
\end{figure}

Now, let us vary the values of the parameters $\Phi$ and $g$, while at the same time
keep all the other parameters fixed, and meanwhile the EOS of PNM at $0.02\,\rm{fm}^{-3}$ and the symmetry
energy at $0.11\,\rm{fm}^{-3}$ are fixed by the fitting scheme (Subsection~\ref{sb_FS}).
Since $\Phi$ and $g$ essentially have no effect on $E_{\rm{n}}(\rho_{\rm{vl}})$ and the parameter $f$ is
determined by $E_{\rm{n}}(\rho_{\rm{vl}})$, thus $f\approx0.43$ remains unchanged.
Moreover, fixing of the symmetry energy at $\rho_{\rm{c}}$ indicates that $
\Phi g\approx\rm{const.}$, since the effects of $\Phi$ and $g$ on the EOS of SNM
and that of PNM are roughly given by
\begin{equation}
\int\Phi\rho^2\d\rho,~~\int\Phi(1-g)\rho^2\d\rho,
\end{equation}
and thus $\Phi$ and $g$ contribute to the symmetry energy roughly as $\int\Phi g\rho^2\d\rho$.
In Fig.\,\ref{fig_R2PNMcom}, we show the
$E_{\rm{n}}(\rho)$ with the upper (lower) black dash line
corresponding to $ \Phi'\approx4.00,g\approx-0.552$ ($\Phi'\approx2.70,g\approx-0.817$),
by maximally expanding the greed band predicted by the ChPT.
In this sense we can study the extra constraints on the $E_{\rm{n}}(\rho)$ from QCDSR.
It is at this time necessary to point out that although the error on $E_{\rm{n}}(\rho_{\rm{vl}})$ is relatively small
by the state-of-the-art microscopic many-body calculations and simulations, it still has some uncertainties (e.g., the uncertainties
generated by the nucleon-sigma term $\sigma_{\rm{N}}$). However in our scheme, the $E_{\rm{n}}(\rho_{\rm{vl}})$
is fixed at a certain value, thus the constraints on the $E_{\rm{n}}(\rho)$ given in the following paragraphs should be thought only as
a roughly estimate. On the other hand, it is useful to study the saturation properties of the EOS of SNM, since fixing the $E_{\rm{n}}(\rho_{\rm{vl}})$ actually gives
a relevant estimate on the uncertainties on the saturation density of $E_0(\rho)$ due to the higher order density terms in Eq.\,(\ref{chiral_cond}).
It is clearly seen from Fig.\,\ref{fig_R2PNMcom} that {in such a way,} a much stronger
constraint on the EOS of PNM in the density region from about $0.04\,\rm{fm}^{-3}$ to
$0.12\,\rm{fm}^{-3}$ is obtained (the inset of Fig.\,\ref{fig_R2PNMcom}). For
example, one obtains $6.8\,\rm{MeV}\,(10.0\,\rm{MeV})\lesssim
E_{\rm{n}}(0.05\,\rm{fm}^{-3})\,(E_{\rm{n}}(0.1\,\rm{fm}^{-3}))\lesssim7.3\,\rm{MeV}\,(12.1\,\rm{MeV})$,
leading to a 68\% (54\%) reduction on the uncertainties on
$E_{\rm{n}}(\rho)$ at $0.05\,\rm{fm}^{-3}\,(0.1\,\rm{fm}^{-3})$, compared with the constraints from ChPT\,\cite{Tew13,Kru13}.
These results indicate that combing the QCDSR and ChPT can significantly improve the predictions on the EOS of PNM.

\begin{figure}[h!]
\centering
  \includegraphics[width=8.5cm]{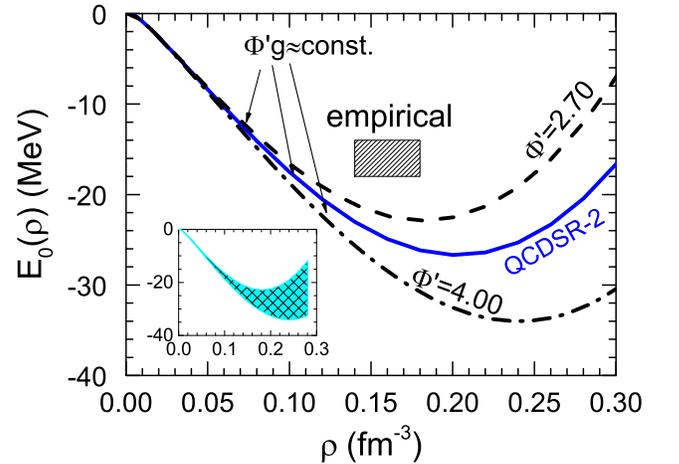}
  \caption{(Color Online) EOS of SNM with different $\Phi$ and $g$. Black dash (dash-dot) line
corresponds to $ \Phi'\approx4.00,g\approx-0.552$ ($\Phi'\approx2.70,g\approx-0.817$).}
  \label{fig_R2SNMcom}
\end{figure}

In Fig.\,\ref{fig_R2SNMcom}, the EOS of SNM $E_0(\rho)$ with different $\Phi$ and $g$ is shown.
It is interesting to find that with a smaller $\Phi'$ (or equivalently a larger $\Phi=\Phi'/\langle\overline{q}q\rangle_{\rm{vac}}
$) the saturation properties are closer to the empirical constraints.
For example, the saturation density (binding energy at that density) is found to be about $0.18\,\rm{fm}^{-3}$ ($-$22.9\,MeV) with $\Phi'\approx2.70$
and $g\approx-0.817$, introducing a relative improvement about 10\% (14\%) on $\rho_0$ ($E_0(\rho_0)$)
compared to the QCDSR-2 prediction ($0.2\,\rm{fm}^{-3}$ and $-$26.7\,MeV). On the other hand, with a larger $\Phi'\approx4.00$ and correspondingly $g\approx-0.552$,
the saturation density (the corresponding binding energy) is found to about $0.24\,\rm{fm}^{-3}$ ($-$34.0\,MeV).
The overall uncertainty on the $E_0(\rho)$ is plotted in the inset of Fig.\,\ref{fig_R2SNMcom}.
Combining the discussions given above, it shows that before the EOS of PNM and the density dependence of the chiral condensates at densities around the saturation density
are well determined, it is hard to make accurate predictions on the EOS of SNM.
Specifically, it could be found that while keeping the EOS of PNM to be consistent with the ChPT predictions within density from zero to about $0.18\,\rm{fm}^{-3}$
and meanwhile the symmetry energy at $\rho_{\rm{c}}$ fixed, the uncertainty
on the nuclear saturation density (binding energy) due to the $\Phi$-term in Eq.\,(\ref{chiral_cond}),
is about $0.06\,\rm{fm}^{-3}$ ($-$11.1\,MeV), a relative uncertainty about 38\% (69\%) compared to $\rho_0\approx0.16\,\rm{fm}^{-3}$ ($E_0(\rho_0)\approx-16\,\rm{MeV}$),
strongly indicating that the higher order terms in density in the chiral condensates
also have sizable impact on the EOS of SNM.
Naturally, if other uncertainties are included, e.g., the uncertainties on $\sigma_{\rm{N}}$,
the twist-four four-quark condensates discussed in the next section, and the uncertainties on the symmetry energy at $\rho_{\rm{c}}$,
etc., it may lead the corresponding uncertainties on $\rho_0$ and $E_0(\rho_0)$ much larger.
A detailed analysis on this issue is beyond the main scope of the present work.

\begin{figure}[h!]
\centering
  \includegraphics[width=8.5cm]{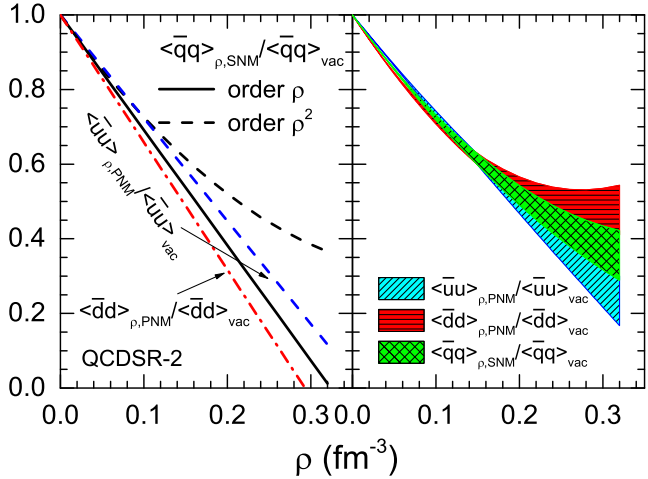}
  \caption{(Color Online) Density dependence
of the chiral condensates both in SNM and PNM.}
  \label{fig_R2QCondcom}
\end{figure}

Finally, we show in Fig.\,\ref{fig_R2QCondcom} the density dependence of the
chiral condensates both in SNM and PNM, including the uncertainties introduced
by the parameters $\Phi$ and $g$.
Besides the chiral symmetry restoration pattern discussed for u and d
quarks in PNM in the above paragraphs (see Fig.\,\ref{fig_R2QCondComparePNM}),
the nonlinear density corrections in the chiral condensates (\ref{chiral_cond})
also make the restoration of the chiral symmetry
in SNM to occur at an even higher density (green band in the right panel of Fig.\,\ref{fig_R2QCondcom}).
More interestingly, the d quark chiral condensate in PNM with the allowable ranges of $\Phi$ and $g$
may even increase at a critical density about $0.25\,\rm{fm}^{-3}$ (red band), indicating that the d quark
in PNM is very stable. Since the main component of neutron stars is the neutrons (roughly 2/3 of the components
of a neutron stars are d quarks),
the hindrance of the chiral symmetry restoration of the d quark may have important
consequences on investigating the quark-involved dynamical processes in these compact objects.

\setcounter{equation}{0}
\section{Twist-four Four-quark Condensates, Symmetry Energy and QCDSR-3}\label{SEC_TWIST4}

In this section, the effects of the twist-four four-quark
six-dimensional condensates on the EOS of ANM are studied and correspondingly the QCDSR-3 is constructed. The twist-four condensates
effects on the symmetry energy were first
studied in ref.\,\cite{Jeo13}. Since the
discussions in this section is very similar as those done in Section~\ref{SEC_HighCond}, here we mainly
focus on the related issues on the EOS of PNM, the EOS of SNM, and the symmetry energy.

The contributions to the QCDSR Eqs.\,(\ref{QSR_EQ_2a})
and (\ref{QSR_EQ_3a}) from the twist-four condensates for proton
are given by\,\cite{Jeo13},
\begin{align}
B_{\rm{tw4}}^{\rm{II}} =&-\frac{1}{4\pi\alpha_{\rm{s}}}
\frac{M}{2}\left[\Omega_1+(\Omega_2+\Omega_3\delta)
-\frac{1}{3}(\Omega_4-\Omega_5\delta)\right]\notag\\
&\times\rho L^{-4/9},\\
B_{\rm{tw4}}^{\rm{III}}=&\frac{4\overline{{e}_{\rm{p}}}}{4\pi\alpha_{\rm{s}}}
\frac{M}{2}\left[\Omega_1+(\Omega_2+\Omega_3\delta)
-\frac{1}{3}(\Omega_4-\Omega_5\delta)\right]\notag\\
&\times\rho L^{-4/9},
\end{align}
the five
parameters, i.e., $\Omega_1\sim\Omega_5 $, characterizing the
twist-four condensates together with three different parameter sets are given and discussed in ref.\,\cite{Jeo13}.
The similar contributions can be written out for the neutron by exchanging
the u and d quarks in the above expressions (i.e., the parameters $\Omega_1\sim\Omega_5$).
In the following, we call them the set 1, set 2, and set 3 parameter sets, respectively.

\begin{figure}[h!]
\centering
  \includegraphics[width=8.5cm]{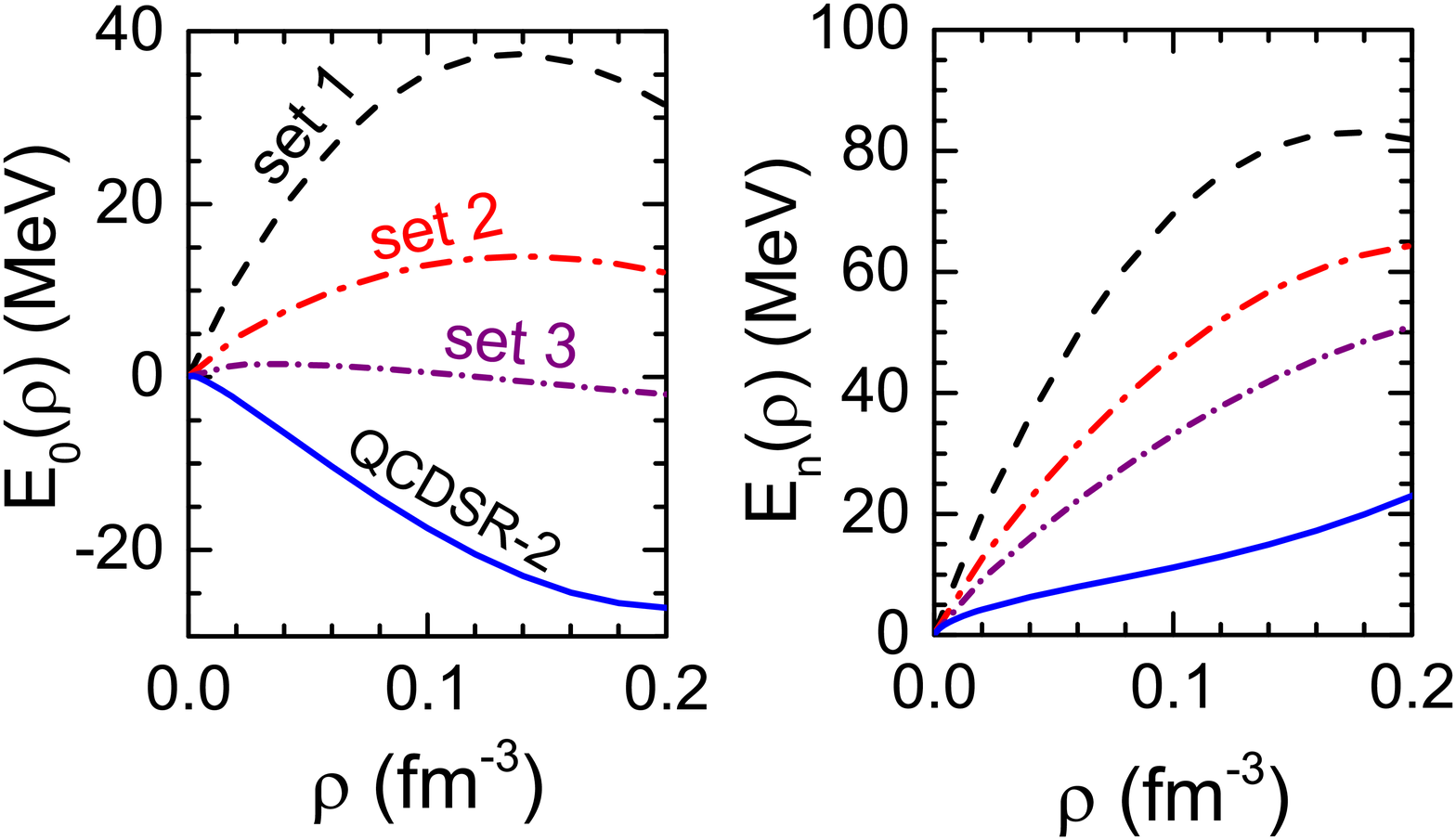}
  \caption{(Color Online) Effects of the twist-four four-quark condensates on the EOS of SNM and the EOS of PNM.}
  \label{fig_TW4SNMPNMEff}
\end{figure}

In Fig.\,\ref{fig_TW4SNMPNMEff}, we show the effects of the twist-four
condensates on the EOS of SNM and the EOS of PNM, where the other parameters are the same as those obtained in
QCDSR-2.
It is clearly shown from Fig.\,\ref{fig_TW4SNMPNMEff} that the twist-four condensates
have large impact both on the $E_0(\rho)$ and $E_{\rm{n}}(\rho)$, as first pointed out in ref.\,\cite{Jeo13}.
For instance, the $E_0(\rho)$ at $\rho_{\rm{c}}$ takes value about
36.3\,MeV (13.3\,MeV, 0.3\,MeV) in the twist-four condensate parameter set 1 (set 2, set 3) compared with the QCDSR-2 prediction about
$-$18.9\,MeV without these condensates, expanding an uncertainty of about 55.2\,MeV.
\begin{figure}[h!]
\centering
  \includegraphics[width=8.5cm]{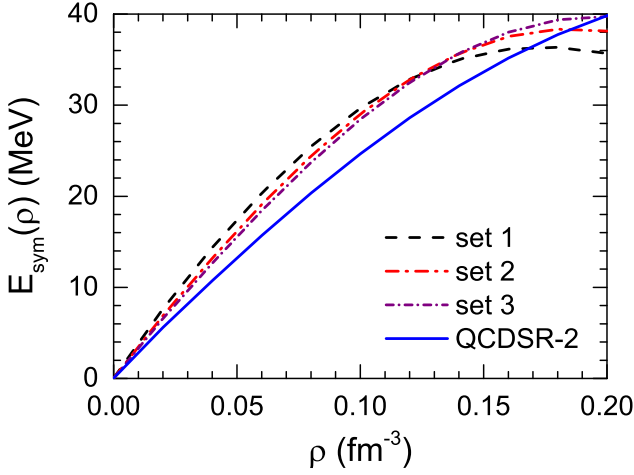}
  \caption{(Color Online) Effects of the twist-four condensates on the symmetry energy.}
  \label{fig_TW4EsymEff}
\end{figure}
Similarly, the $E_{\rm{n}}(\rho)$ at $\rho_{\rm{c}}$ takes value about 72.9\,MeV (49.2\,MeV, 35.4\,MeV)
in parameter set 1 (set 2, set 3) and 12.1\,MeV in QCDSR-2 without the twist-four four-quark condensates, introducing an uncertainty of about 60.8\,MeV.
These sizable influence on the $E_{\rm{n}}(\rho)$ and $E_0(\rho)$ is found to be consistent
with the findings in ref.\,\cite{Jeo13}.
However the symmetry energy obtained by the difference between the $E_{\rm{n}}(\rho)$ and the $E_0(\rho)$ (i.e., the parabolic approximation) directly from the results shown in Fig.\,\ref{fig_TW4SNMPNMEff}
is found to be about 36.6\,MeV (35.9\,MeV, 35.1\,MeV) and 31.0\,MeV in parameter set 1 (set 2, set 3)
and in QCDSR-2 at $\rho_{\rm{c}}$. The uncertainty on the symmetry energy from these four parameter sets is
found to be maximally about 5.6\,MeV at the cross density.
The large uncertainties on the EOS of SNM and EOS of PNM due to the twist-four four-quark
condensates are roughly canceled, leading to a relative smaller impact on the symmetry energy.
In Fig.\,\ref{fig_TW4EsymEff}, the density dependence of the symmetry energy
with the twist-four condensates included and that obtained in QCDSR-2 are shown through Eq.\,(\ref{sec1_Esym}).
Although one uses the QCDSR-2 parameters ($f$, $\Phi$ and $g$), the
density dependence of the symmetry energy including the twist-four condensates is very close
to the prediction by the QCDSR-2, within a wide range of densities.
Specifically, the $E_{\rm{sym}}(\rho_{\rm{c}})$ is found to be about 31.4\,MeV (30.9\,MeV, 30.5\,MeV) in parameter set 1 (set 2, set 3), indicating once again that the $E_{\rm{HO}}(\rho)$ is already non-negligible at densities smaller than $\rho_0$ when compared with the parabolic approximation prediction just given above.
It is also very interesting to notice from Fig.\,\ref{fig_TW4EsymEff} that
the twist-four condensates tend to soften the symmetry energy at densities larger than and/or around the saturation density.
In the following we readjust the parameters $\Phi$, $g$ and $f$ according to the fitting scheme (Subsection \ref{sb_FS}),
i.e., the EOS of PNM at $\rho_{\rm{vl}}$ to be about 4.2\,MeV, the symmetry energy at $\rho_{\rm{c}}$
to be about 26.65\,MeV, and the meanwhile the $E_{\rm{n}}(\rho)$ to be fitted with the APR EOS
as much as possible.

\begin{figure}[h!]
\centering
  \includegraphics[width=8.5cm]{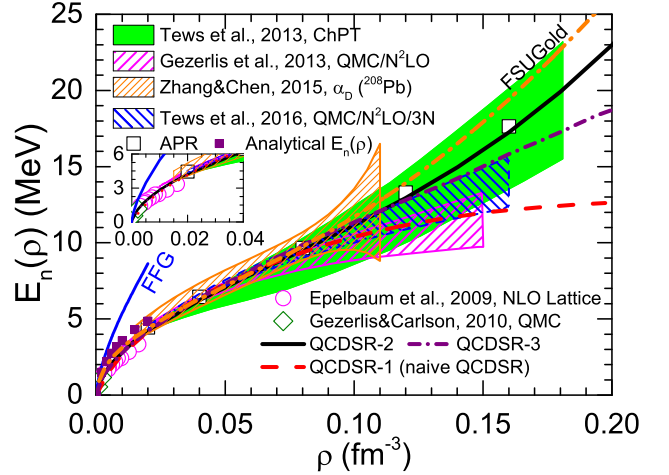}
  \caption{(Color Online) Same as Fig.\,\ref{fig_R1PNM} but for QCDSR-3. The corresponding predictions from QCDSR-1, QCDSR-2 as well as FSUGold are also included for comparison. See the text for details.}
  \label{fig_R3PNM}
\end{figure}

Based on the fitting scheme (Subsection~\ref{sb_FS}), we obtain $f\approx0.360$, $\Phi'\approx4.50$ and $g\approx-0.50$
when the twist-four four-quark condensates are included, and this parameter set
is abbreviated as QCDSR-3. In the meanwhile the Ioffe parameter is still about $-$1.22, which is unaffected
by the twist-four four-quark condensates.
Shown in Fig.\,\ref{fig_R3PNM} is the EOS of PNM obtained in QCDSR-3,
and the corresponding predictions from QCDSR-1, QCDSR-2 as well as the nonlinear RMF model with FSUGold are also included for comparison.
For reference, we note the $E_{\rm{n}}(\rho)$ at $\rho_{\rm{c}}$ ($\rho_0$) in FSUGold is about
12.8\,MeV (19.4\,MeV).
We find that the EOS of PNM from QCDSR-3
at densities $\lesssim0.12\,\rm{fm}^{-3}$ is essentially the same as the one
without the twist-four condensates (i.e., in QCDSR-2),
which is also in nice agreement with the FSUGold prediction at these densities.
And at the nuclear saturation density $\rho_0 = 0.16\,\rm{fm}^{-3}$, the $E_{\rm{n}}(\rho_0)$ changes from about 17.1\,MeV in QCDSR-2
to 15.9\,MeV in QCDSR-3\,\cite{Cai17QCDSR}, introducing a relative reduction about 7\%.
However, the discrepancy from the APR EOS becomes eventually apparent as densities increases $\gtrsim0.12\,\rm{fm}^{-3}$.
Since the high-twist operators have some impacts on several processes in hadronic
physics\,\cite{Col11,Gre07}, the exact knowledge on the density dependence of the $E_{\rm{n}}(\rho)$
provides a novel tool to study these operators.

\begin{figure}[h!]
\centering
  \includegraphics[width=8.5cm]{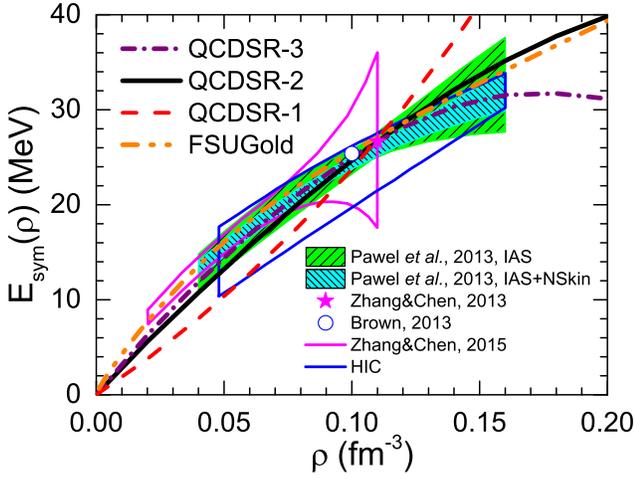}
  \caption{(Color Online) Density dependence of the symmetry energy obtained in QCDSR-3. The corresponding predictions from QCDSR-1, QCDSR-2 as well as FSUGold are also included for comparison. See the text for details.}
  \label{fig_R3Esym}
\end{figure}

We show in Fig.\,\ref{fig_R3Esym} the density dependence of the symmetry energy
obtained in QCDSR-3 as well as the corresponding predictions from QCDSR-1, QCDSR-2 and FSUGold.
The symmetry energy at densities $\lesssim0.1\,\rm{fm}^{-3}$
is improved compared with several constraints, e.g., the $E_{\rm{sym}}(\rho)$
at $0.04\,\rm{fm}^{-3}$ is found to be about 11.9\,MeV (compared with the 10.8\,MeV in QCDSR-2), which is
safely within the IAS constraints.
Moreover, the symmetry energy at $\rho_0$ now is found to be about 31.6\,MeV.
Furthermore, the slope parameter of symmetry energy $L$ at $\rho_{\rm{c}}$ is now about 49.8\,MeV,
which is consistent with the empirical constraints about $46\pm4.5\,\rm{MeV}$\,\cite{Zha13}.
The $\gamma$ parameter at $\rho_{\rm{c}}$ now is about 0.62, which safely fall within the band in Fig.\,\ref{fig_R2gamma}.
On the other hand, the $L$ parameter at the saturation density $\rho_0$ in QCDSR-3 is found to be about 16.4\,MeV.
The softening of the symmetry energy at densities $\gtrsim\rho_0$ is a possible signal
that even higher order terms in density in the condensates beyond the $\Phi$-term need to
be included effectively.

\begin{figure}[h!]
\centering
  \includegraphics[width=8.5cm]{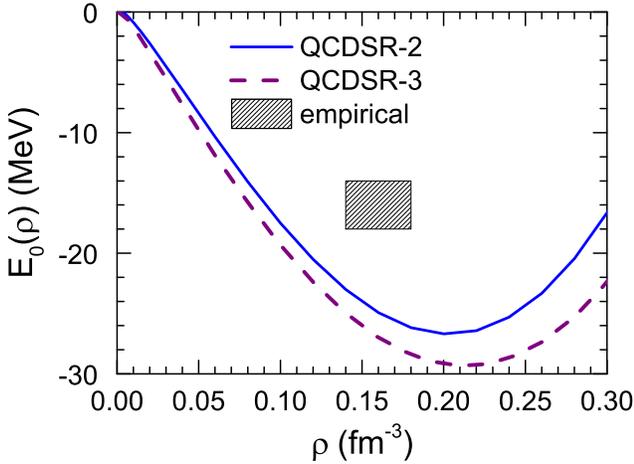}
  \caption{(Color Online) EOS of SNM obtained in QCDSR-3, the prediction on $E_0(\rho)$ from QCDSR-2 is also shown.}
  \label{fig_R3SNM}
\end{figure}

Finally, we show the density dependence of $E_0(\rho)$ obtained in QCDSR-3 in Fig.\,\ref{fig_R3SNM}.
Obviously, the saturation properties obtained in QCDSR-3 and QCDSR-2 are similar.
For instance, the saturation density (binding energy) in QCDSR-3 is now given as
about 0.21\,$\rm{fm}^{-3}$ ($-$29.0\,\rm{MeV}), which is slightly larger (deeper) than the
prediction by the QCDSR-2.
It is also interesting to notice that the uncertainties introduced by the twist-four four-quark condensates
on the saturation properties of the SNM could be covered by the effective ranges of $\Phi$ and $g$ discussed in the last section
(see Fig.\,\ref{fig_R2SNMcom}) although the origins of the uncertainties are different, showing that the effects of the twist-four condensates
on the saturation properties of the SNM are smaller than the higher order density terms in the chiral condensates.
This again shows the particular importance of the $\Phi$-term in the chiral condensates in Eq.\,(\ref{chiral_cond}).
Moreover, other characteristic quantities for the EOS of ANM are also obtained in QCDSR-3, e.g.,
the nucleon Landau/Dirac effective mass $M_{\rm{L/D}}^{\ast}(\rho_0)/M$ in SNM at $\rho_0$ ($\rho_{\rm{c}}$)
is found to be about 1.10/0.67 (1.21/0.74), while the high order effects in the EOS of ANM $E_{\rm{HO}}(\rho)$
is found to be about 10.9\,MeV (6.1\,MeV) at $\rho_0$ ($\rho_{\rm{c}}$).
These features together show again that the parameter set QCDSR-3 is very similar like the QCDSR-2.

\setcounter{equation}{0}
\section{Summary and Outlook}\label{SEC_Summary}

In this work, we have systematically investigated the EOS of isospin asymmetric nucleonic matter
within the framework of QCDSR, and mainly focused on the relativistic self-energy decomposition of the nuclear symmetry
energy and the EOS of PNM.
Based on the fitting scheme that the EOS of PNM at $\rho_{\rm{vl}}=0.02\,\rm{fm}^{-3}$ and the symmetry energy $E_{\rm{sym}}(\rho)$
at $\rho_{\rm{c}}=0.11\,\rm{fm}^{-3}$ are fixed at 4.2\,MeV and $26.65\pm0.2$\,MeV, respectively, and the total $E_{\rm{n}}(\rho)$
to densities about $\rho_0$ is fitted to the APR EOS as much as possible, several interesting results are obtained as following:

1. In the conventional QCDSR, the prediction on the nucleon mass in vacuum is
not necessarily about 939\,MeV\,\cite{Coh95}. In this work the nucleon mass in vacuum is self-consistently determined
via $M_{\rm{D}}^{\ast}(0)\equiv M_0^{\ast}(0)=M\equiv939\,\rm{MeV}$,
leading to the Ioffe parameter $t$ to be about $-1.22$ (which is close to its natural value $t_{\rm{Ioffe}}=-1$) in three QCDSR parameter sets,
and more specifically the Ioffe parameter expressed in terms of the chiral condensate and the gluon condensate in vacuum is obtained, see Eq.\,(\ref{def_Ioffe}).
This paves an important step to the consequent investigations on the EOS.
For instance, the EOS of SNM via $E_0(\rho)=\rho^{-1}\int_0^{\rho}\d\rho[e_{\rm{F}}^{\ast}(\rho)+\Sigma_{\rm{V}}^0(\rho)]-M$ needs an accurate nucleon mass in vacuum.

2. The Lorentz structure based on the nucleon self-energy decomposition of the symmetry energy is
carefully explored. Specifically,
the first-order symmetry scalar self-energy is found to
depend heavily on the nucleon sigma term $\sigma_{\rm{N}}$.
For instance,
in the msQCDSR, this term is given by
$E_{\rm{sym}}^{\rm{1st,S}}(\rho)
=2\pi^2M_0^{\ast}\sigma_{\rm{N}}\alpha\rho/(\mathscr{M}^2m_{\rm{q}}e_{\rm{F}}^{\ast}\beta)$,
where $\alpha/\beta$ characterizes the isospin effects of the chiral condensates at linear order.
This relation establishes a useful connection between the symmetry energy and the
evolution of the quark mass since the nucleon sigma term actually characterizes the evolution of the nucleon mass
as a function of the light quark mass, i.e.,
$\sigma_{\rm{N}}=m_{\rm{q}}\d M/\d m_{\rm{q}}$.
Consequently, the $\sigma_{\rm{N}}$ term is found to largely affect $E_{\rm{sym}}^{\rm{1st,S}}(\rho)$
since $y=M_0^{\ast}/e_{\rm{F}}^{\ast}\approx1-x^2/2$ with $x=k_{\rm{F}}/M_0^{\ast}$, where effective mass $M_0^{\ast}
=M+\Sigma_{\rm{S}}^0$ depends on the $\sigma_{\rm{N}}$ almost linearly, and the final effect is as $\sigma_{\rm{N}}$
increases the $E_{\rm{sym}}^{\rm{1st,S}}(\rho)$ decreases.
Moreover, the kinetic nuclear symmetry energy
$E_{\rm{sym}}^{\rm{kin}}(\rho)=k_{\rm{F}}^2/6e_{\rm{F}}^{\ast}$ is also found to be largely
affected by the $\sigma_{\rm{N}}$ since $e_{\rm{F}}^{\ast}=(M_0^{\ast,2}+k_{\rm{F}}^2)^{1/2}$
could be approximated as $M_0^{\ast}+k_{\rm{F}}^2/2M_0^{\ast}=M+\Sigma_{\rm{S}}^0+k_{\rm{F}}^2/2M_0^{\ast}$,
i.e, $E_{\rm{sym}}^{\rm{kin}}(\rho)$ increases as $\sigma_{\rm{N}}$ increases.
Finally, as the $\sigma_{\rm{N}}$ increases, the total symmetry energy $E_{\rm{sym}}(\rho)$
increases as a result of the competition between the $E_{\rm{sym}}^{\rm{kin}}(\rho)$ and the $E_{\rm{sym}}^{\rm{1st,S}}(\rho)$, see the left panel of Fig.\,\ref{fig_EsymSignMs}.

3. The vector nucleon self-energy contributes to the nuclear symmetry energy as, e.g., $E_{\rm{sym}}^{\rm{1st,V}}(\rho)
=4\pi^2\rho/\mathscr{M}^2$, in the msQCDSR, and this part is originated from the densities of d quarks ($\langle \rm{d}^{\dag}\rm{d}\rangle_{\rho,\delta}=(3+\delta)\rho/2$) and u quarks ($\langle \rm{u}^{\dag}\rm{u}\rangle_{\rho,\delta}=(3-\delta)\rho/2$).
The linear density dependence is almost unchanged in the fQCDSR via the very complicated QCDSR equations.
Moreover, the $E_{\rm{sym}}^{\rm{1st,V}}(\rho)$ becomes the dominant contribution
to the symmetry energy as $\rho\gtrsim 0.1\,\rm{fm}^{-3}$.

4. Contributions to the nuclear symmetry energy due to the momentum
dependence of the nucleon self-energies are found to be
$E_{\rm{sym}}^{\rm{mom,0,S}}(\rho)\sim[\d\Sigma_{\rm{S}}^0/\d|\v{k}|]_{k_{\rm{F}}}<0$
and
$E_{\rm{sym}}^{\rm{mom,0,V}}(\rho)\sim[\d\Sigma_{\rm{V}}^0/\d|\v{k}|]_{k_{\rm{F}}}>0$, at densities smaller or around $\rho_0$.
Moreover, the magnitude of these terms is much smaller than the
kinetic part and the first order symmetry parts, showing the corresponding momentum
dependence in QCDSR is weak.

5. The dependence of the symmetry energy on the strange quark
mass is found to be large.
This large effect can be understood via $E_{\rm{sym}}^{\rm{1st,S}}(\rho)
=2\pi^2M_0^{\ast}\sigma_{\rm{N}}\alpha\rho/(\mathscr{M}^2m_{\rm{q}}e_{\rm{F}}^{\ast}\beta)$ in the msQCDSR since a larger strange quark
mass induces a smaller factor $\alpha/\beta$, and consequently a smaller $E_{\rm{sym}}^{\rm{1st,S}}(\rho)$.
The connection between the strange quark mass and the symmetry energy
provides a useful bridge to understand the possible origins of the uncertainties
on the $E_{\rm{sym}}(\rho)$, see the right panel of Fig.\,\ref{fig_EsymSignMs} and Fig.\,\ref{fig_EsymDecomMs}.

6. A few useful approximations for the EOS of ANM and the related quantities are obtained in the QCDSR, and among which the approximation
for the EOS of PNM, i.e., $E_{\rm{n}}(\rho)
\approx E_{\rm{n}}^{\rm{FFG}}(\rho)+({M\rho}/2{\langle\overline{q}q\rangle_{\rm{vac}}})[(1-\xi)({\sigma_{\rm{N}}}/{2m_{\rm{q}}})-5]
$, already has quantitative predictive power at low densities, where $E_{\rm{n}}^{\rm{FFG}}(\rho)$
is the FFG EOS of PNM.
This low density formula clearly demonstrates how the chiral condensate goes into play in the EOS
of PNM, e.g., the second term characterized by several constants
($\xi,\sigma_{\rm{N}},m_{\rm{q}}$ and
$\langle\overline{q}q\rangle_{\rm{vac}}$) leads to a
reduction on the $E_{\rm{n}}(\rho)$ compared to the FFG prediction.

7. The high order term in the EOS of ANM characterized by the $E_{\rm{HO}}(\rho)\approx E_{\rm{sym,4}}(\rho)+
E_{\rm{sym,6}}(\rho)+\cdots$ is carefully studied in the QCDSR. Specifically, $E_{\rm{HO}}(\rho)$
is found to be sizable at densities $\lesssim\rho_0$, indicating the conventional
parabolic approximation for the EOS of ANM is broken in QCDSR.
For instance, the $E_{\rm{HO}}(\rho)$ at the saturation
density in the fQCDSR calculations with the fitting scheme adopted is found to be about 1.1$\sim$11.9\,MeV, indicating
the uncertainty on $E_{\rm{HO}}(\rho)$ in QCDSR is essentially large compared with the one from phenomenological models.

8. The correlation between the symmetry energy and several quantities
characterizing the quark/gluon condensates is investigated.
Specifically, besides the strong dependence on the chiral condensates in vacuum $\langle\overline{q}q\rangle_{\rm{vac}}$,
the symmetry energy is also found to heavily depend on the five-dimensional mixing
condensate $\langle
g_{\rm{s}}q^{\dag}\sigma \mathcal{G}q\rangle_{\rho,\delta}$. This correlation provides a novel tool to explore
the properties of symmetry energy and even the EOS of ANM via the knowledge on the in-medium quark/gluon condensates from, e.g., hadronic physics.

9. The effects of twist-four four-quark six-dimensional condensates
on the EOS of SNM and the EOS of PNM are found to be large.
For instance, these condensates induce an amount about 50\,MeV to 60\,MeV
on the $E_0(\rho)$ and $E_{\rm{n}}(\rho)$. However, their effects on the symmetry
energy are almost canceled since the $E_{\rm{sym}}(\rho)$ is roughly
the difference between the $E_{\rm{n}}(\rho)$ and $E_0(\rho)$, and as a result the twist-four condensates
induces an uncertainty of about several MeVs on the symmetry energy.

10. The effective higher order terms in density in the chiral condensates, i.e., $\Phi(1\mp g\delta)\rho^2$, are found to strongly
affect the EOS of PNM and the EOS of SNM. By refitting the model parameters in the presence
of the $\Phi$-term, the EOS of PNM at densities around $\rho_0$ is found to be systematically consistent
with the APR EOS, which is selected as the reference EOS in this work.
Moreover, the higher-order density terms in
quark condensates also leads to the stabilization of u/d chiral condensates at higher
densities, which may have important implications
on the QCD phase diagram under extreme conditions at low temperatures, large isospin and large baryon
chemical potentials, which is essential for understanding the physical
degrees of freedom in the core of neutron stars.
Furthermore, using a different higher order density term such as $\Phi(1\mp g\delta)\rho^{5/3}$ gives a very similar
prediction on the $E_{\rm{n}}(\rho)$, see Fig.\,\ref{fig_R2PNM}.

11. Three parameter sets of QCDSR are constructed, i.e.,
QCDSR-1 (naive QCDSR), QCDSR-2 and QCDSR-3, respectively.
The QCDSR-1 includes only the linear approximation for the chiral condensates
without the effective $\Phi$-term in the chiral condensates (\ref{chiral_cond}) and the twist-four four-quark condensates.
Compared with the QCDSR-1, the QCDSR-2 additionally includes the effective $\Phi$-term in the chiral condensates (\ref{chiral_cond}) but without the twist-four four-quark condensates.
The QCDSR-3 includes the linear approximation for the chiral condensates, the effective $\Phi$-term in the chiral condensates (\ref{chiral_cond}) and the twist-four four-quark condensates.
The symmetry energy at $\rho_0\approx0.16\,\rm{fm}^{-3}$ in QCDSR-1, QCDSR-2 and QCDSR-3 is found to be about 45.1\,MeV, 35.2\,MeV and 31.6\,MeV, respectively, while the corresponding slope parameter of the symmetry energy $L$ at $\rho_{\rm{c}}$ ($\rho_0$) is found to be 105.9 (196.8)\,MeV, 64.7 (67.5)\,MeV, and 49.8 (16.4)\,MeV, respectively.
The tendency
of the change in $E_{\rm{sym}}(\rho)$ and $L(\rho)$
shows that the higher order terms in density in the chiral condensates improve the density behavior
of the symmetry energy compared with the empirical constraints,
while the twist-four condensates soften the $E_{\rm{sym}}(\rho)$ at densities larger than about $\rho_{\rm{c}}$.
Moreover, the saturation properties of SNM are largely improved from QCDSR-1 to QCDSR-2 or QCDSR-3, e.g.,
the ($\rho_0,E_0(\rho_0)$) are changed from $(0.6\,\rm{fm}^{-3},-100.0\,\rm{MeV})$ in QCDSR-1
to $(0.2\,\rm{fm}^{-3},-26.7\,\rm{MeV})$ in QCDSR-2 or $(0.21\,\rm{fm}^{-3},-29.0\,\rm{fm}^{-3})$
in QCDSR-3.
Furthermore, the EOS of PNM obtained in three QCDSR parameter sets is consistent with each other at low
densities less than about $0.08\,\rm{fm}^{-3}$, indicating that at these low densities the naive QCDSR is
well-behaved for the $E_{\rm{n}}(\rho)$.
Finally, we have better constrained the
$E_{\rm{n}}(\rho)$ in the density region from about $0.04\,\rm{fm}^{-3}$ to $0.12\,\rm{fm}^{-3}$ by
combining the results from QCDSR and
ChPT, e.g., the
$E_{\rm{n}}(0.05\,\rm{fm}^{-3})\,(E_{\rm{n}}(0.1\,\rm{fm}^{-3}))$ is
constrained to be between 6.8\,MeV\,(10.0\,MeV) and
7.3\,MeV\,(12.1\,MeV), leading to an uncertainty about 0.5\,MeV and 2.1\,MeV, respectively.

Besides the above results we have obtained from the QCDSR, a few interesting issues that are closely related to our present work should be pointed out and need further exploration in the future QCDSR calculations on dense nucleonic matter, i.e.,

1. In this work as well as in many conventional QCDSR calculations\,\cite{Coh95}, the four-quark condensates effects
are incorporated by the effective parameter $f$, i.e., using $
(1-f)\langle\overline{q}q\rangle_{\rm{vac}}^2
+f\langle\overline{q}q\rangle_{\rho,\delta}^2$ to account for the four-quark condensates
at finite densities. In this work, the $f$ is found to largely influence several quantities, such as the EOS of PNM and the $E_0(\rho)$, and the value of $f$ is essentially determined by the $E_{\rm{n}}(\rho)$
at the very low density $\rho_{\rm{vl}}\approx0.02\,\rm{fm}^{-3}$ in the present work.
From the more fundamental viewpoint, it is important to explore the density behavior
of the four-quark condensates, in order to make further progress in applying QCDSR to
dense nucleonic matter calculations.
For instance, in ref.\,\cite{Tho07}, more phenomenological parameters are introduced into
the QCDSR equations, and they are determined by nuclear quantities and/or other information
from, e.g., hadronic physics.

2. The three-body forces (TBF) are found to be important for the saturation properties of SNM,
e.g., in Brueckner--Hartree--Fock (BHF) calculations (see, e.g., ref.\,\cite{ZHLi06}) and in phenomenological approach.
For instance, in the SHF model,
a traditional two-body force contributes a term proportional to $\rho$ to the EOS, and a $\rho^{1+\alpha}$
term emerges once the effective three-body force is considered\,\cite{Zha16xx} with $\alpha$ the parameter characterizing the three-body force.
Recently, three-body forces even the four-body forces are included in the QCDSR calculations
for nucleonic matter\,\cite{Dru17a,Dru17b,Dru19}, and it is really also interesting and important
to see how these many-body forces influence, for example, the $E_{\rm{n}}(\rho)$ and/or the nuclear symmetry energy.
Investigations on these problems will help us better understanding
the origins of the uncertainties on the symmetry energy and/or the difficulties
to produce reasonable saturation properties of the symmetric matter, and
they are also important for making further progresses in the nucleonic matter calculations,
such as to explore the incompressibility property of the ANM with any isospin asymmetry, and/or the single nucleon optical potential\,\cite{Ham90,Hol16xx,LiX13,LiX15}.
It is also interesting to investigating the three-body force and its connection to the higher order density
terms in the chiral condensates.

3. The Borel transformation with the Borel mass $\mathscr{M}$ as a real parameter is the standard treatment in QCDSR to deal with the high order states including the continuum excitations. Recently, a generalization to the complex-valued Borel mass $\mathscr{M}$ was introduced in ref.\,\cite{Ara14} (see also ref.\,\cite{Iof01}).
It was demonstrated that the complex-valued sum rules approach allows one to extract the spectral function with a significantly improved resolution, and thus provides a useful tool to study more detailed structures of the hadronic spectrum\,\cite{Ara14}. To our purpose, it would be interesting to investigate whether the complex-valued Borel transformation could improve the calculations on the EOS of dense nucleonic matter in the high density region where the high mass-dimensional condensates and continuum effects are important.

4. Finally, the neutron matter at sub-saturation even to very low densities
composed of spin-down and -up neutrons with a large s-wave
scattering length shows several universal properties\,\cite{Tan08},
such as the simplicity of its EOS characterized by a few universal
parameters\,\cite{CaiLi15a,Sch05,Kru15,Kol16,ZhaNB17}. Moreover, the high momentum
tail above the Fermi surface of the single nucleon momentum
distribution function in cold PNM is also found to be very similar
as that in ultra-cold atomic Fermi gases\,\cite{Hen14} although the
magnitude of the density in the two systems differs by an
amount about 25 orders\,\cite{Hen15}.
Naturally, the cold PNM at low densities
provides a perfect testing bed to explore novel ideas in the
unitary region\,\cite{Gio08,Blo08}, helping to find
deep physical principles behind these quantum many-body
systems in this so-called unitary region\,\cite{Zwe12}.
Recently, the sum rule approach with the help of maximum entropy (ME) method
was applied to investigate the imaginary part of the particle self-energy
in the unitary Fermi gas\,\cite{Gub15,Gub18}.
Thus it is interesting to explore, e.g., the imaginary part of the neutron self-energies in PNM under the QCDSR+ME method,
which will be extremely useful for exploring the transport properties of the PNM, or even
to generalize the method to a general isospin asymmetric nucleonic  matter to better understanding the quantum many-body properties
of the system.

Our results in the present work have demonstrated that the QCDSR approach can be used to explore the properties of ANM
in a quantitative manner, at least in lower density region.
The QCDSR approach establishes a bridge connecting the EOS of ANM and the non-perturbative QCD vacuum, and thus provides a useful way to understand the properties of dense nucleonic matter from non-perturbative QCD vacuum. These studies are helpful to investigate the QCD origins about the uncertainties of nucleonic matter properties, e.g., the uncertainties of the symmetry energy.
On the other hand, the exact knowledge on the EOS of ANM extracted from experiments, observations and model-independent calculations is also very useful for understanding the quark/gluon condensates in nuclear medium, which can provide important information on the chiral symmetry restoration phase transition in nuclear matter as well as the in-medium effects of hadron properties.

\section*{ACKNOWLEDGEMENTS}
This work was supported in part by the National Natural Science
Foundation of China under Grant Nos. 11625521 and 11805118, the Major State Basic Research
Development Program (973 Program) in China under Contract No.
2015CB856904, the Program for Professor of Special Appointment (Eastern
Scholar) at Shanghai Institutions of Higher Learning, Key Laboratory
for Particle Physics, Astrophysics and Cosmology, Ministry of
Education, China, and the Science and Technology Commission of
Shanghai Municipality (11DZ2260700).


\begin{references}
\bibitem{Footnotexx}
A similar expression for the nucleon mass in vacuum was obtained by Ioffe\,\cite{Iof81}
\[
M=\frac{2a\mathscr{M}^4(1+m_0^2/2\mathscr{M}^2)}{\mathscr{M}^6+
4a^2/3},\] where $m_0$ is another parameter. Ioffe
assumed that if $M=\mathscr{M}$ then the above expression gives
\[
M=\left[(2\pi)^2\langle\overline{q}q\rangle_{\rm{vac}}^2
\left[1+\chi+\sqrt{(1+\chi)^2-\frac{4}{3}}\right]\right]^{1/3}\]
with $\chi={m_0^2 }/{2M^2}$.
\bibitem{Footnote3} This relation itself gives an estimate for the nucleon mass,
\[
M=-\frac{24\mathscr{M}^4\pi^2\langle\overline{q}q\rangle_{\rm{vac}}}{3\mathscr{M}^6+64\pi^4
\langle\overline{q}q\rangle_{\rm{vac}}^2},\]
using, e.g., $\mathscr{M}^2=1.05\,\rm{GeV}^2$ and $\langle \overline{q}q\rangle_{\rm{vac}}
=-(252\,\rm{MeV})^3$, we find $M\approx824\,\rm{MeV}$.
However, in this estimate, the Ioffe parameter $t=-1$ is used for qualitative illustration, the obtained $M$
may have deviation from the physical one,  see Eq.\,(\ref{for_Ioffe_1}) and the discussion in Section~\ref{SEC_MassVacuum}.
\end{references}
\end{document}